\documentclass[aps,amsfonts,nofootinbib,preprintnumbers,nobalancelastpage]{revtex4}
\usepackage[english]{babel}
\usepackage{amsmath,amssymb}
\usepackage{graphicx}
\usepackage[sort&compress]{natbib}
\usepackage{bbm}
\usepackage{color}
\usepackage{footnote}
\usepackage{slashed} 
\usepackage{pdflscape} 
\usepackage{multirow}
\usepackage{multirow}
\usepackage{epsfig}
\usepackage{soul} 
\usepackage{comment}

\makeatletter
\def\endfmffile{%
  \fmfcmd{\p@rcent\space the end.^^J%
          end.^^J%
          endinput;}%
  \if@fmfio
    \immediate\closeout\@outfmf
  \fi
  \IfFileExists{\thefmffile.mp}{\immediate\write18{mpost \thefmffile}}{}
  \let\thefmffile\relax
}
\makeatother
\usepackage[normalem]{ulem}

\newcommand {\blue} {\color{blue}}

\newcommand{\ie} {{\it i.e.}}

\newcommand {\beq} {\begin{equation}}
\newcommand {\eeq} {\end{equation}}
\newcommand {\bea} {\begin{eqnarray}}
\newcommand {\eea} {\end{eqnarray}}

\newcommand{\cf}{{\it cf. }}
\newcommand{\eg}{{\it e.g. }}

\definecolor{red1}{cmyk}{0,1,1,0.1}
\definecolor{blue1}{cmyk}{1,0,0,0}

\newcommand{\ignore}[1]{}
\newcommand{\nn}{\nonumber} \renewcommand{\bf}{\textbf}

\newcommand{\MET}{\sl{E_T}\,\,\,}

\newcommand{\GeV}{{\rm\ GeV}}

\newcommand{\TeV}{{\rm\ TeV}}
\newcommand{\fb}{{\rm\ fb}}

\def\sl#1{#1 \!\!\! \!\!   \!\! \slash}

\allowdisplaybreaks

\begin{document}

\begin{titlepage}
\begin{flushright}
CP3-15-21
\end{flushright}
\vskip 1.0cm

\begin{center}
{\Large Search Strategies for TeV Scale Fermionic Top Partners with Charge 2/3}\\
 \vskip 1.0cm
{\large Mihailo Backovi\'{c}}\\
{\it  Center for Cosmology, Particle Physics and Phenomenology - CP3,\\ Universite Catholique de Louvain, Louvain-la-neuve, Belgium}
 \vskip 0.5cm
{\large Thomas Flacke}\\
{\it  Department of Physics, Korea Advanced Institute of Science and Technology, \\
335 Gwahak-ro, Yuseong-gu, Daejeon 305-701, Korea}
 \vskip 0.5cm
{\large Jeong Han Kim}\\
{\it  Department of Physics, Korea Advanced Institute of Science and Technology, \\
335 Gwahak-ro, Yuseong-gu, Daejeon 305-701, Korea and\\
Center for Theoretical Physics of the Universe, IBS, Daejeon, Korea}
 \vskip 0.5cm 
{\large Seung J. Lee}\\
{\it  Department of Physics, Korea Advanced Institute of Science and Technology, \\
335 Gwahak-ro, Yuseong-gu, Daejeon 305-701, Korea and\\
School of Physics, Korea Institute for Advanced Study, Seoul 130-722, Korea}
\vskip 1.0cm
{\it E-mail:}\\ 
{\blue
\href{mailto:mihailo.backovic@uclouvain.be}{mihailo.backovic@uclouvain.be}, \href{mailto:flacke@kaist.ac.kr}{flacke@kaist.ac.kr}, \href{mailto:jeonghan.kim@kaist.ac.kr}{jeonghan.kim@kaist.ac.kr}, \href{mailto:sjjlee@kaist.ac.kr}{sjjlee@kaist.ac.kr}  }\\
\vskip 1.0cm
\textbf{Abstract}
\end{center}
 \vskip 0.5cm  
 
 Searches for fermionic top partners at the TeV scale will bring forward a new final state kinematic regime and event topologies,  where Run I search strategies will inevitably fail.
We propose concrete search strategies for singly produced charge 2/3 fermionic top partners ($T'$) adequate for LHC Run II. Our analysis spans over all of the $T'$ decay modes ($i.e.$ $tZ$, $th$ and $Wb$) where we present detailed discussion of the search performances, signal efficiencies and backgrounds rates. Our LHC Run II search proposals utilize signatures with large missing energy and leptons, as well as jet substructure observables for tagging of boosted heavy SM states, customized $b$-tagging tactics  and forward jet tagging. 
We analyze the prospects for discovery and exclusion of  $T'$ models within the framework of partially composite quarks at the LHC Run II. Our results show that the LHC Run II has good prospects for observing $T'$ models which predict single production cross section of $\sigma_{T'} \sim 70 -140 \, (30 -65) \fb$ for $M_{T'} = 1\, (1.5) \TeV$ respectively with 100 $\fb^{-1}$ of integrated luminosity, depending on the branching ratios of the $T'$. Similarly, we find that cross sections of $\sigma_{T'} \sim 27 -60\, (13 - 24) \fb$ for $M_{T'} = 1\, (1.5) \TeV$ respectively can be excluded with the same amount of data. Our results are minimally model dependent and can be applied to most $T'$ models where $\Gamma_{T'} \ll M_{T'}$.

\end{titlepage}

\newpage
\setcounter{tocdepth}{3}
\setcounter{page}{1}
\section{Introduction}\label{sec:intro}

Naturalness has long been one of the main guiding principles of theoretical particle physics. While there is nothing fundamentally wrong with the concept of finely tuned Higgs mass, it is theoretically difficult to understand why the Higgs particle itself has the mass of $\sim 100 \GeV$ ($i. e.$ the Hierarchy Problem). 
The two most common approaches to address the Higgs mass Hierarchy Problem within the framework of Naturalness are either to introduce Supersymmetry (SUSY) into the Standard Model (SM), which protects the Higgs mass from large corrections, or to make the Higgs boson a composite state, so that above some scale $\Lambda$ the Higgs mass becomes irrelevant \footnote{Here we do not consider twin Higgs models~\cite{Chacko:2005pe} (a.k.a. neutral naturalness~\cite{Craig:2014aea}), which can be implemented in both supersymmetric and composite setups~\cite{Geller:2014kta,Burdman:2014zta,Craig:2014roa,Craig:2015pha,Barbieri:2015lqa,Low:2015nqa} to address the Little Hierarchy Problem and hide colored top partners by pushing their scale higher without worsening the level of fine-tuning.}. Both in Supersymmetry and Composite Higgs models the existence of top partners is crucial for natural solutions to the Hierarchy Problem, the search for which present an important aspect of the LHC program. 

The prospects for collider phenomenology in SUSY and Composite Higgs scenarios are qualitatively quite different. The results of  LHC Run I searches for scalar top partners pushed the mass limits on most SUSY models to $ m_{\tilde{t}} \gtrsim 600-700 \GeV$ \cite{Aad:2015iea}, and hence severely constrained the available phase space of natural SUSY models. While the prospect of LHC Run II to test SUSY at higher mass scales seems limited  with a modest improvement over the existing searches, the prospects of LHC Run II to probe Composite Higgs models at the TeV scale seem quite promising.

Composite Higgs models were already constrained from the LEP and Tevatron data, with the resulting limits on symmetry breaking scale $f \gtrsim  800 \GeV$~\cite{Grojean:2013qca,Ciuchini:2013pca} for the case of Higgs as a pseudo-Nambu-Goldstone boson. As the mass of states in composite theories is typically expected to be $\gtrsim f$, it is hence not surprising no fermionic top partner has been observed with mass less than the scale $f$. Past studies of ATLAS \cite{ATLASnotes1,Aad:2014efa,Aad:2015kqa,Aad:2015gdg} and CMS \cite{Chatrchyan:2013uxa,Khachatryan:2015axa,CMS:2014rda,CMS:1900uua,CMS:2014dka}  established bounds on mass of the vector-like top partners, excluding states with mass lighter than $\sim 700 - 800$ GeV with the precise bound depending on the top partner branching fractions. The studies focused on pair production of charge 2/3 vector-like top partners ($T'$) with subsequent decays of $T'$ into 3rd generation quarks and  $h$, $W$, or $Z$ bosons. Bounds from single-$T'$ searches also exist~\cite{Aad:2014efa}, and result in similar bounds~\footnote{The bounds on single production of $T'$ depend not only on the $T'$ mass and its branching ratios but also on the size of the electroweak couplings of the $T'$ to SM quarks which themselve depend on the detailed model realization and Beyond the Standard Model (BSM) coupling parameters. }. Just like SUSY scenarios, the Composite Higgs models with top partners at the TeV scale suffer from a higher degree of fine tuning. Yet, phenomenologically, these models are very interesting both because of the limits on the scale of compositness, as well as the LHC Run II has great prospects to probe these models well into the TeV mass regime ~\cite{DeSimone:2012fs, Azatov:2013hya, Backovic:2014uma, Matsedonskyi:2014mna,Backovic:2015lfa}.

As pointed out in Refs.~\cite{DeSimone:2012fs, Azatov:2013hya, Backovic:2014uma, Matsedonskyi:2014mna,Backovic:2015lfa}, single production of top partners starts to dominate over pair production for  $M_{T'} \gtrsim 1 \TeV, $ (in most of the  parameter space of typical Composite Higgs models), making this channel of utmost importance for the upcoming LHC Run II. In addition, single production of $T'$ presents a unique opportunity to study the Composite Higgs models, as the unique event topology of singly produced fermionic top partners offers a myriad of useful handles on large SM backgrounds. 

Past effort to study the phenomenology of charge 5/3 fermionic top  partners ~\cite{Azatov:2013hya, Backovic:2014uma},  as well as studies of charge 2/3 production ~\cite{Reuter:2014iya,Backovic:2015lfa}  have pointed out the major differences in the search strategies for fermionic top partners at LHC Run II, compared to the previous searches of Run I. The exploration of the TeV mass scale comes with a different kinematic regime compared to the existing searches, where we expect the TeV scale top partners to decay exclusively into highly boosted heavy SM states with $p_T \approx M_{T'} /2 \sim O(100) \GeV$. The event selection strategies suitable for non-boosted Run I analyses will inevitably fail at LHC Run II and new (boosted event topology appropriate) methods will have to be employed. Signatures of large missing energy, tagging of boosted heavy SM objects and $b$-tagging of boosted jets will hence become indispensable tools in searches for fermionic top partners at the LHC Run II.  Furthermore, as singly produced vector-like top partners are typically accompanied by a single high energy light jet in the forward detector region, forward jet tagging will also play an important role in efficient identification of signal events.

In this paper, we investigate search strategies for vector-like top partners at the LHC Run II in the TeV mass regime. For concreteness, we focus on colored $T'$, which decays into $tZ$, $th$, and $bW$, where we consider a simplified model of a singlet partner, with all the other possible top partners being decoupled.  In the simplified case, the top partner couplings to the SM particles are rather severely constrained by electroweak constraints as well as by the direct measurement of $V_{tb}$ \cite{Chatrchyan:2012ep}. However, such constraints on the top partner couplings can be significantly altered in Composite Higgs models with different top partner representations~\cite{Grojean:2013qca} or other vector-like quark models with two or more partner multiplets~\cite{Cacciapaglia:2015ixa}.

We propose concrete event selections appropriate for the TeV scale mass regime, including tagging of boosted objects, $b$-tagging strategies and forward jet tagging. Furthermore, we perform comprehensive studies of signal sensitivities and prospects for the LHC Run II to discover or rule out $T'$ partners, both in the regime where the top partner has a dominant branching ratio to each of the allowed final states, as well as a function of branching ratio to each channel~\footnote{Our results assume a narrow width approximation.}. As we will show shortly, the LHC Run II has great prospects for studying the TeV mass scale of $T'$ partners with the first 100 $\fb^{-1}$ of integrated luminosity. 

Note that vector-like top partners are typically present in multiplets of a large class of models involving strong dynamics. The details depend on the realization of the top-partner sector, but among them at least one colored state ($T'$) with electric charge of $2/3$ and decays into $tZ$, $th$, and $bW$ appears as a generic feature. Hence, our results, which we present in terms of the signal production cross section necessary for discovery at a fixed luminosity, does not exclusively apply for the Composite Higgs models, but rather for a larger class of vector-like quark models which are able to accommodate the signal event topology of singly produced $T'$.

We organized the paper in several sections which deal with a description of the search strategies for singly produced $T'$ partners in the most sensitive final states and resulting LHC Run II sensitivities, as well as an extensive Appendix which deals with example analyses in channels with sub-dominant sensitivity. In Section \ref{sec:model} we provide a brief introduction to the benchmark model which serves to illustrate our analysis. Section \ref{sec:PreCuts} deals with a brief discussion of the Monte Carlo setup, as well as a discussion of the technology we use for tagging of boosted objects, $b$-tagging and forward jet tagging. We present a detailed analysis of signal/background kinematic properties as well as signal/background efficiencies in the final states which result in the highest signal sensitivity for the three decay modes of $T'$ individually in Section \ref{sec:results}, while we present a combined analysis without the assumptions on the branching ratios in Section \ref{sec:combres}. We reserve the Appendix for a discussion of other final states, which even though are less sensitive, could provide useful insight for TeV scale $T'$ searches.
\section{Models}\label{sec:model}
Many extensions of the SM contain vector-like quark partner (and in particular top-partner) multiplets. Examples include Composite Higgs models \cite{Agashe:2004rs, Contino:2006qr, Giudice:2007fh, Matsedonskyi:2012ym}, Little Higgs models \cite{ArkaniHamed:2002qx, ArkaniHamed:2002qy}, models with extra dimensions \cite{Antoniadis:1990ew, Antoniadis:2001cv,Csaki:2003sh, Hosotani:2004wv} and others \cite{Choudhury:2001hs, Panico:2008bx, Agashe:2006at, Chivukula:2011jh}.
The partner multiplets can be classified according to their $SU(2)\times U(1)$ quantum numbers, where at least one of the partners needs to have the same electro-magnetic charge $2/3$ as the corresponding SM quark. However, in most realistic models, the existing constraints often require to group the partner multiplets in even larger units. For example, corrections to the electroweak $T$ parameter crucially 
depend on whether the top partner multiplets are introduced in full multiplets of a custodial $SU(2)_L\times SU(2)_R\simeq SO(4)$ symmetry.
As a consequence, many SM extensions come with a top partner sector which contains several vector-like $SU(2)\times U(1)$ multiplets which enter at the same mass
scale. Several of these partners can have electro-magnetic charge $2/3$ and thus they can mix with each other and the SM-like top quark. The mixing in turn influences the production cross sections and branching fractions of the individual charge $2/3$ top partners on one hand, and constraints from precision measurements on the other hand ({\cf Ref.~\cite{Cacciapaglia:2015ixa}  for an overview on the interplay of pairs of top partner multiplets in different $SU(2)\times U(1)$ representations on electroweak precision bounds).

The phenomenology of vector-like charge $2/3$ top partners can thus be studied for individual models (and in terms of the individual model parameters), in terms of an effective model description \footnote{{\it Cf.} {\it e.g.} Refs. \cite{delAguila:2000rc,Cacciapaglia:2010vn,Okada:2012gy,Buchkremer:2013bha}. In case of several top partners, sufficient freedom in the effective parameters is necessary to include different partner masses $M_i$, the couplings to SM quarks and electroweak bosons $W,Z,h$.}, or in terms of a simplified model which does not make strict assumptions about model parameter relations. Use of effective and simplified models has a benefit that the results of the studies  could be used to make predictions for more complex models with larger top-partner field content (\cf \eg \cite{Pappadopulo:2014qza,Matsedonskyi:2014mna}), and here we follow a version of this approach. For the concrete Monte-Carlo simulations we use a simple, consistent, gauge-invariant model with only one $SU(2)$ singlet top partner which we outline in the following section. However, we choose to represent our results in terms of the production cross section of $T'$ times its branching ratio into the individual signal final states  -- \ie \, in terms of the physical observables relevant for $T'$ searches -- rather than in terms of the model parameters of the specific model. This way, even though we obtain our results for one specific model, they can be re-interpreted for a larger class of $T'$ models, while we will comment on borders of validity of such re-interpretations along the course of this article.


\subsection{Example Benchmark Model}

For the purpose of event simulation, we use the Minimal Composite Higgs Model based on the $SO(5)\rightarrow SO(4)$ symmetry breaking with a partially composite top embedded in the $\bf{5}$ of $SO(5)$ ({\it cf.} Ref.~\cite{Backovic:2014uma} for the model Lagrangian, parameter definitions and the detailed derivation of the interactions.) The top-partners in this model form an $SU(2)_L$ bi-doublet as well as an $SU(2)_L$ singlet. In the singlet-partner-limit, in which the bi-doublet is decoupled, the model contains only one light vector-like  top partner: an $SU(2)_L$ singlet with charge $2/3$. The top-partner sector of the model is described by the effective Lagrangian\\
\begin{eqnarray}
\mathcal{L} &\supset& \bar{T}\left(i \slashed{D}- M_1\right)T+\bar{q}_{L} i \slashed{D}q_{L}+\bar{t}_{R} i \slashed{D}t_{R} \label{eq:Leff}
 -  \left(\lambda_R f \cos (\bar{h} /f) \bar{t}_R T_L -  \frac{\lambda_L f \sin (\bar{h} /f)}{\sqrt{2}} \bar{t}_L T_R+\mbox{h.c.}\right)\, , \label{Lmodel}
\end{eqnarray}
where $T$ is the vector-like $SU(2)_L$ singlet top partner in the gauge eigenbasis $\bar{h}= v+h$, $f$ is the Higgs compositeness scale and $M_1$ is the singlet mass scale. When expanding around the vacuum $v$, The Yukawa-type terms induce mixing between the chiral (elementary) $t_{L,R}$ and the vector-like (``composite'') partner $T$. Diagonalizing the mass matrix yields the mass eigenstates
\beq
\left(\begin{array}{c} t'_{L,R} \\ T'_{L,R} \end{array} \right) 
=
\left(\begin{array}{cc} \cos(\phi_{L,R}) &  \sin(\phi_{L,R}) \\ -\sin(\phi_{L,R}) &  \cos(\phi_{L,R}) \end{array} \right) 
\left(\begin{array}{c} t_{L,R} \\ T_{L,R} \end{array} \right) \, ,
\eeq
with \footnote{Here, for illustration, we give results to leading order in $v/f$. For the numerical simulation, we use the exact results obtained from the diagonalization.}
\bea
\tan(\phi_R) = - \frac{\lambda_R f}{M_1}+\mathcal{O}\left(\frac{v^2}{f^2}\right) \quad , \quad \tan(\phi_L) = - \frac{\lambda_L v M_1}{\sqrt{2}M^2_{T'}}+\mathcal{O}\left(\frac{v^2}{f^2}\right) = \frac{m_{t'}}{m_{T'}} \tan^{-1}(\phi_R) +\mathcal{O}\left(\frac{v^2}{f^2}\right) \, ,
\label{LRangles}
\eea
and the eigenmasses 
\bea
m_{t, {\rm phys}}\equiv m_{t'} = \quad \frac{v}{\sqrt{2}} \frac{\lambda_L\lambda_R f}{\sqrt{M_1^2+\lambda_R^2 f^2}}+ \mathcal{O}\left(\frac{v^3} {f^3}\right) \quad &\mbox{and}& \quad M_{T'} = \sqrt{M_1^2+\lambda_R^2 f^2}+\mathcal{O}\left(\frac{v^2}{f^2}\right).
\eea
  
Requiring the lightest mass eigenvalue to be the physical top mass fixes one combination of the three input parameters $\lambda_Lf, \lambda_R f, M_1$ \footnote{$v/f$ represents a further input parameter which affects, for example, the Higgs physics of the model, but with respect to the top sector Lagrangian in Eq.~(\ref{Lmodel}) and at the level of dimension 4 operators it can be absorbed into the other parameters.}. The top sector of this simple model can thus be parameterized in terms of only two BSM parameters which we choose to be $\tan(\phi_R)$ and $M_T'$.

In the mass eigenbasis, the Lagrangian in Eq.~(\ref{Lmodel}) reads
\bea
\mathcal{L}  =  \mathcal{L}_{SM,t,b} &-& c^{ttZ}_L\frac{g}{2 \cos(\theta_w)}\bar{t}'_L\slashed{Z}t_L-\left(c^{tbW}_L\frac{g}{\sqrt{2}}\bar{t}'_L\slashed{W}b_L+\mbox{h.c.}\right) - c^{tth} \frac{m_t'}{v} h \bar{t}'t'\nonumber\\
& + & \bar{T}' \left(i\slashed{\partial} - m_{T'} + g_3 \slashed{G} +\frac{2}{3}e\slashed{A}+c^{T'T'Z}_{L,R}\frac{g}{2 \cos(\theta_w)}\slashed{Z}P_{L,R}+c^{T'T'h}h\right)T' \nonumber\\
& + & \left(c^{T'tZ}_L \frac{g}{2 \cos(\theta_w)} \bar{T}'_L\slashed{Z} t_L+c^{T'bW}_L \frac{g}{\sqrt{2}} \bar{T}'_L\slashed{W} b_L  - c^{T'th}_{L,R}  h  \bar{T}'_{L,R}t'_{R,L} + \mbox{h.c.}\right),
\label{Lmeb}
\eea
where
\bea
c^{ttZ}_L =\sin^2(\phi_L) \quad , \quad & c^{t'bW}_L = 1 - \cos(\phi_L) & \quad , \quad c^{tth} = \mathcal{O}\left(\frac{v^2}{f^2}\right)\,, \nonumber\\
&c^{T'T'h}_L = \mathcal{O}\left(\frac{v}{f}\right)&  \quad , \quad  \,  c^{T'T'Z}_{L,R} =-\frac{4}{3}\sin^2(\theta_w)+\mathcal{O}\left(\frac{v^2}{f^2}\right)\delta_L \nonumber\, ,\\
 c^{T'tZ}_L =- \sin(\phi_L)\cos(\phi_L) \quad , \quad  & c^{T'bW}_L = - \sin(\phi_L) & \quad , \quad  c^{T'th}_R = - \frac{y_L}{\sqrt{2}}\frac{M_1}{M_{T'}}+\mathcal{O}\left(\frac{v}{f}\right) \quad , \quad c^{T'th}_L =\mathcal{O}\left(\frac{v}{f}\right) .
 \label{coeffs}
\eea

As can be seen from the first line of Eq.~(\ref{Lmeb}), the model predicts deviations of the SM top couplings to electroweak gauge bosons and the Higgs which implies bounds on the model parameters from measurements of single top production, top decay width, electroweak precision measurements, as well as from Higgs searches. A full study of the precision bounds of this particular model is beyond the scope of this article ($cf.$ \eg Refs.\cite{Aguilar-Saavedra:2013qpa,Grojean:2013qca} for related studies), as we only use the model as illustration for our $T'$ search strategies, and the precision bounds strongly depend on the particular realization of the top-partner embedding \footnote{{\it Cf.} Ref.\cite{Cacciapaglia:2015ixa} for a recent study of precision bounds on top partner models with several top partner multiplets.}.

 Here, we only give one representative bound which arises from the modification of the $t'bW$ coupling, parameterized by $V_{tb} =1- c^{t'bW}_L$. The strongest current bound arises from a CMS measurement of single top cross sections at 7~TeV \cite{Chatrchyan:2012ep}, which gives $|V_{tb}| > 0.92$ at 95\% CL, implying $\cos(\phi_L) > 0.92$ in our simplified model.

The second line of Eq.~(\ref{Lmeb}) shows the kinetic term of the $T'$ as well as interactions of two $T'$ with gauge bosons and the Higgs and is just given for completeness. Charge and color of the $T'$ fix the couplings to gluons and photons and in particular show that the $T'$ can be pair produced via QCD interactions. The couplings to the $Z$ and Higgs depend on the model parameters, but these couplings play a minor role in $T'$ production due to the larger QCD coupling.

Finally, the third line of Eq.~(\ref{Lmeb}) shows the $T'$ couplings which determine $T'$ single production as well as the $T'$ branching fractions of the decays to $ht$, $Zt$, and $Wb$ final states. Before discussing the various production and decay channels and their prospects for detection at LHC Run II in more detail, let us comment on several features specific to the above discussed model and how they generalize to other $T'$ models. 

Our simplified model contains only one $T'$ partner and is described by only two BSM parameters which we can take to be $\sin(\phi_L)$ and $M_{T'}$. Demanding $|V_{tb}| > 0.92$ within this model puts a direct constraint of $|\sin(\phi_L)| = |c_L^{T'bW}| < 0.4$. Fig.~\ref{fig:PXS} shows the resulting production cross section at $\sqrt{s}= 8\, \mbox{TeV}$ (left) and 14~TeV (right) for different values of $c_L^{T'bW}$ as a function of $M_{T'}$ within the model. For reference, we also show the QCD pair production cross section. In more generic models with top partners, the bound on the $T'$ production cross section can be altered, in particular if the top mixes with several top partners~\cite{Cacciapaglia:2015dsa}. 

\begin{figure}[!]
\includegraphics[scale=0.6]{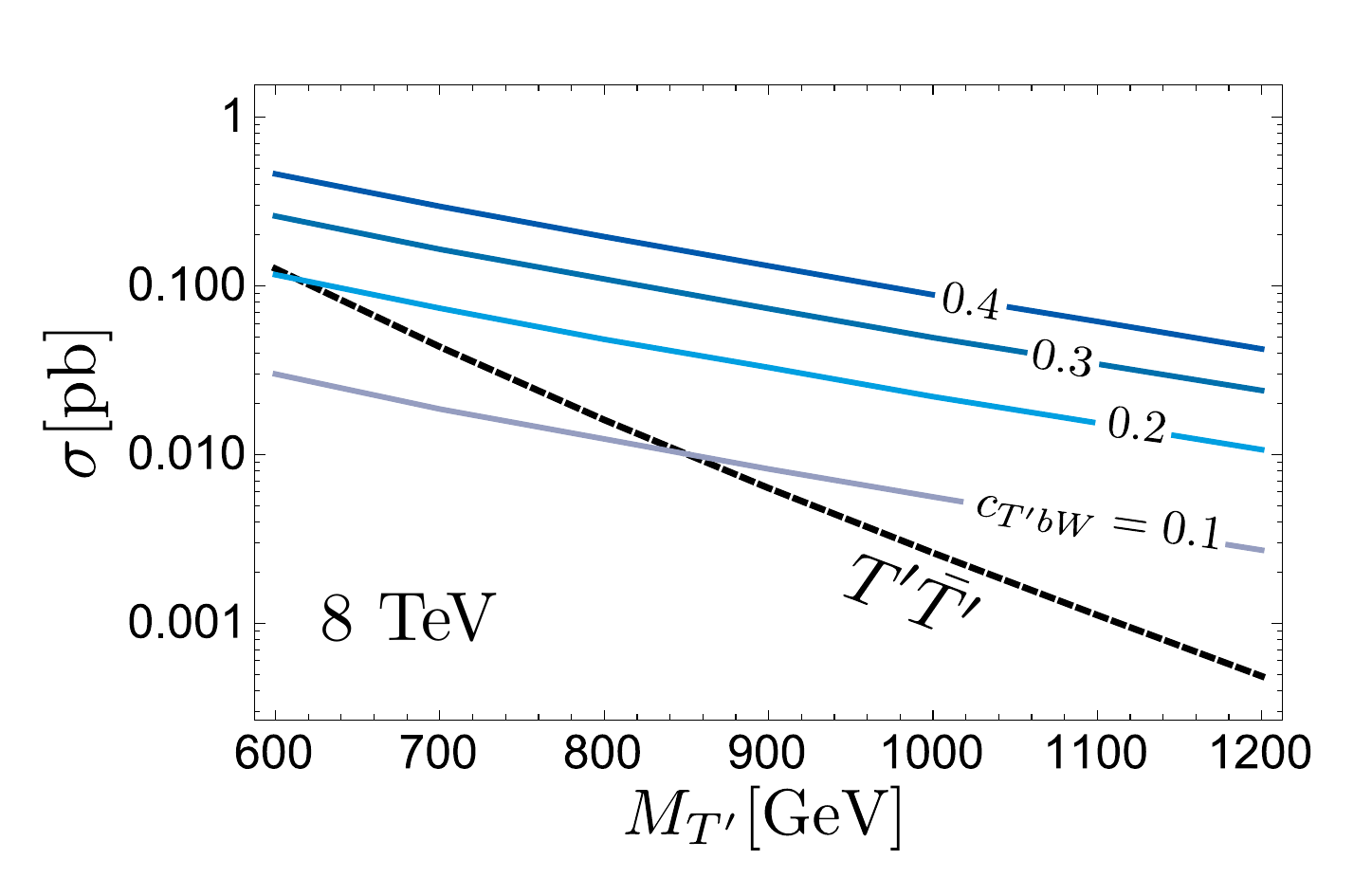}
\includegraphics[scale=0.6]{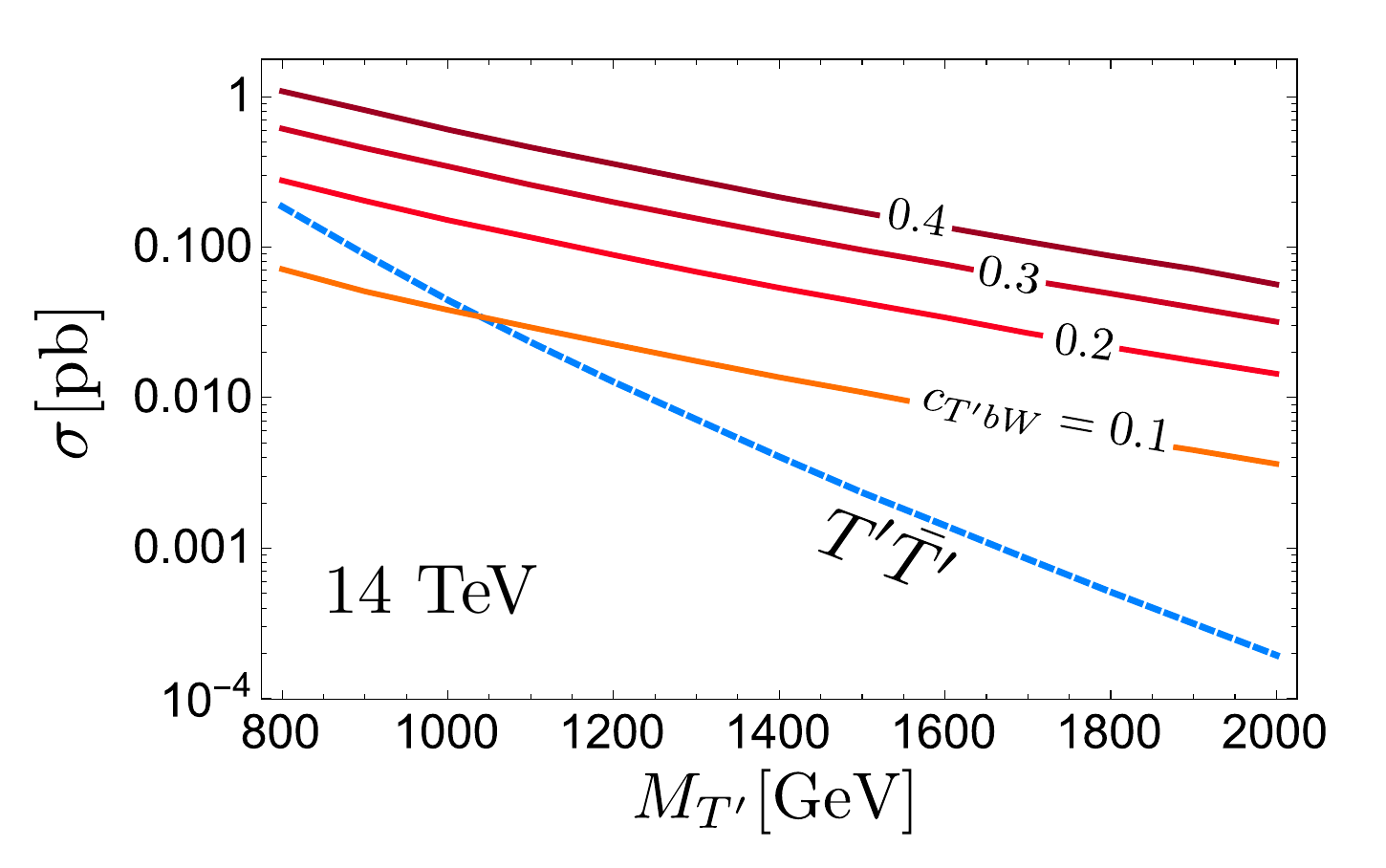}

\caption{  $T'$ production cross section at 8 TeV (left) and 14 TeV (right). The solid lines represent the single $T'$ and $\bar{T}'$ production cross section ($\sigma_{T'}+\sigma_{\bar{T}'}$) for several values of the $c^{T'bW}_L$ coupling, while the dashed curves show the $T'\overline{T}'$ pair production cross section for reference.}
\label{fig:PXS}
\end{figure}

The branching fractions of $T'$ into $th$, $tZ$ and $bW$ within the framework of our simplified model are a to good approximation given by ratios $1:1:2$ as expected by the Goldstone boson equivalence theorem for a heavy singlet top partner. The branching ratios can be altered when considering more general models with more than one top partner or top partners in a different $SU(2)$ representation. In the remainder of this article, we will therefore perform our analysis for the sample model described above, but parameterize our results in terms of the production cross section(s) times the branching ratios of the top partner(s) in order to provide a description which can be applied to more generic top partner models.

\subsection{Production and Decay}
\label{sec:Production}

The recent ATLAS \cite{ATLASnotes1} and CMS \cite{Chatrchyan:2013uxa} studies, which focused on QCD pair production of $T'$,  excluded the mass region below $\sim 700 - 900$ GeV depending on the $T'$ branching ratios. One advantage of studying pair produced top partners is that the QCD pair production cross section only depends on $M_{T'}$ while other underlying model dependences enter only into the branching ratios of the $T'$. Contrary to pair production, the $T'$ single production cross section depends also on the weak couplings parameterized in the Lagrangian of Eq.~(\ref{Lmeb}) by $c^{T'tZ}$, $c^{T'bW}$ and $c^{T'th}$, implying that single production of $T'$ introduces additional model dependence and occurs via weaker couplings.  However, benefits of studying single production lie in less phase-space suppression at large $T'$ masses (as only one heavy particle is produced and not two). By simple kinematics, there always exists an $M_{T'}$ at which single $T'$ production will dominate the QCD pair production, where the exact $M_{T'}$ value at which the transition occurs is model and parameter dependent.  Fig.~\ref{fig:PXS} shows an example, where the single production cross section becomes dominant at or below $M_{T'} \sim 1 \TeV$ $\sqrt{s} = 14 \TeV$ for $c^{T'bW}_L\gtrsim 0.1$.

Due to the potentially larger production cross sections at the TeV mass scale, in the following sections we will focus on $T'$ single production. Fig.~\ref{fig:Diagrams} shows the final states for singly produced $T'$ searches which we find to be most promising for discovery of TeV scale $T'$. The $T'$ partner can be produced through fusion of $Wb$ (as shown), $Zt$, or $ht$.  Production via $ht$ fusion is a very rare process because it requires an initial sate radiated top (instead of the $b$ in Fig.~\ref{fig:Diagrams}) as well as a Higgs (instead of the $W$) in the intermediate state, deeming the production process irrelevant for LHC Run II. The production from $Zt$ fusion requires a $Z$ to be radiated off from an initial state quark (instead of the $W$ in Fig.~\ref{fig:Diagrams}) which yields a suppression by a factor $\sim 2$. More importantly, a top from $g\rightarrow t\bar{t}$ splitting is required instead of the $g\rightarrow b\bar{b}$ splitting in Fig.~\ref{fig:Diagrams} which yields a more substantial suppression. In principle, the above argument is model dependent and it is possible to tune the couplings so that production via $Wb$ is not the dominant mode. In our simplified model, the coupling parameters $c^{T'bW}$ and $c^{T'tZ}$ are related and of the same order, such that production via $Wb$ fusion always dominates~\footnote{In more general models, $Zt$ fusion could only become relevant if the  $c^{T'bW}$ coupling is substantially suppressed which in turn implies a small production cross section. In the following we hence neglect this option.}.

\begin{figure}[t]
\includegraphics[scale=0.2]{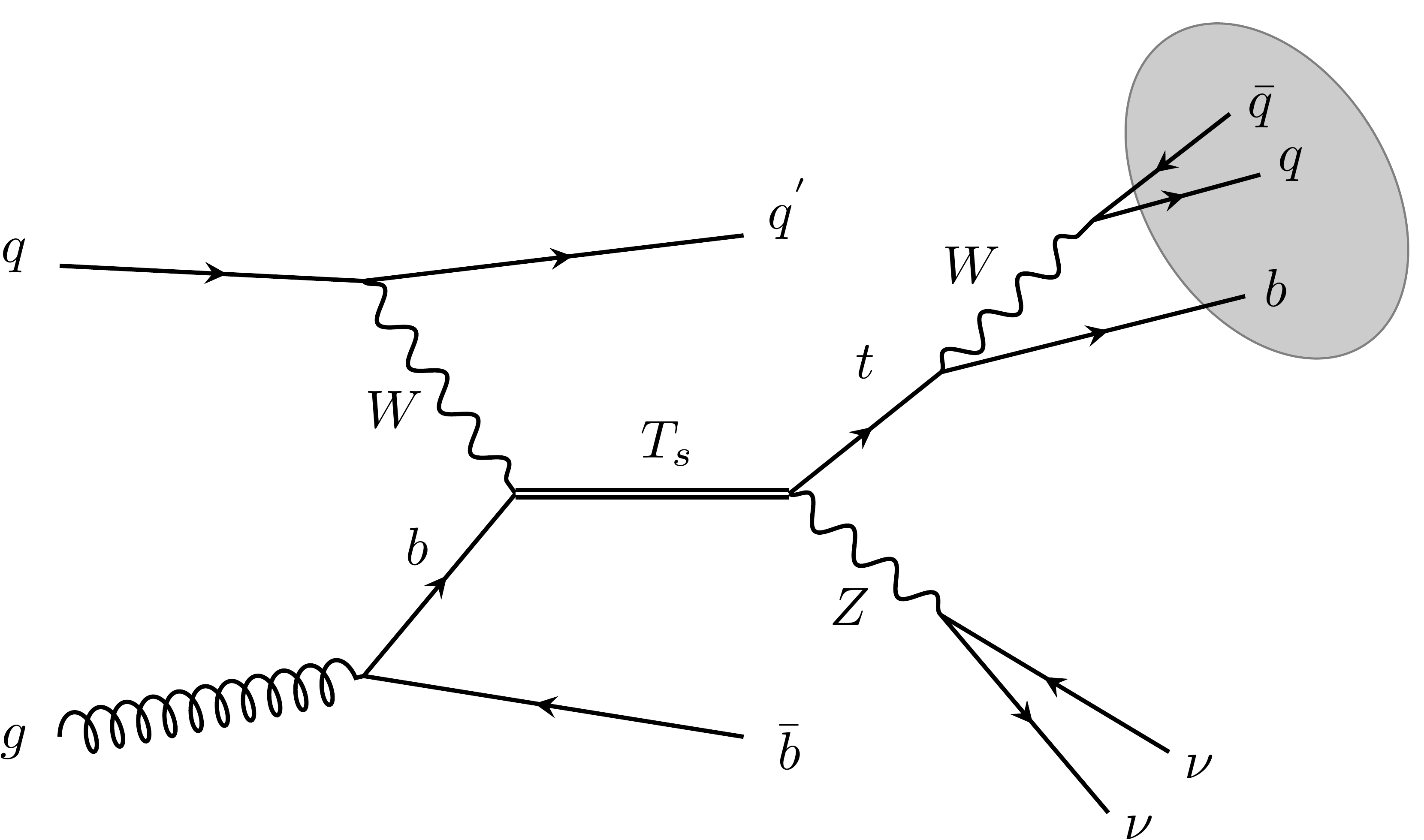}
\includegraphics[scale=0.2]{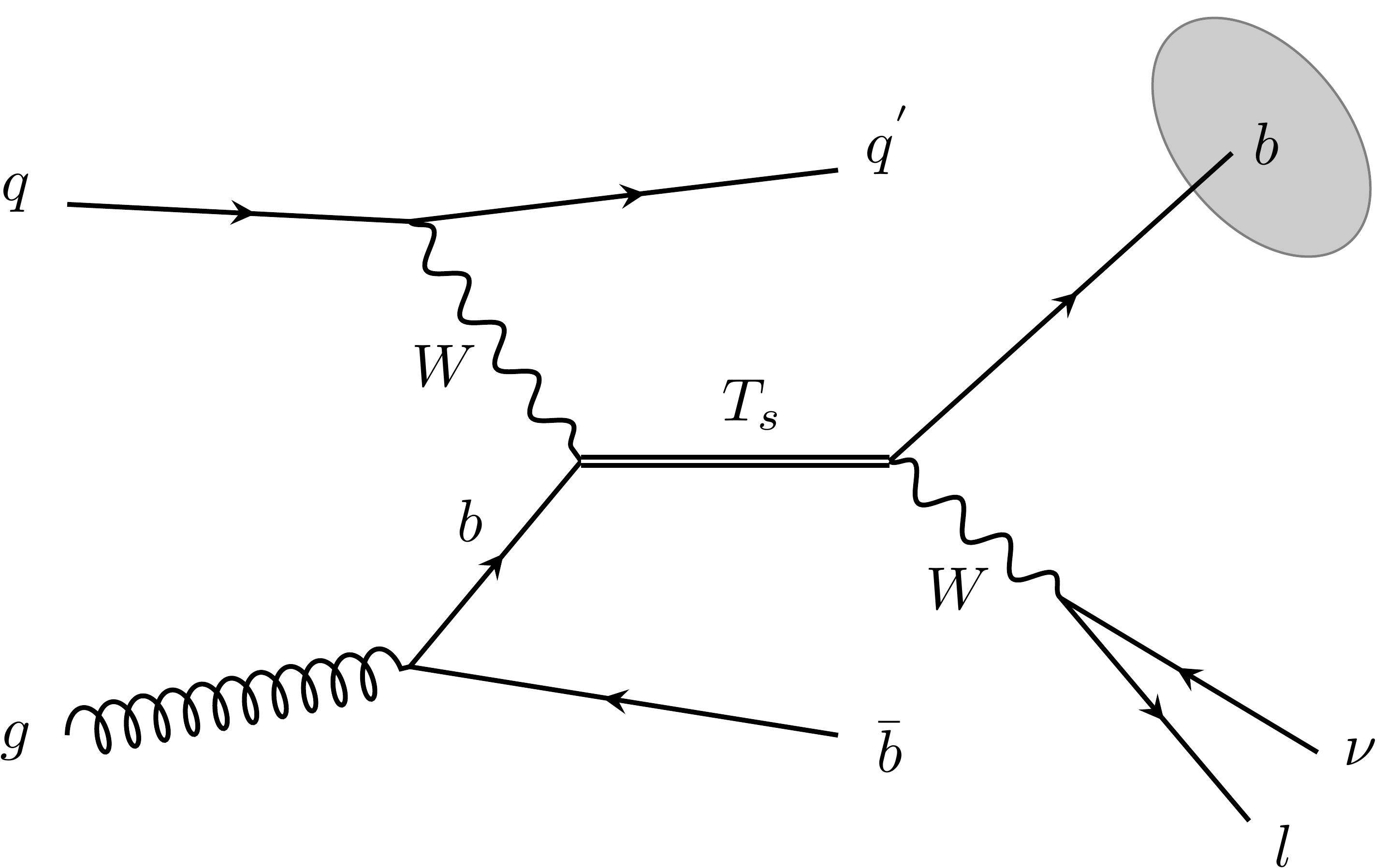}
\includegraphics[scale=0.2]{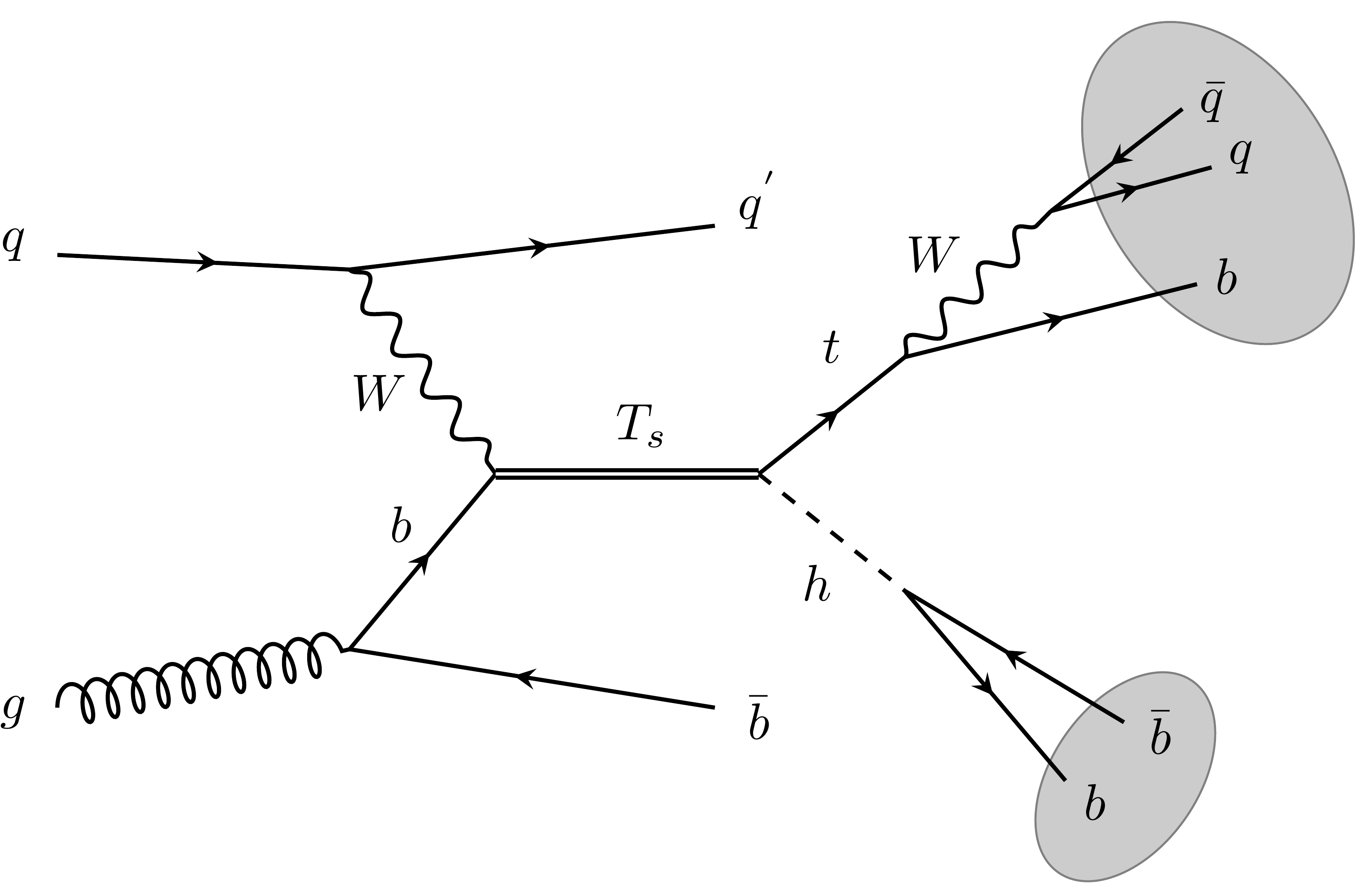}
\caption{The single production of $T'$ decaying into various final states. We show only the final states which show greatest potential for discovery at LHC Run II.}
\label{fig:Diagrams}
\end{figure}	

The singly produced $T'$ can decay to $Zt$, $Wb$ or $ht$ where the respective $Z,W,t$, or Higgs themselves have various decay channels, yielding different final states for the experimental searches. We distinguish $Z$ decays into leptonic ($e^+e^-, \mu^+\mu^-$), invisible ($\bar{\nu}\nu$), and hadronic ($jj, b\bar{b}$) final states with branching ratios of $\sim 6.7\%,\sim 20\%, \sim 70 \%$, $W$ decays into leptonic ($e^-\nu, \mu^-\nu$) and hadronic ($jj$) final states with branching ratios of $\sim 21\%$ and $67\%$, $t$ decays into semi-leptonic or hadronic final states (with the same branching ratios as the $W$) and Higgs decays into $b\bar{b}, W W^*, Z Z^*, \gamma\gamma$ where for the branching ratios we assume the $m_h= 125$ GeV SM Higgs branching ratios according to \cite{Heinemeyer:2013tqa}: $58 \%, 22 \%, 2.6 \%, 0.23 \%$, respectively. In the following we discuss the main options for the $T'$ decays to $Zt$, $Wb$ or $ht$ to identify which final states are most promising for $T'$ detection. 
\bigskip

\underline{\emph{$T' \rightarrow Zt$}}

\begin{itemize}
\item $Z_{ll} t_{\rm had}$ (relative branching ratio (rBR) with respect to $Zt$: $\sim 4.5\%$):\\
The hadronic $t$ decay comes with a large branching ratio while the di-leptonic $Z$ decay yields a very clean $Z$ signature, though at a possibly severe price on the branching ratio. The $Z_{ll}$ decay channel has been the most sensitive channel in the 8 TeV $T'\rightarrow t Z$ pair searches \cite{Aad:2014efa} and its prospects for LHC at 14 TeV  have been studied in Ref \cite{Reuter:2014iya} for singly produced $T'$ in the di-lepton channel. We investigate this channel in more detail in the Appendix, and show that while it is relevant at $M_{T'} \sim 1 \TeV$ it becomes a sub-dominant channel at higher masses.
\item $Z_{\rm inv}t_{\rm had}$ (rBR: $\sim 14\%$; \cf Fig.~\ref{fig:Diagrams} top panel, left):\\
Compared to the $Z_{ll}$ channel, the invisible $Z$ decay comes with roughly three times the branching ratio, but it does not allow to reconstruct the $Z$ boson directly. The final state is a top fat jet with a large amount of missing transverse energy ($\slashed{E}_T$) as the $Z$ boson results from a decay of a heavy $T'$ and thus has high $p_T$. Ref.~\cite{Backovic:2015lfa} showed that at LHC 14 and for $M_{T'}\gtrsim 1 \mbox{ TeV}$, cutting on $\slashed{E}_T$ allows to eliminate QCD backgrounds and substantially reduce other SM backgrounds, making this channel competitive to the $Z_{ll}$ at $M_{T'} \sim 1 \TeV$ and superior at $M_{T'} \gtrsim 1.5 \TeV$. We present a detailed discussion of this channel in Sec.~\ref{sec:Zt}.
\item$Z_{\rm had}t_{\rm had}$ (rBR: $\sim 45\%$):\\
The fully hadronic $tZ$ decay channel comes with the largest branching fraction, but suffers from very large QCD background which cannot be reduced by demanding final state leptons or large $\slashed{E}_T$. One way to reduce the QCD background would be to focus only on $Z_{b\bar{b}}t_{\rm had}$ part and rely on $b$-tagging. However,  $Z_{b\bar{b}}t_{\rm had}$ comes with a BR of only $\sim 10\%$. Although we do not explicitly consider the $Z_{bb} t_{\rm had}$ final state in this paper, we can infer the performance of the analysis in this channel based on the detailed  analysis of the $h_{b\bar{b}}t_{\rm had}$ final state in Sec.~\ref{sec:th}.   The main difference of the analysis in Sec.~\ref{sec:th} compared to $Z_{b\bar{b}}t_{\rm had}$ is the jet substructure tagging efficiency, where we expect the boosted $Z$ tagging to perform at least slightly worse compared to boosted Higgs tagging. The effect is primarily due to the $Z$ mass being lower than the Higgs mass, hence resulting in the jet sub-structure which resembles that of a light jet more closely.  However, even if we assume that boosted $Z$ tagging performs at the same level as the boosted $h$ tagging, the $h_{bb}t_{\rm had}$ channel has an rBR of 39\% and yields  $S/\sqrt{B}$ of 7.7 (3.2) for 1 TeV (1.5 TeV) with the best b-tagging strategy (see Sec.~\ref{sec:th} for more details), while $Z_{bb}t_{\rm had}$ has a rBR of only 10\%. Hence, the overall signal significance in $Z_{bb} t_{\rm had}$ channel will be reduced by a factor of 4, resulting in $S/\sqrt{B} \sim 1.9\, (0.8)$ for 1 TeV (1.5 TeV) with the best $b$-tagging strategy, in the optimistic scenario in which $Z$ and $h$ tagging perform at the same level.  We conclude that $Z_{bb} t_{\rm had}$ channel does not yield sufficient signal sensitivity Run II LHC with 100 $ {\fb^{-1}}$ of data compared to the $Z_{\rm inv} t_{\rm had}$ and $Z_{ll}  t_{\rm had}$ channels,  and we do not include it into our analysis. 
\item$Z t_{\rm lep}$ channels:\\
The $Z_{\rm had} t_{\rm lep}$ channel (rBR $\sim 15\%$) has a similar BR compared to the $Z_{\rm inv}t_{\rm had}$ channel, but we expect it to perform worse for two reasons. First, the amount of  $\slashed{E}_T$ expected from the top is less than $Z_{\rm inv}$, and hence a cut on missing energy has much worse prospects to suppress SM backgrounds. Second, the boosted hadronic $Z$-tagging performs significantly worse than hadronic top-tagging (combined with the $b$-tagging). Similarly, we expect the $Z_{ll}t_{\rm lep}$ and $Z_{\rm inv}t_{\rm lep}$ channels to achieve similar signal efficiencies and background rejection rates as their counter parts with hadronically decaying tops, but as their BR's are smaller by a factor of $\sim 1/3$ they will lead to a smaller overall number of signal events and hence likely to a lower overall signal significance at high $T'$ masses.\\
\end{itemize}

\underline{\emph{$T' \rightarrow Wb$ }}

\bigskip

The $Wb$ channel (top right panel of Fig.~\ref{fig:Diagrams}) has a relatively simple structure in that it contains a $b$ quark and a boosted $W$ which can decay either leptonically or hadronically. The channel in which $W$ decays into a lepton ($W_{\rm lep} b$, BR: $\sim 21 \%$) has been studied for a hypothetical top partner mass below 1 TeV in Ref.~\cite{Ortiz:2014iza} for 8 and 14 TeV and Ref.~\cite{Li:2013xba} for 8 TeV. Following the same strategy presented in Ref.~\cite{Ortiz:2014iza}, we study this channel in Sec.~\ref{sec:Wb} and find that $W_{\rm lep}b$ is still the most relevant channel for $M_{T'}$ above 1 TeV. The channel where $W$ decays hadronically ($W_{\rm had} b$,  BR: $\sim 67\%$) has not been studied yet in the literature. We present a sample analysis of $W_{\rm had} b$ in the Appendix but find that  QCD multi-jet background poses a problem with no clear strategy to resolve and hence a poor signal sensitivity.  
\bigskip

\underline{\emph{$T' \rightarrow ht$}}

\begin{itemize}
\item $h_{bb}t_{\rm had}$ (rBR: $\sim 39\%$; Fig.~\ref{fig:Diagrams} bottom panel):\\
The all-hadronic $ht$ channel is characterized by a large branching ratio, but also suffers from the enormous contamination of multi-jet QCD bacgkrounds.
 All-hadronic $ht$ final states have first been studied in Ref.\cite{CMS:2014aka} for pair-produced vector-like T quarks where the authors made use of a large $H_T$ cut along with top-tagging and Higgs-tagging techniques to exclude the masses below $\sim 750$ GeV at 8 TeV. So far, however, no search strategy has been proposed to explore $T'$ single production channel in all hadronic final states. Here we will show that a multi-dimensional Higgs- and top-tagging technique \cite{Backovic:2014ega} combined with b-tagging is able to suppress QCD and $t\bar{t}$ backgrounds sufficiently to make this channel the most promising $T' \rightarrow ht$ discovery channel.
\item $h_{bb}t_{\rm lep}$ (rBR: 12\%):\\
Ref.~\cite{Li:2013xba} has studied this channel in the context of LHC at $\sqrt{s} = 8 \TeV$.  As compared to the fully-hadronic $h_{bb} t_{\rm had}$ channel, this channel's BR is reduced by a factor of $\sim 1/3$, with the advantage that the QCD background can be effectively suppressed due to the existence of a hard isolated lepton in signal events. However, in our $h_{bb}t_{\rm had}$ analysis we find that -- after applying b-tagging -- the dominant SM background arises from SM $t\bar{t}$ events which also constitute significant backgrounds in case of the leptonic $t$ decay. As boosted hadronic top tagging typically performs better than leptonic $t$ tagging, and because the signal is diminished by a small branching ratio, we conclude that this channel is less sensitive to TeV scale $T'$ compared to $h_{bb} t_{\rm had}$ (see Appendix for more details on the $h_{bb} t_{\rm lep}$ final state).

\item $h_{WW^*}$ channels:\\
Before decaying further, the $h_{WW^*}$ channels come with a BR reduced by $\sim 22 \% / 58 \% \sim 0.4$  compared to the corresponding $h_{bb}$ channels. The $h_{WW^*}$ can then further decay fully hadronically ($jjjj$), semi-leptonically ($jjl\nu$) or fully leptonically ($ll\nu\nu$). The fully hadronic decay has substantially larger backgrounds compared to the $h_{bb}$ while for the semi-leptonic and fully leptonic decays, the final state contains one or two neutrinos, complicating the reconstruction of the invariant mass and momentum of the Higgs and therefore substantially making the differentiation from SM backgrounds more difficult. We thence do not consider $h_{WW^*}$ channels further in this analysis.
\item $h_{ZZ^*}$ channels:\\
The fully hadronic decay of the $h_{ZZ^*}$ is difficult to identify due to a large jet multiplicity and has larger background than the $h_{bb}$ final state, due to a lower number of $b$-jets in the final state (on average). The $h_{ZZ^*}$ decay $llll$ comes with a suppression  of $\sim 2.3\% * (6.7\%)^2\sim 1\times 10 ^{-4}$ relative to the $h_{bb}$ final state. Requiring $\sim 5$ events at a luminosity of 100~fb$^{-1}$ thus requires a production cross section of $\sim 500 \fb$ which is about one order of magnitude larger than the projected upper bound on production cross sections we establish in this paper. Finally, the $h_{ZZ^*}$ decay in to $jjll$ comes with a BR of $\sim 0.22\%$. This channel is unlikely to provide a competitive bound because it would require to identify a boosted $h\rightarrow jjll$ with high signal efficiency. Standard separation criteria between leptons and jets could not be applied as the signature consists of a collimated $jjll$ system, rendering the search problematic, and here we do not investigate this option further. 

\item $h_{\gamma \gamma}t\, ({\rm rBR}: \sim 0.23\%)$:\\
The $h_{\gamma\gamma}$ channel comprises a very clean final state which, however, has a strongly suppressed signal cross section. To give an estimate on how competitive this channel can be compared to the $h_{bb}t_{\rm had}$ channel, we anticipate some of the results of our study. In Sec.~\ref{sec:combres} we find that assuming a BR of 100\% for $T' \rightarrow ht$, the discovery reach for LHC14 at a luminosity of 100~fb$^{-1}$ for the $T'$ single production cross section is 80~fb (30~fb) for a 1 TeV (1.5~TeV) top partner. For an $h_{\gamma\gamma}$ signal, these cross sections amount to signal cross sections of $\sim 0.2$ fb ($\sim 0.07$ fb) or 20 (7) events at a luminosity of 100 fb$^{-1}$ before any cuts and signal efficiencies are applied. In spite of the low background, testing cross sections at this level appears challenging with 100 $\fb^{-1}$ of data \footnote{For comparison, the CMS search on the 8 TeV data set with 19.7~fb$^{-1}$ for the $T'\rightarrow h_{\gamma\gamma}t$ search established a bound of the order of 1 fb for a 900 GeV $T'$ \cite{CMS:2014rda}.}. Rather than focussing on the specifics of the $T'$ single production topology, a dedicated search for highly boosted $h\rightarrow\gamma\gamma$ signals which minimizes signal loss could be an alternative for this channel. Here we do not study the $h_{\gamma \gamma}$ channel, but we do not exclude it from the list of possibly useful final states in searches for $T'$. 

\end{itemize}

\section{Event Samples and Analysis Method for LHC14} \label{sec:PreCuts}
\label{sec:Simulation}

\subsection{Event Generation \& Preselection Cuts}

We generate all event samples in this analysis using the leading order \verb|MadGraph 5| \cite{Maltoni:2002qb} at a $\sqrt{s} = 14$ TeV $pp$ collider with a \verb|nn23nlo| parton distribution functions \cite{Ball:2012cx}. At generation level, we require all final state patrons to pass cuts of $p_T > 15$~GeV, ~$| \eta |< 5$, except for hard level leptons, for which we require $p_T > 10$~GeV, ~$| \eta |< 2.5$. In order to improve the statistics in the SM backgrounds, we generate all event samples with an $H_T$ cut at generator level, where $H_T$ denotes the scalar sum of the transverse momenta of all final state quarks and gluons. We choose the generator level $H_T$ cuts for each individual channel based on the mass scale of $T'$ and specify the numerical values in the tables which summarize the background cross sections ($e.g.$ Table \ref{tab:TotalBackGroundsZt}).

Next, we shower the events with \verb|PYTHIA 6| \cite{Sjostrand:2006za}  using the modified MLM-matching scheme~\cite{Maltoni:2002qb,Mangano:2006rw}, and cluster all showered events with the \verb|FastJet| \cite{Cacciari:2011ma} implementation of the anti-$k_T$ algorithm \cite{Cacciari:2008gp}. We use a cone size $R = 1.0$ to cluster the decay products of boosted heavy particles, such as boosted Higgses, top quarks and $Z/W$ bosons, while we use $r = 0.4$ for non-forward light jets ($i.e.$ $|\eta| < 2.5$) including the $b$-jets. 

\subsection{Boosted Heavy Jet Tagging}
\label{sec:TOM}

As the energy frontier is being pushed to ever higher mass scales, tagging of boosted heavy objects is becoming a central topic of new physics searches at LHC Run II. 
Lower limits on new physics mass scales in most vanilla scenarios already exceed 1 TeV, implying that if new physics is to be found, it will likely be in the highly boosted regime. 

Recently, there has been an extensive effort on designing, improving and understanding jet substructure observables in order to classify and distinguish boosted heavy objects such as Higgs, top and $W/Z$ bosons from the QCD backgrounds \cite{Butterworth:2008iy, Almeida:2011aa, Backovic:2012jj, Schlaffer:2014osa, Ellis:2007ib, Abdesselam:2010pt, Salam:2009jx, Nath:2010zj, Almeida:2011ud, Plehn:2011tg, Altheimer:2012mn, Soper:2011cr, Soper:2012pb, Jankowiak:2011qa, Krohn:2009th, Ellis:2009me, Backovic:2012jk, Backovic:2013bga, Hook:2011cq, Thaler:2010tr, Thaler:2011gf, Thaler:2008ju, Almeida:2008yp,Almeida:2010pa, Rentala:2014bxa, Cogan:2014oua, Larkoski:2014wba}.

In this paper,  we use the \verb| TemplateTagger v.1.0 |\cite{Backovic:2012jk} implementation of the Template Overlap Method (TOM) \cite{Almeida:2010pa, Almeida:2011aa,  Backovic:2012jj, Backovic:2013bga} as a tagger of massive boosted objects. We opted to use TOM due to the flexibility of the method to tag any type of heavy SM states as well as the method's weak susceptibility to pileup contamination \cite{Backovic:2013bga}. 
The TOM algorithm for boosted jet tagging utilizes an overlap function as an estimate of likelihood that a jet energy distribution matches a parton level model (template) for a decay of a heavy SM state. The procedure of matching the templates to jets is performed by minimizing the difference between the calorimeter energy depositions within small angular regions around template partons and actual parton energies, over the allowed phase space of the template four-momenta. The output of the TOM algorithm is an overlap score $Ov$ which measures the likelihood that a given boosted jet is a Higgs, top or a $W/Z$ boson as well as the template which maximizes the overlap. 

In the following sections we will use TOM primarily as a top or a Higgs tagger, while we find that channels in which a $W$ or a $Z$ boson decay hadronically are typically not the most sensitive to $T'$ searches at LHC Run II. Following the proposal of Ref. \cite{Backovic:2014ega} we define a multi-dimensional TOM tagger as a vector of overlap scores:
\begin{equation}
\overrightarrow{Ov} = (Ov_2^i, Ov_3^t)\,,
\end{equation}
where $i =  W,Z h, t$ \footnote{For simplicity, we only use $(Ov_2^h, Ov_3^t)$ in searches for $h_{bb}$ in the final state and  $(Ov_2^W, Ov_3^t)$ in searches for a hadronic $W$ in the final state while neglecting the other  two-body templates in the respective searches.}. 

The multi-dimensional TOM analysis significantly improves the tagging efficiency of top and Higgs jets. As the three prong decay of a boosted top is a more complex object than the typical two prong decay of a boosted Higgs/$W$/$Z$, it is possible for a top fat jet to fake the two-body Higgs/W/Z template tagging procedure. The converse, however,  is not very likely. We hence define a fat jet to be a top candidate if it passes the requirement 
\begin{equation}
	Ov_3^t >0.6\, ,
\end{equation}
 while we define a fat jet to be a Higgs/$W$/$Z$ candidate, if it passes the requirement 
\begin{equation}
	Ov_2^i > 0.5, \,\,\,\,\, Ov_3^t <0.6\, ,
\end{equation}
where $i = h, W, Z$. In the following sections we will demonstrate the capability of multi-dimensional TOM analysis to reduce background contaminations of $T'$ signal events with a reasonable cost to signal efficiency. 

For the purpose of our analysis, we generate 17 sets of both two body Higgs/$W$/$Z$ and three body top templates at fixed $p_T$, starting from $p_T = 325 \GeV$ in steps of $50 \GeV$, while we use a template resolution parameter $\sigma = p_T /3$ and scale the template sub-cones according to the rule of Ref. \cite{Backovic:2012jj}.

\subsection{$b$-tagging}
\label{sec:b-tagging}

Efficient tagging of $b$-jets is a crucial tool in BSM studies at the LHC. 
Identifying $b$ quarks inside of strongly collimated decay products of new particles is challenging as a dense environment of tracks degrades the $b$-tagging efficiency. Yet, a recent ATLAS study \cite{ATLAS:2014} has shown that performance of various $b$-tagging algorithms in boosted topologies as well as fake rates of light-flavour and charm jets performs at a level relatively comparable to the standard, non-boosted $b$-tagged strategies.

\begin{table}[htb]
\begin{tabular}{|c|c|c||c|c||}
\hline
	$b$-tagged score	                 &  Efficiency  (at least 1 $b$-tag)                                                                                                                               & value                     &  Efficiency  (exactly 2 $b$-tags)                                                                                                    & value        \\ \hline
 	 0 (jet: u,d,s,g)				& 	 	$\epsilon_j$				                                                                                                                &  0.01                            &                ${\epsilon_j}^2$                                                                                                                                     &      0.00026              \\  \hline
	1 (1c)					& 	 	$\epsilon_c$					                                                                                                        &  0.18	                          &                $\epsilon_j \epsilon_c$                                                                                                                                    &     0.0029                 \\  \hline
	2 (2c)					& 	 	$2\, \epsilon_c (1-\epsilon_c)+ {\epsilon_c}^2$		                                                                               &  0.33		                 &   ${\epsilon_c}^2$                                                                                                                          &     0.032          \\  \hline
	3 (1b)					& 	 	$\epsilon_b$					                                                                                                         &  0.75		                 &                $\epsilon_b \epsilon_j$                                                                                                                                     &      0.011                 \\  \hline
	4 (1b+1c)					& 	 	$\epsilon_b (1-\epsilon_c) + \epsilon_c (1-\epsilon_b)+\epsilon_b \epsilon_c$	                                     & 	0.80		                  &  $\epsilon_b \epsilon_c$                                      	                                                                      &      0.14           \\  \hline
	\multirow{2}{*}{ 5 (1b+2c)}		& 	 	$\epsilon_b (1-\epsilon_c)^2 + 2 (1-\epsilon_b) (1-\epsilon_c) \epsilon_c$ 		                                      & \multirow{2}{*}{0.83}	 &  \multirow{2}{*}{ $2 \epsilon_b \epsilon_c (1- \epsilon_c) + \epsilon_c^2 (1-\epsilon_b)$}	          &    \multirow{2}{*}{0.23}   \\  
							&		$ +  2 \epsilon_b \epsilon_c (1- \epsilon_c) + \epsilon_c^2 (1-\epsilon_b) +\epsilon_b {\epsilon_c}^2$	    &	                                  &                                                                                                                                                      &                       \\ \hline
	6 (2b)					& 	 	$2 \epsilon_b (1-\epsilon_b)+ {\epsilon_b}^2$			                                                                       &   0.94		                 &   ${\epsilon_b}^2$                                                                                                                         &    0.56          \\  \hline
	7 (2b+1c)					& 	 	$1 - (1-\epsilon_c) (1-\epsilon_b)^2$				                                                                                &   0.95		                 &  $2 \epsilon_b \epsilon_c (1-\epsilon_b) + {\epsilon_b}^2 (1-\epsilon_c)$                                   &    0.53          \\  \hline
	\multirow{2}{*}{8  (2b+2c)}		& 	 	\multirow{2}{*}{$1 - (1-\epsilon_c)^2 (1-\epsilon_b)^2$}				                                              & \multirow{2}{*}{0.96}	&  ${\epsilon_b}^2 {(1-\epsilon_c)}^2 + 4 \epsilon_b \epsilon_c (1-\epsilon_b) (1-\epsilon_c)$       &    \multirow{2}{*}{ 0.49 }    \\  
	  					         & 	 				                                                                                                                                            & 	                          &  $ + {\epsilon_c}^2 {(1-\epsilon_b)}^2$                                                                                         &              \\  \hline
	9  (3b)					& 	 	$1 - (1-\epsilon_b)^3$						                                                                                 &  0.98                          &   $3 {\epsilon_b}^2 (1-\epsilon_b)$                                                                                                &   0.42        \\  \hline					 
\end{tabular}\par
\caption{Efficiencies for at least 1 $b$-tag (left) and exactly 2 $b$-tags (right) of a fat jet which contains a specific number of light, $c$ or $b$ jets within $\Delta R = 1.0$ from the fat jet axis. $\epsilon_j$, $\epsilon_c$ and $\epsilon_b$ are $b$-tagging efficiencies for light, $c$ and $b$ jets respectively.} \label{tab:btag} 
\end{table}

In our semi-realistic $b$-tagging procedure, where we assign to each $r=0.4$ jet a $b$-tag if there is a parton level $b$ or $c$ quark within $\Delta r=0.4$ from the jet axis, we adopt the ATLAS benchmark $b$-tagging efficiency point of
\begin{equation}
	\epsilon_b = 0.70, \,\,\,\, \epsilon_c = 0.18, \,\,\,\,\, \epsilon_j = 0.016\,, 
\end{equation}
where $\epsilon_{b, c, j}$ are the efficiencies that a $b$, $c$ or a light jet will be tagged as a $b$-jet. Table \ref{tab:btag} shows the $b$-tagging efficiencies for all relevant possibilities of jet $b$-tags, assuming at least one $b$-tag (left) and exactly 2 $b$-tags (right). 

For the fat jet to be $b$-tagged, we require that a $b$-tagged $r=0.4$ jet land within $\Delta R = 1.0$ from the fat jet axis. We take into account that more than one $b$-jet might land inside the fat jet, whereby we reweigh the $b$-tagging efficiencies depending on the $b$-tagging scheme described in Table \ref{tab:btag}.

\subsection{Forward Jet Tagging}
\label{sec:FWD-tagging}

Single production of composite vectorlike quark partners is typically accompanied by a single high energy forward jet, a useful handle on the SM backgrounds. In this analysis, we define forward jets as in Ref. \cite{Backovic:2014uma}, whereby we cluster the event with a jet cone radius $r=0.2$ and demand that it satisfies the following criteria:
\begin{equation}
			p_T^{\rm fwd} > 25 \GeV, \,\,\,\,\,\,\,  2.5 < \eta^{\rm fwd} < 4.5\, . \label{eq:fwd_jet_def}
\end{equation}
We then define an event to be ``forward jet tagged'' if  $N^{\rm fwd} \ge 1$. 

Clustering the event in the forward region with a small jet cone ($i.e.$ $r=0.2$) is beneficial when considering the large pileup environment expected at LHC Run II \cite{Backovic:2014uma}. Note also that since we wish to only tag a forward jet and not measure it, the detector effects on our forward jet tagging strategy should be mild \cite{Backovic:2014uma}. 

\section{Searches for $T^\prime$ at LHC14 } \label{sec:results}
\label{sec:results}
As discussed in the previous sections, searches for TeV scale $T^\prime$ partners at LHC Run II at $\sqrt{s} = 14 \TeV$ will be characterized by  different kinematics as compared to previous searches at LHC Run I.  Event selections deemed suitable for mostly non-boosted final states relevant for searches at LHC Run I lack efficiency in the detection of highly boosted final states and reconstruction techniques need to be modified for TeV scale $T'$ partner searches. 

In the following we explore ways to optimize strategies to detect $T^\prime$ partners in its various decay channels and present detailed Monte-Carlo based analyses of Run II sensitivity to partners of mass O(1 TeV). In this section, we show results on searches which we found to be most sensitive. Results on other candidate channels as well as further details on alternative cuts are summarized in Appendix~\ref{app:Zt} - \ref{app:ht} .  

There are significant differences in optimal strategies for discovering different decay modes of singly produced $T'$ partners at LHC Run II. However, several features will appear in all channels. First, a high energy forward jet and a spectator heavy flavor quark are present in all final states.\footnote{The dominant $T'$ production via $Wb$ fusion creates a final state $\bar{b}$ while $Zt$ fusion creates a final state $\bar{t}$. In all our proposed search strategies we do not specifically cut on the spectator $\bar{b}$ (or $\bar{t}$) such that modifications in signal efficiencies for $T'$ production from $Zt$ fusion would be minor.} Second, signal events will be characterized by final states of high transverse boost ($i.e.$ $p_T \sim M_{T'}/2$), be it in the form of a fat jet, high energy lepton or large $\MET$. In the following we will discuss several useful handles for tagging and reconstructing the $T'$ partners, as well as search strategies useful to suppress the large SM backgrounds. 

\subsection{$T' \rightarrow Z_{{\rm inv}} t_{{\rm had}}$ Channel }
\label{sec:Zt}

As a precursor to this more complete overview article, we studied the $T' \rightarrow Z_{{\rm inv}} t_{{\rm had}}$ channel in Ref. \cite{Backovic:2015lfa}, where we showed that for $M_{T'}\gtrsim 1 \mbox{TeV}$ a tight cut on $\slashed{E}_T$ renders this channel more sensitive to $T'$ partner searches than (more commonly employed) searches using the  $Z\rightarrow l^+ l^-$ decay channel. For completeness, here we give an overview of our results in the $T' \rightarrow Z_{{\rm inv}} t_{{\rm had}}$ channel, while we refer the reader to Ref.~\cite{Backovic:2015lfa} for more details \footnote{Note that as compared to Ref.~\cite{Backovic:2015lfa} we refined our assumed $b$-tagging efficiencies to mimic efficiencies of Ref.~\cite{ATLAS:2014} which study $b$-tagging efficiencies for non-isolated $b$'s as they occur in boosted searches. Therefore, the numerical results on the $T' \rightarrow Z_{{\rm inv}} t_{{\rm had}}$ channel presented in this article marginally differ from Ref.~\cite{Backovic:2015lfa}.}.

The main SM backgrounds for the $Z_{\rm inv} t_{\rm had}$ channel are SM processes containing a $Z$ boson in the final state, as well as semi-leptonically decaying $t\bar{t}$.  The ``$Z$-containing'' backgrounds include $Z + t$, where we include $Z t\bar{t}$ and $Z t/\bar{t}$ (with up to two extra jets) into our simulation. Other $Z+X$ backgrounds contain jets, hadronically decaying gauge bosonos or $b$ quarks which can ``fake'' a (hadronic) top signal. In this class, we include $Z$, $Z b\bar{b}$, $Z+Z/W$ with up to  two additional jets.  Finally, we include  $t\bar{t}$ background with up to two additional jets.

We simulate all the backgrounds with the preselection cuts described in Section \ref{sec:PreCuts} where we demand $H_T > 500~ (750)$ GeV at event generation level for a hypothetical mass of the top partner of $M_{T'} = 1$ TeV (1.5 TeV), which we choose as two benchmark scenarios. Table \ref{tab:TotalBackGroundsZt} summarizes the background cross sections which we simulate at leading order and then multiply it by a (conservative) K-factor of 2.

\begin{table}[h]
\begin{center}
\begin{tabular}{|c|c|c|c|}
\hline
		Signal Channel								&	Backgrounds 						& $\sigma (H_T > 500 \GeV) [{\rm fb}]$& $\sigma (H_T > 750 \GeV) [{\rm fb}]$  \\ \hline
\multirow{8}{*}{$T' \rightarrow Z_{\rm inv} t_{{\rm had}}$}  &$t \bar{t}$(semi-leptonic) + jets 	 		&  $2.6 \times 10^4$				 &	$5.23 \times 10^3 $ 				\\
											&$Z_{\nu \nu}$ + jets 			&  $1.88 \times 10^4$ 			&	 $4.65 \times 10^3 $				\\
											&$Z_{\nu \nu}$+ $b \bar{b}$ + jets 	&  $4.59 \times 10^2$ 			&	 $1.0 \times 10^2 $				\\
											&$W_{l \nu}$ + jets 				&  $4.76 \times 10^4$ 			&	 $1.21 \times 10^4$				\\
											&$W_{l \nu}$ + $b \bar{b}$ + jets 	&  $4.17 \times 10^2$ 			&	 $1.08 \times 10^2$				\\
											&$Z_{\nu \nu}$ + $t \bar{t}$ + jets 	& $16.6$ 				&	$4.6$						\\
											& $Z_{\nu \nu}$ + $t$/$\bar{t}$ + jets &  $27.6$   			&	$7.3$						 \\
											&$Z_{\nu \nu}$ + $Z/W$+ jets 								& $1.77 \times 10^3 $ 			&	 $5.69 \times 10^2 $				\\ 
\hline		
\end{tabular}
\end{center}
\caption{The simulated cross sections of SM backgrounds (Simulated at leading order and multiplied by a conservative  K-factor estimate of 2 after preselection cuts described in Sec.~\ref{sec:PreCuts}.}
\label{tab:TotalBackGroundsZt}
\end{table}

\begin{table}[h]
\setlength{\tabcolsep}{2em}
{\renewcommand{\arraystretch}{1.8}
\begin{tabular}{c|c}

							&$T' \rightarrow Z_{\rm inv} t_{\rm had}$                                                      \\ \cline{1-2}
\multirow{2}{*}{ \textbf{Basic Cuts} }	&$N_{\rm fj} \geq 1$ ($R=1.0$), $N^{\rm iso}_{\rm lepton}= 0$ ,     \\ 
		 					&$p_T^{\rm fj} > 400 \;(600) \GeV$, $|\eta_{\rm fj} |< 2.5$ .	 \\
\end{tabular}} \par
\caption{Summary of Basic Cuts for $T' \rightarrow Z_{\rm inv} t_{\rm had}$ channel . ``fj" stands for the fat jet with $|\eta_{\rm fj}| < 2.5$ and $p_T > 400\, (600) \GeV$ for $M_{T'} = 1 (1.5) \TeV$. $N^{\rm iso}_{\rm lepton}$ represents the number of isolated leptons with mini-ISO $> 0.7$, $p_T^{l} > 25\GeV$ and $|\eta_{l} |< 2.5$.} \label{tab:BasicCutstZ} 
\end{table}
Next, we select the events in a  series of customised \emph{Basic Cuts} (see Table~\ref{tab:BasicCutstZ} for a summary). We begin by requiring the absence of any isolated leptons (mini-ISO $> 0.7$) \cite{Kaplan:2008ie} with $p_T^{l} > 25 \GeV$ and $|\eta_{l} |< 2.5$. As the final state contains a single top, we demand at least one fat jet ($R=1.0$) with $p_T^{\rm fj} > 400 \;(600)\GeV$ and $|\eta_{{\rm fj}} |< 2.5$.

The events  which pass the Basic Cuts are subject to a set of  \emph{Complex Cuts} (see Table \ref{tab:ComplexCuts2}) to further suppress the background channels. Since we expect large $\MET$ from the boosted $Z$ boson, we require $\MET > 400 \;(600)\GeV$ for $M_{T'} = 1 \,(1.5) \TeV$. To identify the top quark we demand the hardest fat jet to pass the jet substructure selection criterion of $Ov_{3}^t > 0.6$. We require at least one $r=0.2$ forward jet ($p_T^{\rm fwd} > 25 \GeV$ and $2.5 < \eta^{\rm fwd} < 4.5$) to be present in the event, as well as at least one $r=0.4$ $b$-tagged jet inside the fat jet, (defined by $\Delta R_{\rm fj, \,b} < 1.0$). The $b$-tagging procedure employs the simplified $b$-tagging scheme described in  Section \ref{sec:b-tagging}. In order to further eliminate the large $t\bar{t}$ background, we require that all $r=0.4$ jets ($p_T^{\rm j} > 50 \GeV$ and $|\eta_j| < 2.5$)  in the event which are also isolated from the top tagged fat jet by $\Delta R > r + R$ to satisfy $\Delta \phi_{\slashed{E}_T,j} > 1.0$.
\begin{table}[h]
\setlength{\tabcolsep}{2em}
{\renewcommand{\arraystretch}{1.8}
\begin{tabular}{c|c}
							   &$T' \rightarrow Z_{\rm inv} t_{\rm had}$                   \\ \cline{1-2}
\multirow{4}{*}{ \textbf{Complex Cuts} }&$\MET > 400 \; (600)\GeV$ ,                   \\ 
		 					  &$Ov_3^t > 0.6$ ,                                        \\ 
		 					  &$N^{\rm fwd} \ge 1$, $\Delta \phi_{\slashed{E}_T,j}>1.0$ ,			   \\ 
							  & fat jet $b$-tag .                     	    	\\ 

\end{tabular}}\par
\caption{Summary of Complex Cuts for $T' \rightarrow Z_{\rm inv} t_{\rm had}$ channel. $Ov_3^t$ refers to the top tagging score with Template Overlap Method, $N^{\rm fwd}$ is the multiplicity of forward jets ($p_T^{\rm fwd} > 25 \GeV$ and $2.5 < \eta^{\rm fwd} < 4.5$), $b$-tag refers to presence of at least one $b$-tagged $r=0.4$ jet inside the fat jet which is tagged as a top. $\Delta \phi_{\slashed{E}_T,j}$ is the azimuthal distance between missing energy and all $r=0.4$ jets in the central region ($i.e. \, |\eta| <2.5$) which are also outside of the top tagged fat jet.} \label{tab:ComplexCuts2} 
\end{table}

Fig. \ref{fig:ZtOv3MET} shows examples of several observables relevant for the $Z_{\rm inv} t_{\rm had}$ event selection. The first row of Fig. \ref{fig:ZtOv3MET} displays the top template overlap distributions ($Ov_3^t$) for the $M_{T'} = 1\, (1.5) \TeV$ selection on the left (right). Both signal and true top containing backgrounds display a prominent peak at $Ov_3^t \sim 1, $ while we find that TOM is very efficient at discriminating the non-top containing backgrounds such as $Z+X$. Notice also that even though $t\bar{t}$ and $Z+t$ do contain a true top quark, a lower cut on $Ov_3^t$ still provides some background discriminating power. This can be attributed to the effects of higher order corrections on the SM backgrounds (partly included via MLM matching) which are not significant in the signal events (for a detailed discussion of higher order effects on TOM distribution see Ref. \cite{Backovic:2013bga}.). 

\begin{figure}[!]
\includegraphics[scale=0.35]{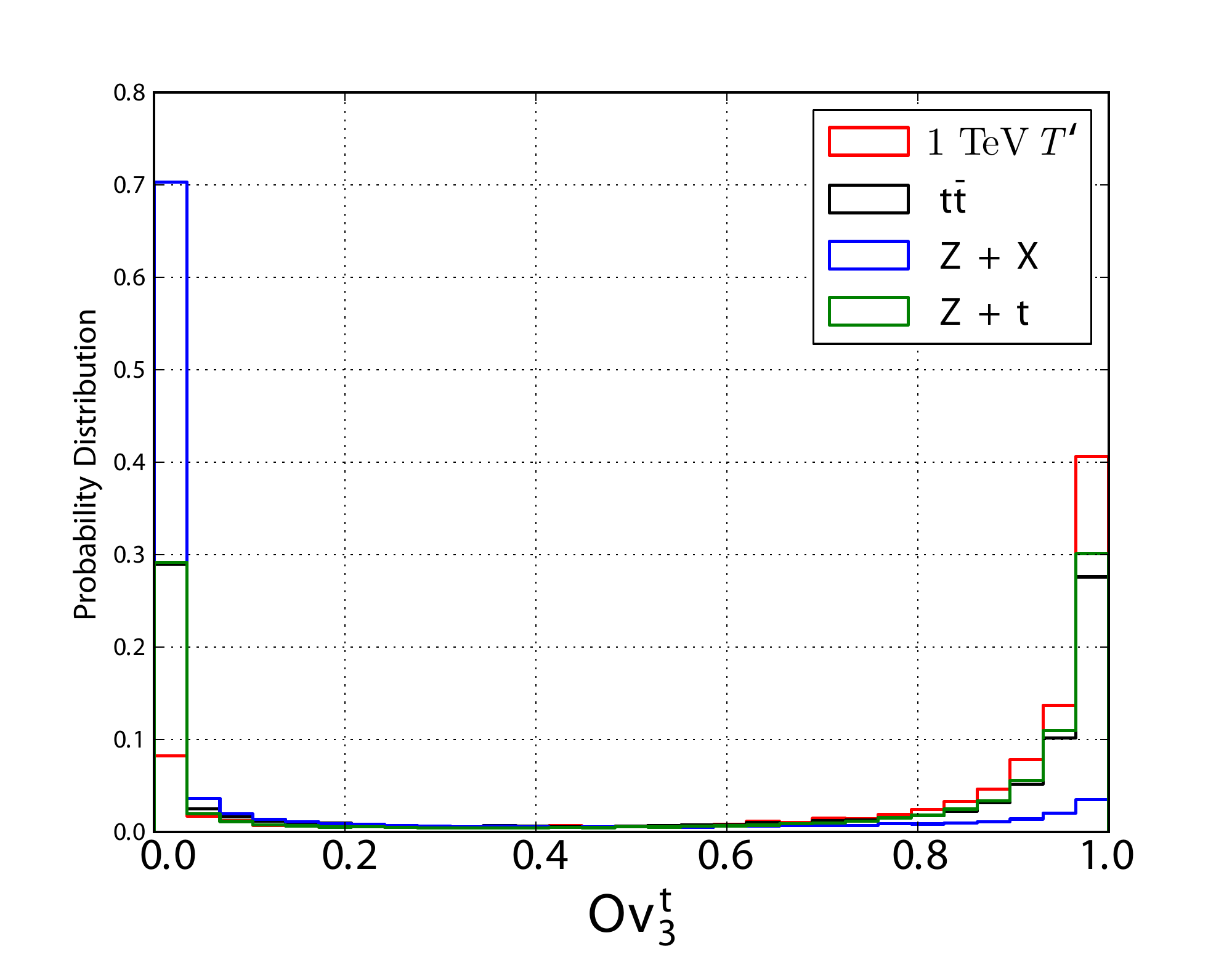}
\includegraphics[scale=0.35]{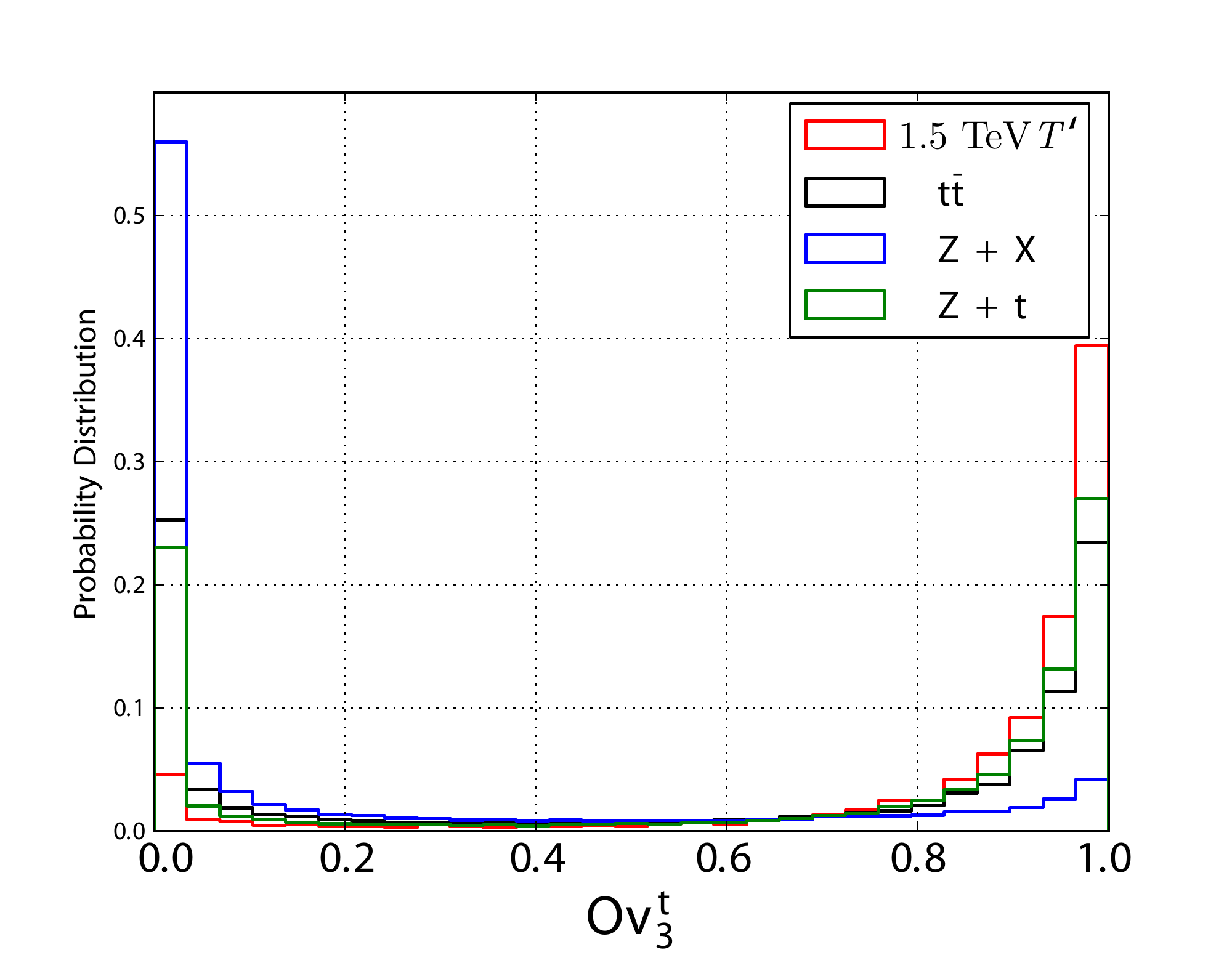}
\includegraphics[scale=0.35]{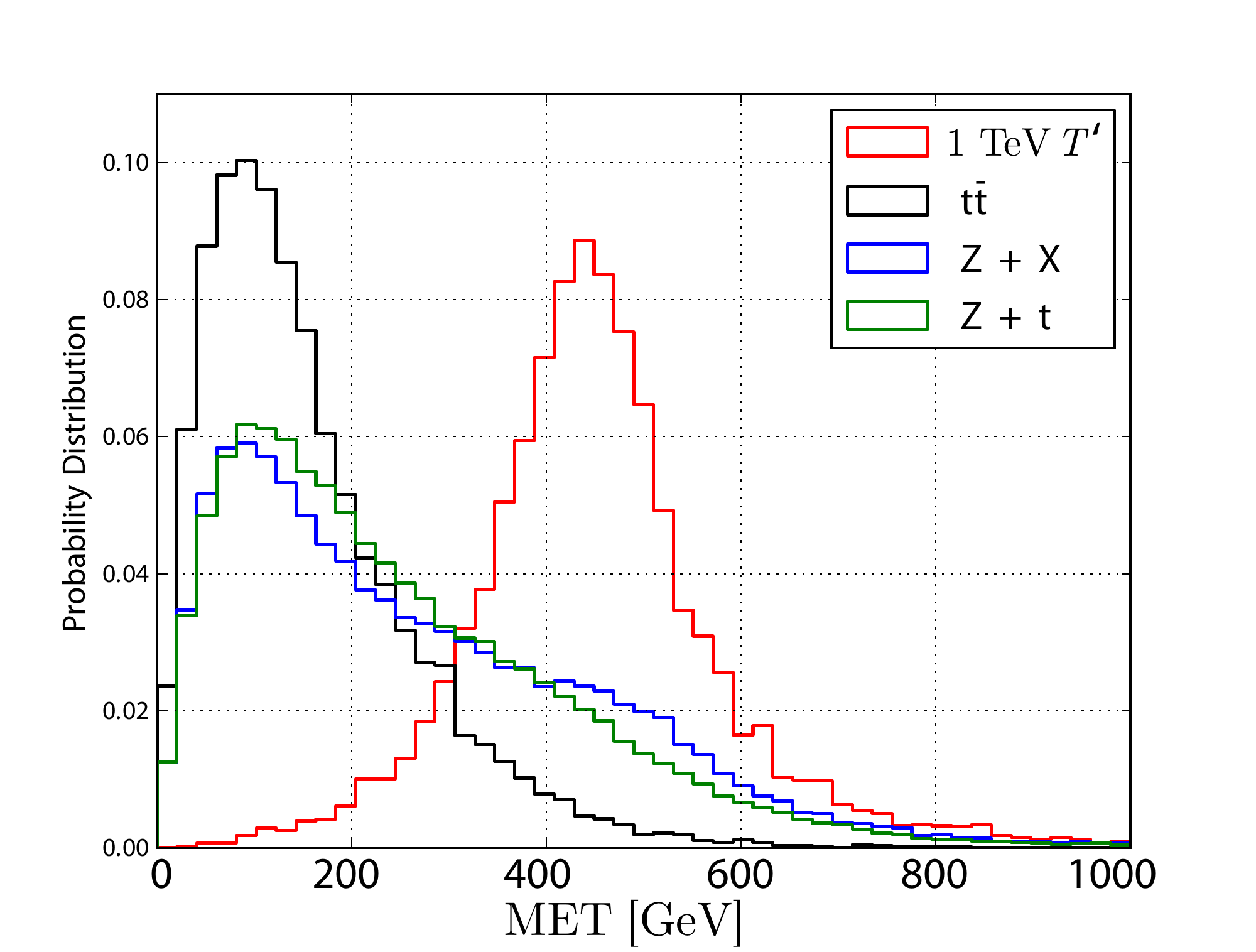}
\includegraphics[scale=0.35]{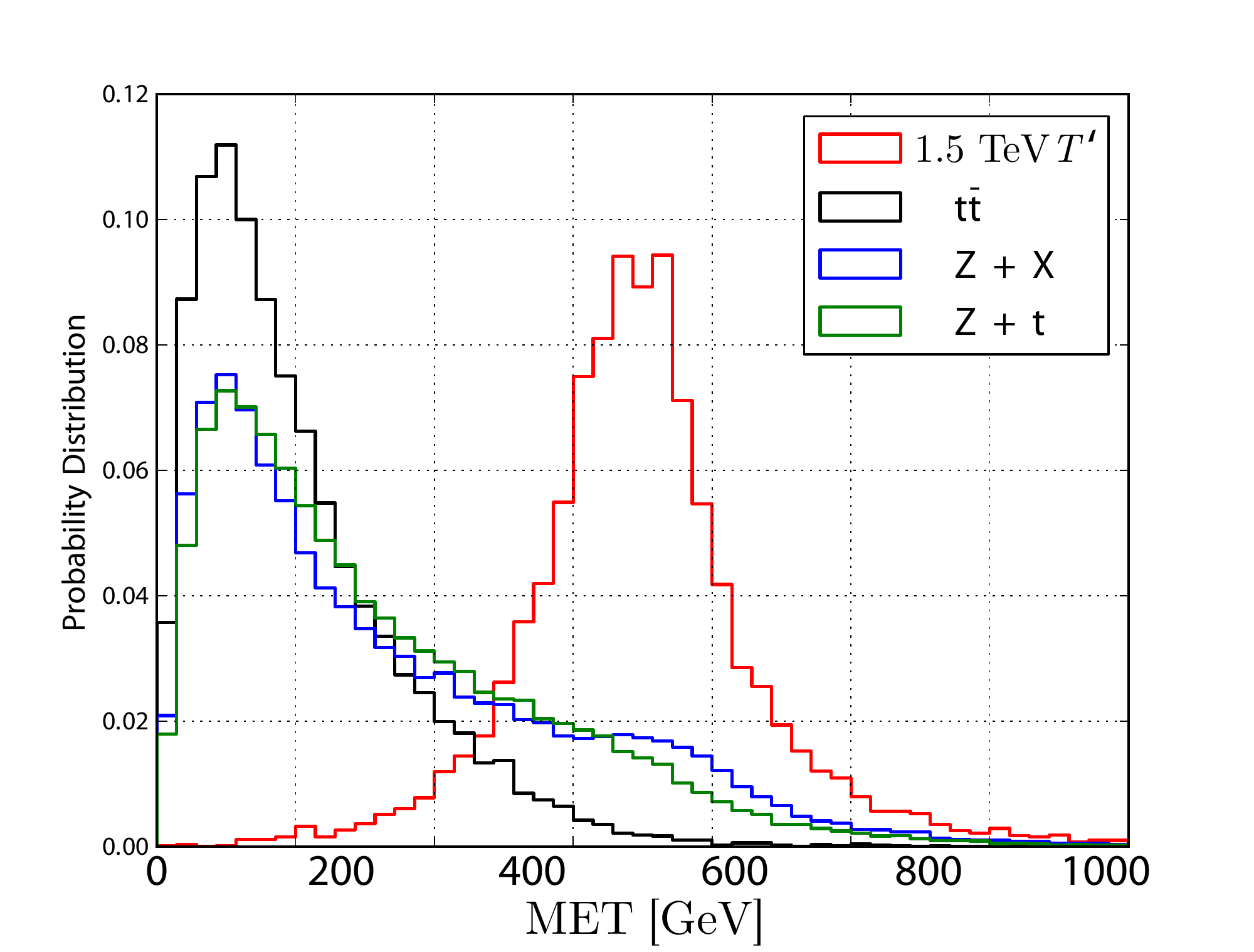}
\includegraphics[scale=0.35]{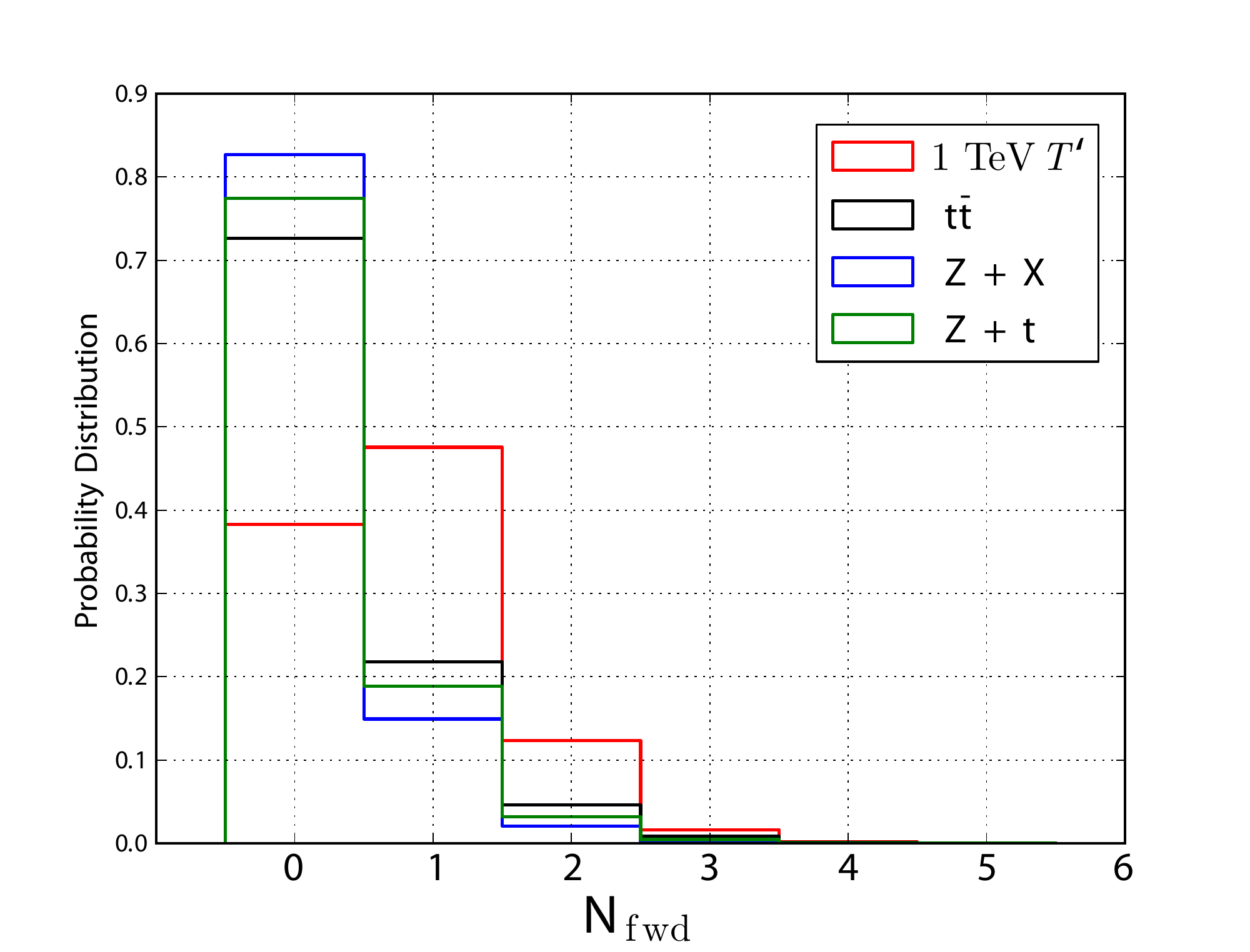}
\includegraphics[scale=0.35]{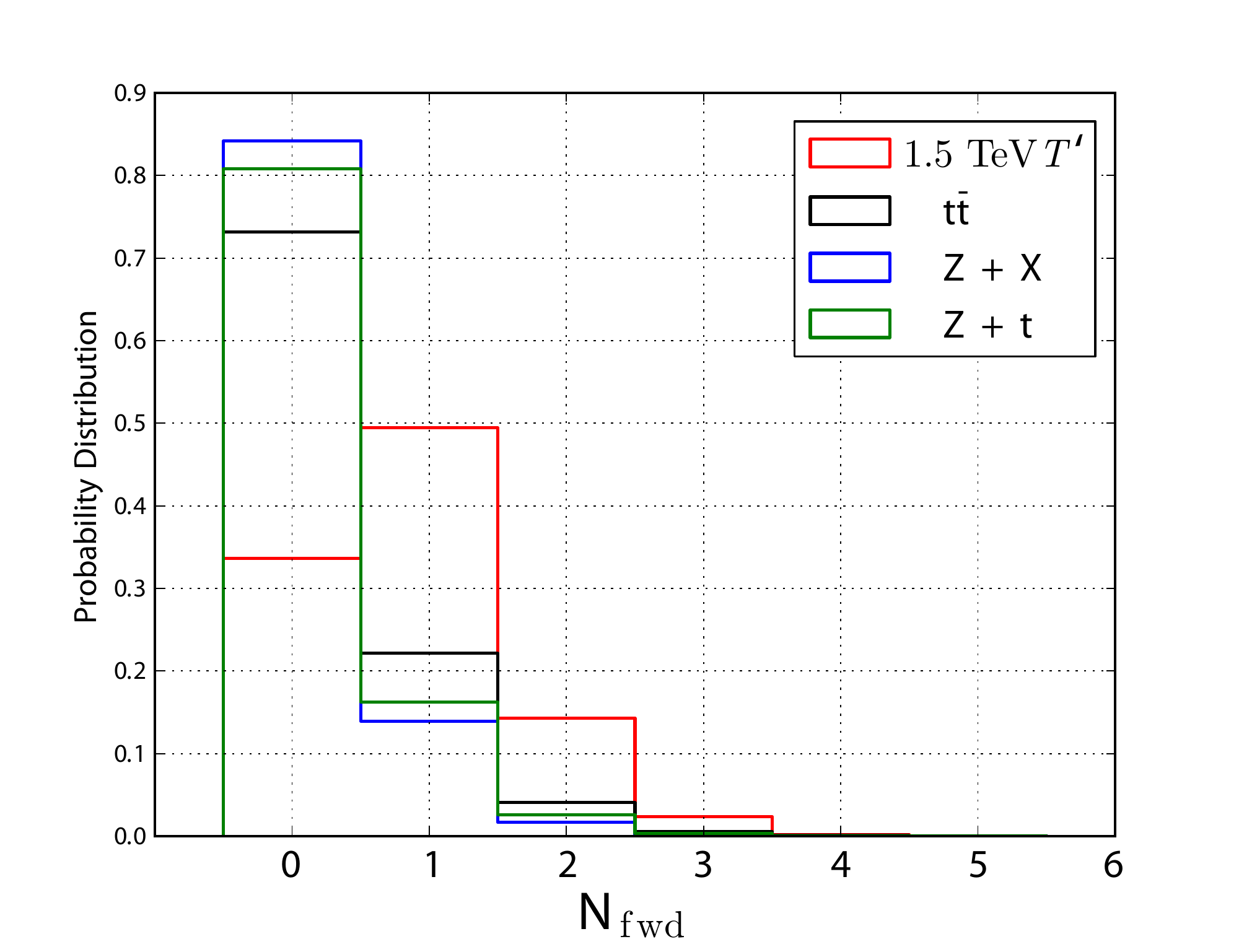}
\caption{Various kinematic distributions of (left column) 1 TeV and (right column) 1.5 TeV data on $T' \rightarrow Z_{\rm inv} t_{\rm had}$ channel after \textbf{Basic Cuts}. Panels in the first and second rows represent the top template overlap score of the hardest fat jet ($Ov^t_3$) and $\MET$ distributions respectively after \textbf{Basic Cuts}, while the third row shows the forward jet multiplicity distributions. }
\label{fig:ZtOv3MET}
\end{figure}

The second row of Fig. \ref{fig:ZtOv3MET} shows the characteristic distributions of $\MET$. The signal events show a prominent peak around $\MET \sim M_{T'} /2$, which is a direct consequence of the $Z_{\rm inv}$ being highly boosted. The background channels, on the other hand, are characterized by much lower $\MET$. The $Z$ containing backgrounds display long $\MET$ tails which extend deep into the signal region, while the $t\bar{t}$ background falls of with the increase of $\MET$ much more sharply. 

Finally, we show example distributions of the forward jet multiplicities (tagged by the prescription of Section \ref{sec:FWD-tagging}) in the third row of Fig. \ref{fig:ZtOv3MET}. 
While 70-80\% of background events contain no forward jets, we find that about 60\% of signal events contain at least one forward jet, illustrating the usefulness of forward jet tagging in the $T'$ event selection. Note also that we expect distributions of forward jet multiplicities similar to ones in Fig. \ref{fig:ZtOv3MET} to appear in all $T'$ decay channels. 

We proceed to show our first result in the cutflow Table~\ref{tab:cutflowZt}. For the purpose of illustration, we only present a parameter point with $M_{T'}=1$~TeV and $c^{T'bW}_L=0.3$
 which in our sample model gives  branching ratios of 0.51~:~0.21~:~0.28 to $Wb$, $Zt$, and $ht$, respectively. Similarly, for a 1.5 TeV partner search, we use the point with $M_{T'}=1.5$~TeV and $c^{T'bW}_L=0.3$
which results in the branching ratios of 0.52~:~0.22~:~0.25 to $Wb$, $Zt$, and $ht$.

\begin{widetext}
\begin{center}
\begin{table*}[h]
\begin{tabular}{|c||cccc|cc||cccc|cc|}
\hline
\multirow{2}{*}{$T' \rightarrow Z_{\rm inv} t_{\rm had}$}&\multicolumn{6}{c||}{$M_{T'}=1.0$ TeV search}&\multicolumn{6}{c|}{$M_{T'}=1.5$ TeV search}\\ \cline{2-13}
&signal& \hspace{5pt}$t\bar{t} \hspace{5pt}$&$Z+X$&$Z+t$&$S/B$&$S/\sqrt{B}\, (100 \, \mathrm{fb}^{-1})$& signal &\hspace{5pt}$t\bar{t}$ \hspace{5pt}&$Z+X$&$Z+t$&$S/B$&$S/\sqrt{B} \,(100 \, \mathrm{fb}^{-1})$\\\hline
preselection                 &4.9                    &26000              &21000     &44        &0.00011    &0.23                &1.3               &5200       &5300    &12         &0.00012    &0.12   \\
Basic Cuts                       &3.5                    & 900                  &6100      &11        &0.00050      &0.42              &1.0               &140        &1200      &2.4       &0.00074       &0.27\\
$Ov^t_{3} > 0.6$            &2.7                    &510                   &840        &6.5        &0.0020      &0.75               &0.87             &81         &230         &1.6       &0.0028   &0.49 \\
$b$-tag                          &1.8                    &300                    &28          &4.1        &0.0055      &1.0               &0.51              &42         &6.7        &0.9      &0.010     &0.72   \\
$\slashed{E}_T >400 \,(600)\, \GeV$ 	    &1.2              &13                     &8.3         &0.84       &0.055    &2.6            & 0.39      &0.95      &1.4       &0.13       & 0.16 & 2.5 \\
$N_{\mathrm{fwd}}\geq 1$             &0.75     &2.5           &1.2        &0.25      &0.19 &3.8           & 0.26     &0.19     &0.23        &0.039     &0.58 & 3.9\\

$| \Delta \phi_{\slashed{E}_T,j} | >1.0$& 0.62  &0.89          &0.91        &0.21       & 0.31     & 4.4        & 0.21     &0.072     &0.17     &0.031    & 0.78   & 4.1\\ \hline
\end{tabular}
\caption{Example cutflow for signal (simulated for $c^{T'bW}_L = 0.3$ with $M_{T'}=1.0$ TeV and $M_{T'}=1.5$ TeV)  and background events in the $T' \rightarrow Z_{\rm inv} t_{\rm had}$ channel for $\sqrt{s} = 14 \, \mbox{TeV}$ LHC. The entries show cross sections after the respective cuts for signal and background channels in fb. The $S/\sqrt{B}$ values are given for a luminosity of $100\, \mathrm{fb}^{-1}$ for the $M_{T'}=1.0$ TeV and $M_{T'}=1.5$ TeV searches. The label $\MET > 400 (600) \GeV$ refers to 1 (1.5) TeV partners respectively. Efficiencies for $b$-tagging are included in the results. }
\label{tab:cutflowZt}
\end{table*}
\end{center}
\end{widetext}

For a 1 TeV partner, Table \ref{tab:cutflowZt} shows that boosted top tagging techniques combined with $b$-tagging can efficiently suppress the background channels which do not contain a true top quark ($Z+X$), where we find an overall improvement in $S/B$ by a factor of $\sim 11$ at a $70\%$ signal efficiency relative to the Basic Cuts. $b$-tagging efficiently reduces the $Z$+X background, where we see and improvement of a factor of $\sim 3$ in $S/B$ at a total $65\%$ $b$-tagging efficiency. Further requirements on large $\MET$ prove to be powerful discriminants of the $t\bar{t}$ backgrounds, with an overall improvement by a factor of $\sim 15$ at an additional $70\%$ signal efficiency relative to the $b$-tagging selection. Demanding a presence of a forward jet in the event delivers an extra improvement in $S/B$ by a factor of $\sim 3$, while requiring $\MET$ to be isolated from other jets in the event improves $S/B$ by an additional factor of $\sim 2$. Our results show that the benchmark points used in Table~\ref{tab:cutflowZt} are nearly discoverable at LHC14 with as little as $100 \fb^{-1}$.

Moving to 1.5 TeV partners, we find that boosted top tagging, $b$-tagging, forward jet tagging and $\MET$ isolation selections  result in similar efficiencies and overall improvement in $S/B$ compared to a search for 1 TeV top partner. However, because of a larger $Z$ boost in the signal events, and hence a higher expected $\MET$, we are able to suppress the $t\bar{t}$ background  more efficiently at higher top partner masses by increasing the cut on missing energy. The final background composition in the case of a 1.5 TeV partner  appears quite different compared to searches for $T'$ of lower masses. Upon the $ \Delta \phi_{\slashed{E}_T,j}$ cut, the $Z+X$ background contributes twice as much compared to SM $t\bar{t}$, while $t\bar{t}$ contribution to the total background was comparable to $Z+X$ in the case of a 1 TeV partner. The effect is mainly due to the tighter $\MET$ cut we apply in case of the 1.5 TeV $T'$ search, which results in an increase in $t\bar{t}$ rejection power of roughly a factor of 2.

Comparison of these results with the results for $T' \rightarrow Z_{ll} t_{{\rm had}}$ presented in Appendix \ref{app:Zt} show that the $T' \rightarrow Z_{\rm inv} t_{{\rm had}}$ is a viable discovery channel for singly produced $T'$ which in our sample study performs comparable (slightly better) than the dilepton channel already for $M_{T'} = 1$~TeV and gains more advantage at higher $M_{T'}$. 

\subsection{$T' \rightarrow W_{\rm lep} b$ Channel }
\label{sec:Wb}

The $W_{\rm lep} b$ channel \footnote{Throughout the paper, we refer to ``leptons'' as muons and electrons only.} is perhaps the simplest $T'$ decay mode to analyze, due to the limited number of reconstructed objects in the final state, and the lack of need for boosted heavy jet tagging on signal events. 
The main SM backgrounds for the $W_{\rm lep} b$ channel are SM $W_{\rm lep}$ + jets and $t\bar{t}$(semi-leptonic) + jets, where we included up to 3 extra jets for $W_{\rm lep}$+jets  and up to 1 additional jet to $t\bar{t}$ in the simulations. We have checked that other background processes, such as the single top or di-boson production are negligible at the $H_T$ characteristic of TeV scale top partner searches.

As in the previous section, we simulate all the background channels with the preselection cuts described in Section \ref{sec:PreCuts} where we demand $H_T > 500~ (750)$~GeV for a hypothetical  the top partner with mass of $1~(1.5)$ TeV. Table~\ref{tab:TotalBackGroundswb} summarizes the background cross sections including a conservative K-factor of 2.

\begin{table}[h]
\begin{center}
\begin{tabular}{|c|c|c|c|}
\hline
		Signal Channel								&	Backgrounds 						& $\sigma (H_T > 500\GeV) [{\rm fb}]$& $\sigma (H_T > 750\GeV) [{\rm fb}]$  \\ \hline
\multirow{2}{*}{$T' \rightarrow W_{\rm lep} b$}                           & $W_{l \nu}$ + jets 	                                           &  $ 4.1  \times 10^4 $	                  &	$ 1.0     \times 10^4 $		           \\
											&$t\bar{t}$(semi-leptonic) + jets                          &	$ 2.1    \times 10^4 $		            &   $  4.2    \times 10^3 $			   \\ 
\hline		
\end{tabular}
\end{center}
\caption{The simulated cross sections of SM backgrounds (including a conservative  NLO K-factor of 2 after preselection cuts described in Section~\ref{sec:PreCuts}.}
\label{tab:TotalBackGroundswb}
\end{table}

We begin the event selection with a set of Basic Cuts, customized to exploit the unique event topology and kinematic features of the $l$+$\MET$+$b$ channel (see Table \ref{tab:BasicCutswb} for a summary). First, we require exactly one isolated lepton in the event (mini-ISO $> 0.7$  with $p_T^{l} > 25 \GeV$ and $|\eta_{l} |< 2.5$), as well the presence of at least one $p_T^{\rm fj} > 200 \, (400) \GeV $ fat jet with $|\eta_{\rm fj} | < 2.5$, in case of 1 (1.5) TeV $T'$ respectively.

\begin{table}[h!]
\setlength{\tabcolsep}{2em}
{\renewcommand{\arraystretch}{1.8}
\begin{tabular}{c|c}

							 &  $T' \rightarrow W_{\rm lep} b$	   			                                     \\ \cline{1-2}
\multirow{2}{*}{\textbf{Basic Cuts} }	 &  $N_{\rm fj}  \geq 1$ ($R=1.0$),	$N^{\rm iso}_{\rm lepton} = 1$ ,   \\ 
		 					 &  $p_T^{\rm fj} > 200 \;(400) \GeV$, $|\eta_{\rm fj} |< 2.5$ .            \\  
\end{tabular}}\par
\caption{Summary of Basic Cuts for $T' \rightarrow W_{\rm lep} b$ channel. ``fj" stands for the fat jet with $|\eta_{\rm fj}| < 2.5$ and $p_T > 200 \,(400) \GeV$ for $M_{T'} = 1 (1.5) \TeV$ and $N^{\rm iso}_{\rm lepton}$ represents the number of isolated leptons with mini-ISO $> 0.7$, $p_T^{l} > 25\GeV$ and $|\eta_{l} |< 2.5$. } \label{tab:BasicCutswb} 
\end{table}

\begin{figure}[h!]
\includegraphics[scale=0.35]{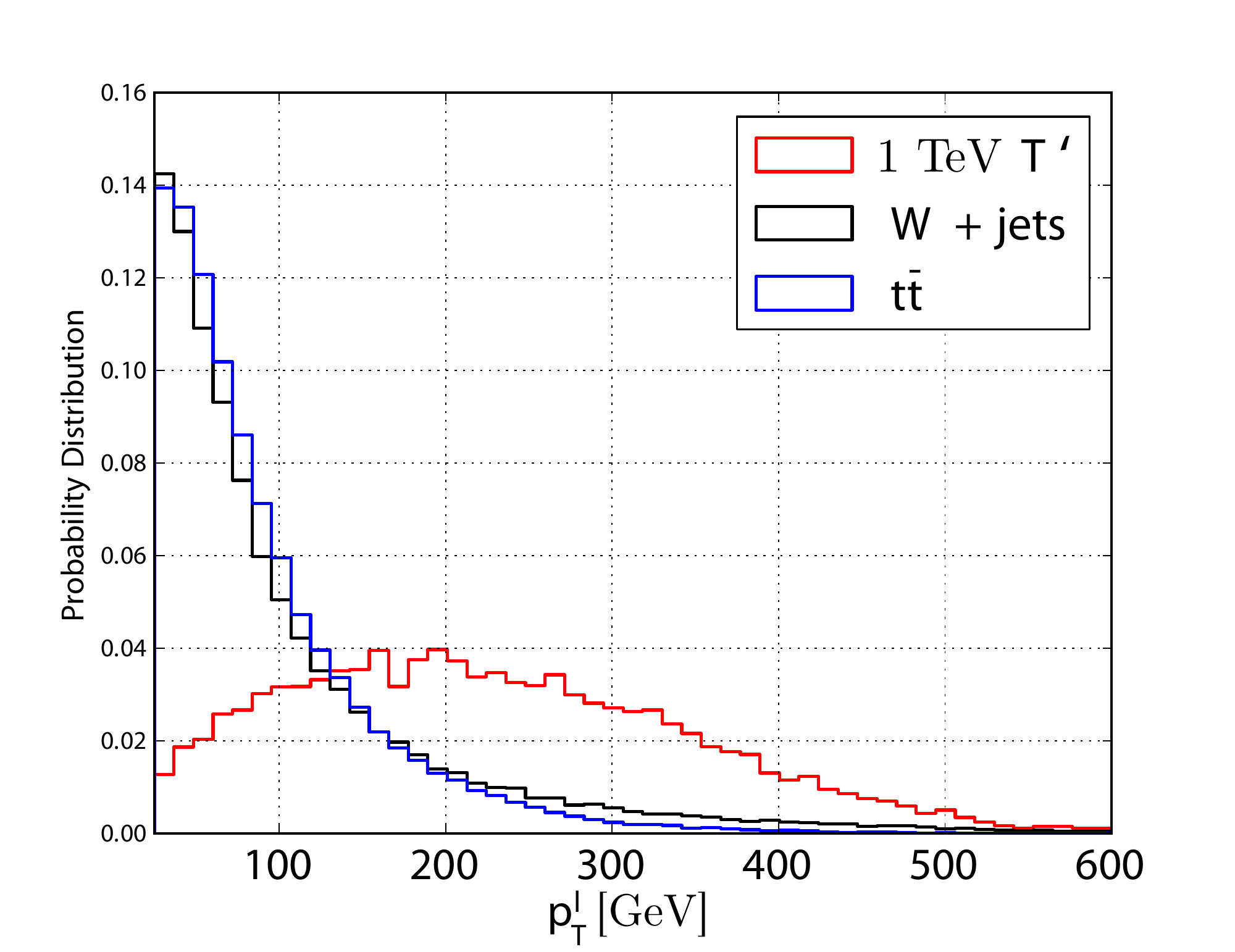}
\includegraphics[scale=0.35]{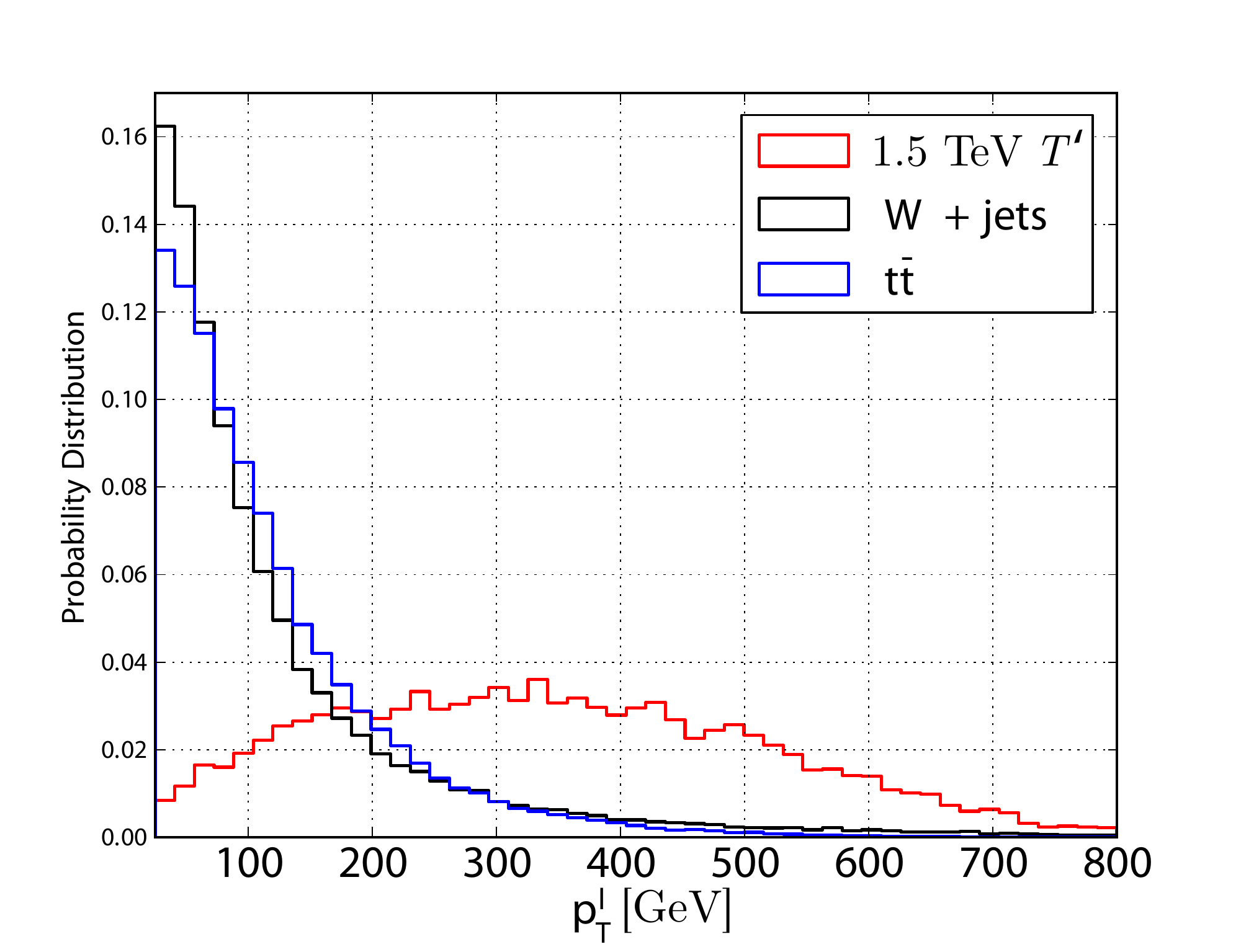}
\includegraphics[scale=0.35]{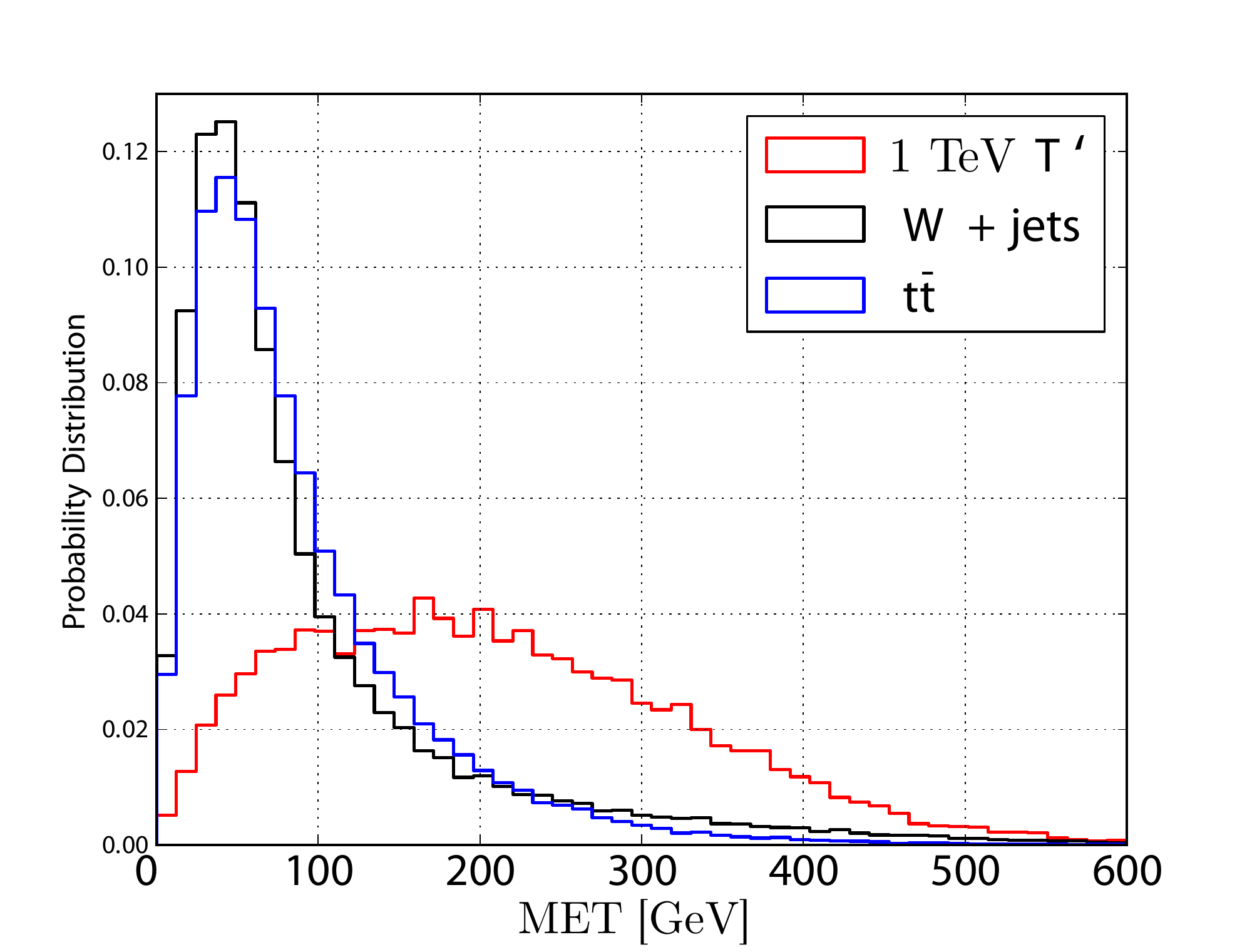}
\includegraphics[scale=0.35]{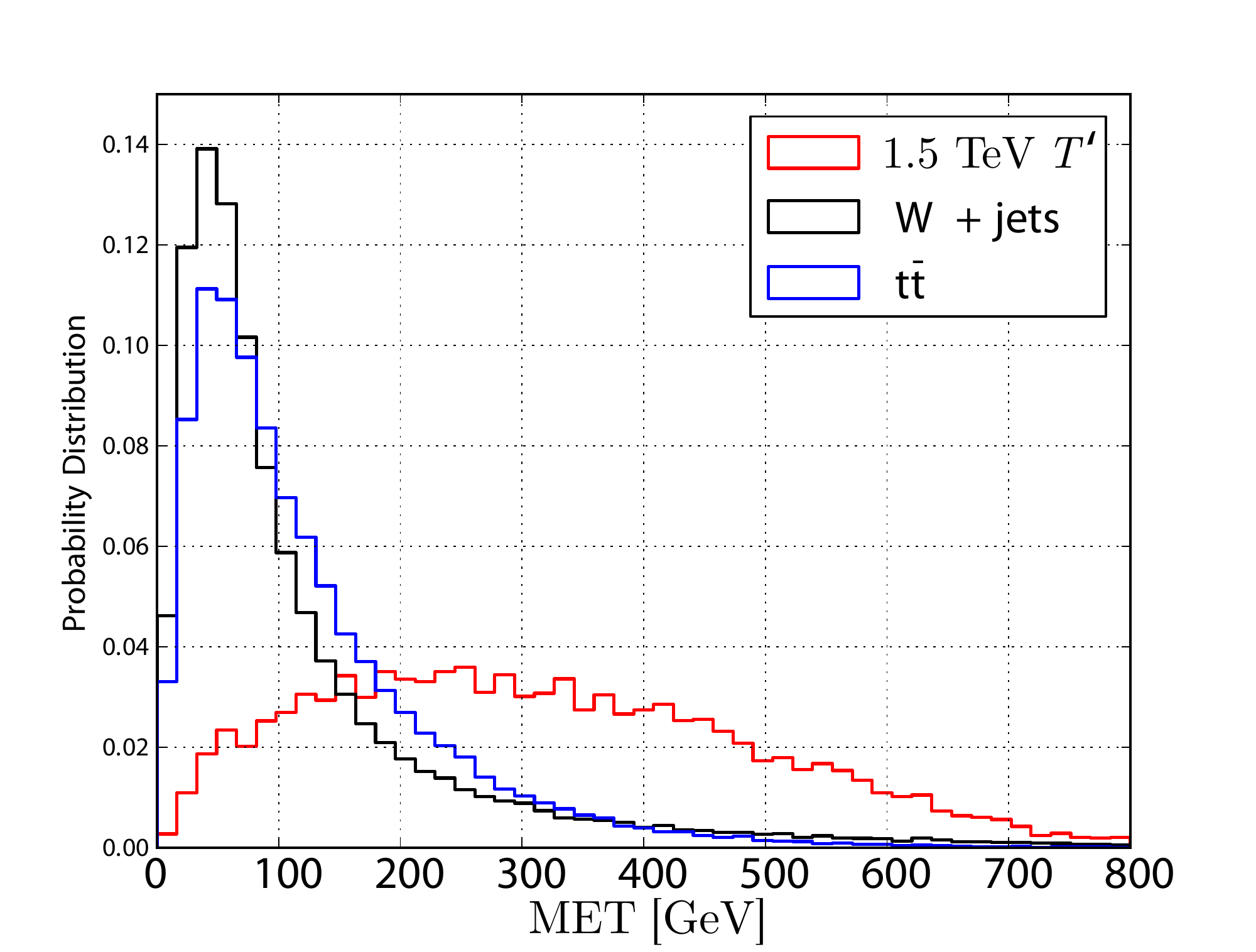}
\caption{Distributions of hardest isolated lepton $p_T$ and $\MET$ in searches for $T' \rightarrow W_{\rm lep} b$ channel after Basic Cuts. The left (right) panels show the distributions relevant for 1 (1.5) TeV searches respectively.}
\label{fig:WbpTlMET}
\end{figure}

\begin{table}[h!]
\setlength{\tabcolsep}{2em}
{\renewcommand{\arraystretch}{1.8}
\begin{tabular}{c|c}

							    &  $T' \rightarrow W_{\rm lep} b$   			  \\ \cline{1-2}
\multirow{5}{*}{ \textbf{Complex Cuts} }&  $p_T^{l}, \MET >100 \; (150)$ GeV ,                       \\[-.1cm]
							  &   $\Delta R_{{\rm fj},\, b} < 1.0$ , $p_T^b > 50 \GeV$ , $\Delta R_{b,\,l} > 1.4$ ,\\[-.1cm]
		 					   & $p_T^{\rm fj} > 350  \; (500)$ GeV, $m_{\rm fj} < 130$   GeV , \\ [-.1cm]
		 					  &   $M_{T'} > 750 \; (1000)$  GeV ,                  		\\ [-.1cm]
							  &   $N_{\mathrm{fwd}}\geq 1$ .         			\\[-.1cm]

\end{tabular}} \par
\caption{The summary of Complex Cuts for $T' \rightarrow W_{\rm lep} b$ channels. The label ``$b$'' refers to a $b$-tagged $r=0.4$ jet, while the values outside (inside) parenthesis show the choice of cuts for 1 (1.5) TeV $T'$ searches. } \label{tab:ComplexCuts} 
\end{table}

Next, we apply a set of Complex Cuts defined in Table \ref{tab:ComplexCuts} to all signal candidate events. Fig. \ref{fig:WbpTlMET} shows example distributions of some observables useful in discriminating the large SM backgrounds. 
As we expect signal events to contain a boosted $W$ boson, we impose a cut of $p_T^{l}, \MET >100 \; (150)$ GeV, for $M_{T'} = 1\,(1.5) \TeV$ respectively. As we show in Fig. \ref{fig:WbpTlMET}, the  $p_T^{l}$ and $\MET$ distributions show significant differences between the signal and background events (both in case of $W$+jets and $t\bar{t}$), where the leptons and $\MET$ originating from signal events are on average much harder compared to the background channels.

\begin{figure}[htb]
\includegraphics[scale=0.35]{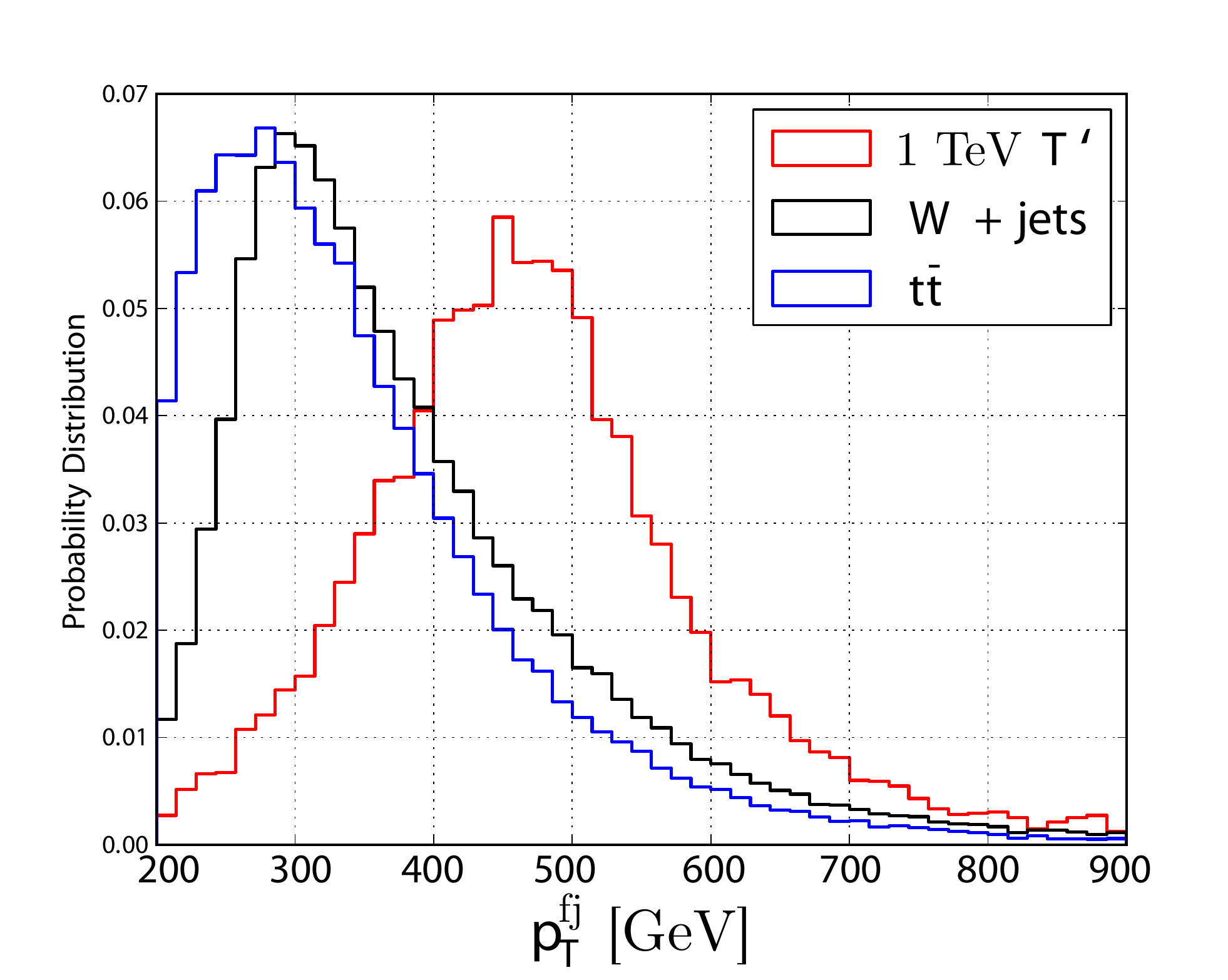}
\includegraphics[scale=0.35]{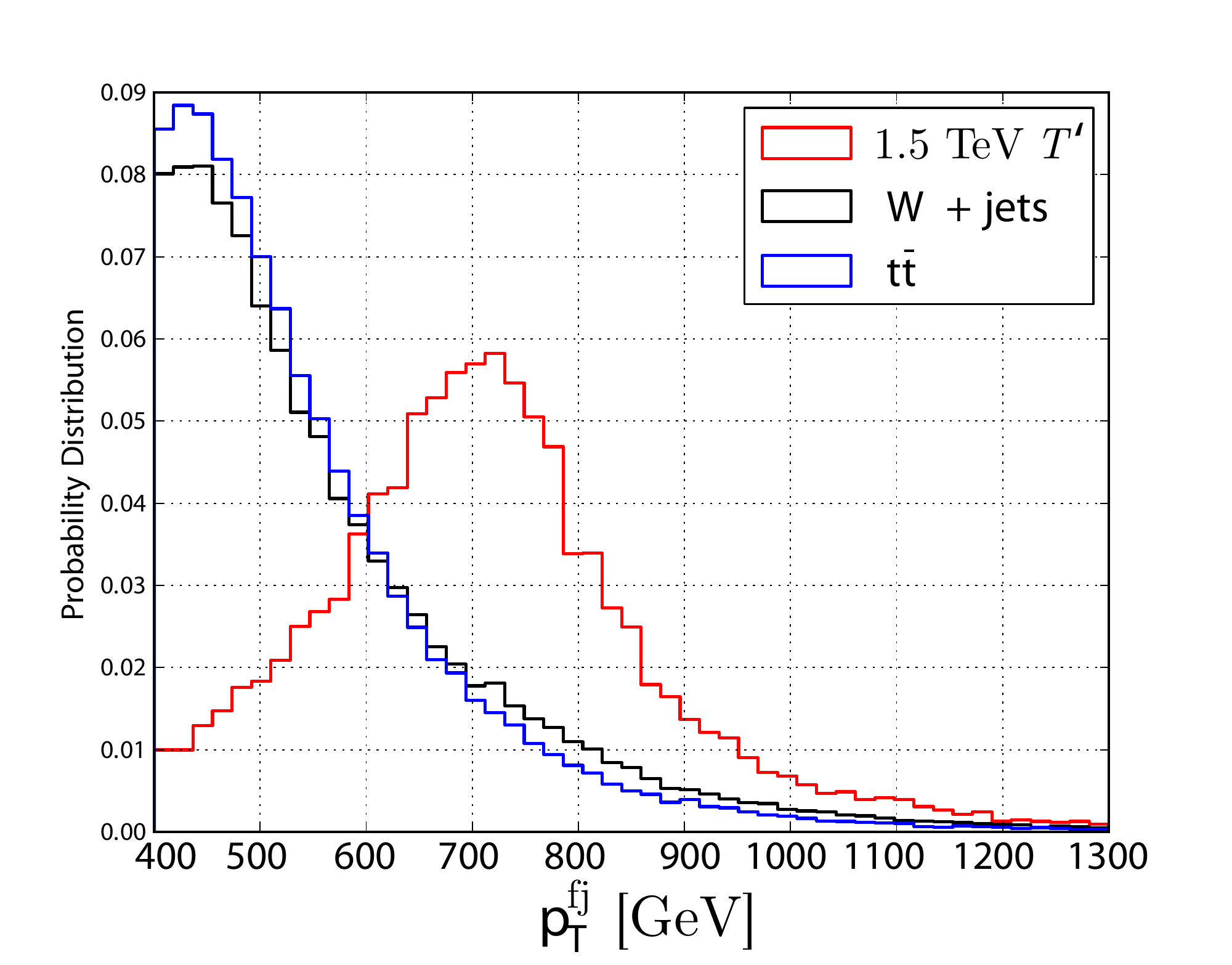}
\includegraphics[scale=0.35]{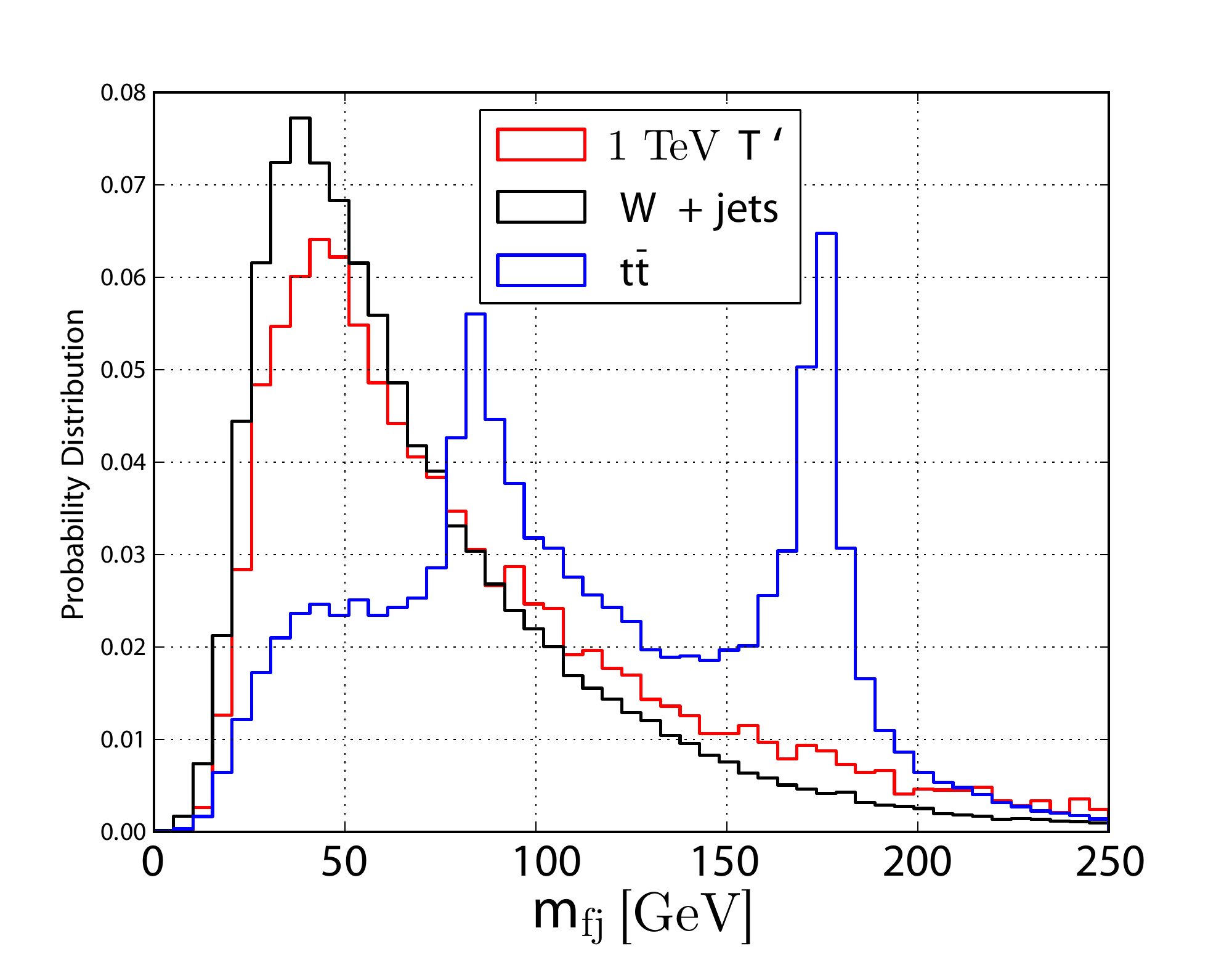}
\includegraphics[scale=0.35]{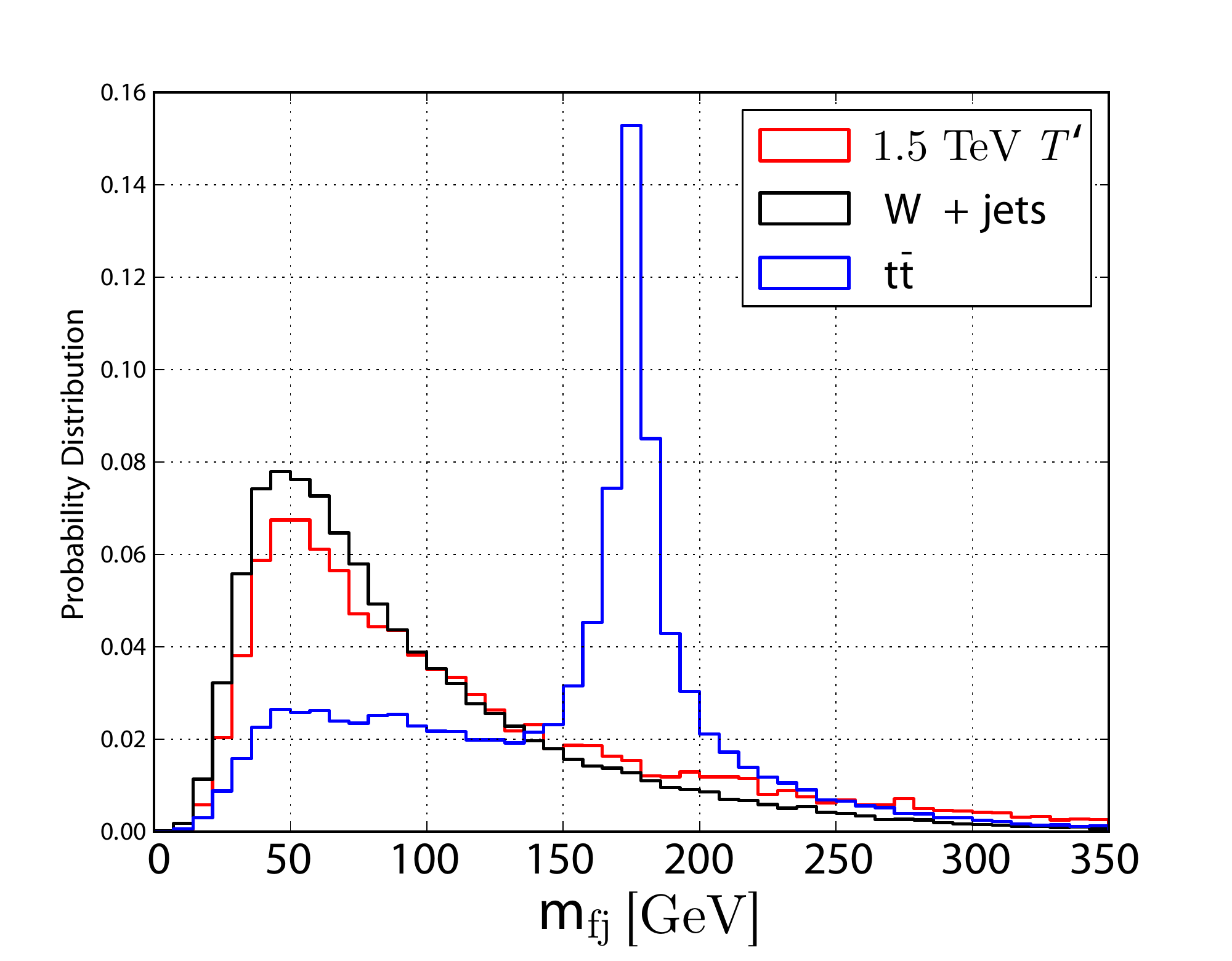}
\caption{Fat jet $p_T$ and mass for the $T'$ searches in the $T' \rightarrow W_{\rm lep} b$ channel after Basic Cuts. The first row shows the $p_T$ distribution of the highest $p_T$ fat jet ($R=1.0$), while the second row shows the mass of the same jet. The results relevant for 1 (1.5)  TeV $T'$ searches are on the left (right) panels.}
\label{fig:WbpTmfj}
\end{figure}

Our Complex Cuts also include elements of the $Wb$ final state analysis proposed in Ref. \cite{Ortiz:2014iza}. We expect the hardest fat jet in the signal events to be the $b$ quark from the resonance decay. However, $t\bar{t}$ background can easily mimic this configuration unless additional features of the fat jet are taken into consideration.  Following the prescription of Ref. \cite{Ortiz:2014iza} we consider the fat jet  $p_T$ mass as well as the $b$-tagging properties as an additional handle on the $t\bar{t}$ background.  As the $b$-jet in the signal events originates from a heavy resonance, we expect its $p_T$ distribution to peak around $M_{T'}/2$, while the background events should display a much softer spectrum. In addition, the fat jets of $t\bar{t}$ processes are expected to peak around the top mass, modulo issues with a finite fat jet cone size which could result in significant leakage of radiation outside the fat jet cone and hence an overall low mass tail in the fat jet mass distribution.

 We find that we can efficiently ``anti top tag'' the fat jet by requiring $p_T^{\rm fj} > 350  \; (500) \GeV$ for 1 (1.5)~TeV top partners, presence of a $b$-tagged jet with $p_T^b > 50$ GeV inside the fat jet (while simultaneously being isolated from the hard lepton in the event by $\Delta R_{b,\,l} > 1.4$) and $m_{\rm fj} < 130 \GeV$. Fig. \ref{fig:WbpTmfj} shows the kinematic distribution of the hardest fat jet relevant for the anti top tagging selection. The transverse momentum of the fat jet originating from the signal events is determined by the mass of the top partner, while the background channels display a much softer fat jet $p_T$ spectrum. The upper cut on the fat jet mass serves as a good discriminant of the $t\bar{t}$ background in the boosted regime (note that we imposed an $H_T$ cut on the background channels at generation level), as seen in the bottom panels of Fig.~\ref{fig:WbpTmfj}. Considerations of fat jet cone size which varies with $h_T^W \equiv p_T^l + \MET$ could further improve the performance of the mass cut, as larger fat jet cones at lower $h_T^W$ could reduce the fraction of $t\bar{t}$ events in the low mass tail. Fig. \ref{fig:WbpTmfj} partly illustrates this effect, where it is evident that the fraction of events with $m_{\rm fj} < 130 \GeV$ is significantly lower for $1.5 \TeV$ $T'$ compared to the $1 \TeV$ top partner using a fixed cone of $R=1.0$. 

 Note that when considering realistic implementations of the above mentioned anti top tagging selections, one needs to take into account the fact that jet mass will be susceptible to the intense pile-up environment expected at LHC14. However, current literature suggest that the effects of pileup on the jet mass could (at least to a good degree) be mitigated by existing pileup subtraction/ correction techniques \cite{Krohn:2009th, Ellis:2009me, Alon:2011xb, Aad:2012meb,Chatrchyan:2011ds, Cacciari:2007fd}.

The boosted  topology of signal events offers simple ways to reconstruct the mass of $T'$ and further suppress the SM backgrounds. The fact that signal events are characterized by a boosted $W$, and hence a lepton and $\MET$ which are highly collimated,  allows for an efficient use of the simple collinear approximation $\eta_\nu = \eta_l, $ where $\nu$ is the total missing transverse momentum in the event and $l$ is the isolated lepton. The collinear approximation offers a way to reconstruct the top partner mass by simple addition of the small-radius $b$ quark, an isolated lepton and missing $\MET$, where $\eta_{\MET} = \eta_l$. Fig. \ref{fig:WbMT} shows example distributions of the reconstructed top partner mass. In both cases, we find that the collinear approximation reproduces the top partner mass in the signal events to an excellent degree, while the background distributions peak at much lower values, mainly due to the fact that they are characterized by significantly softer leptons and $\MET$. 
In order to further suppress the SM backgrounds, we impose a lower mass bound  of $M_{T'} > 750\; (1000) \GeV$ for 1 (1.5) TeV top partners respectively.

Finally, as in all other $T'$ decay modes, we conclude the event selection in the $W_{\rm lep }b$ channel by requiring the presence of at least one $r=0.2$ forward jet ($p_T^{\rm fwd} > 25 \GeV$ and $2.5 < \eta^{\rm fwd} < 4.5$) .

\begin{figure}[t]
\includegraphics[scale=0.35]{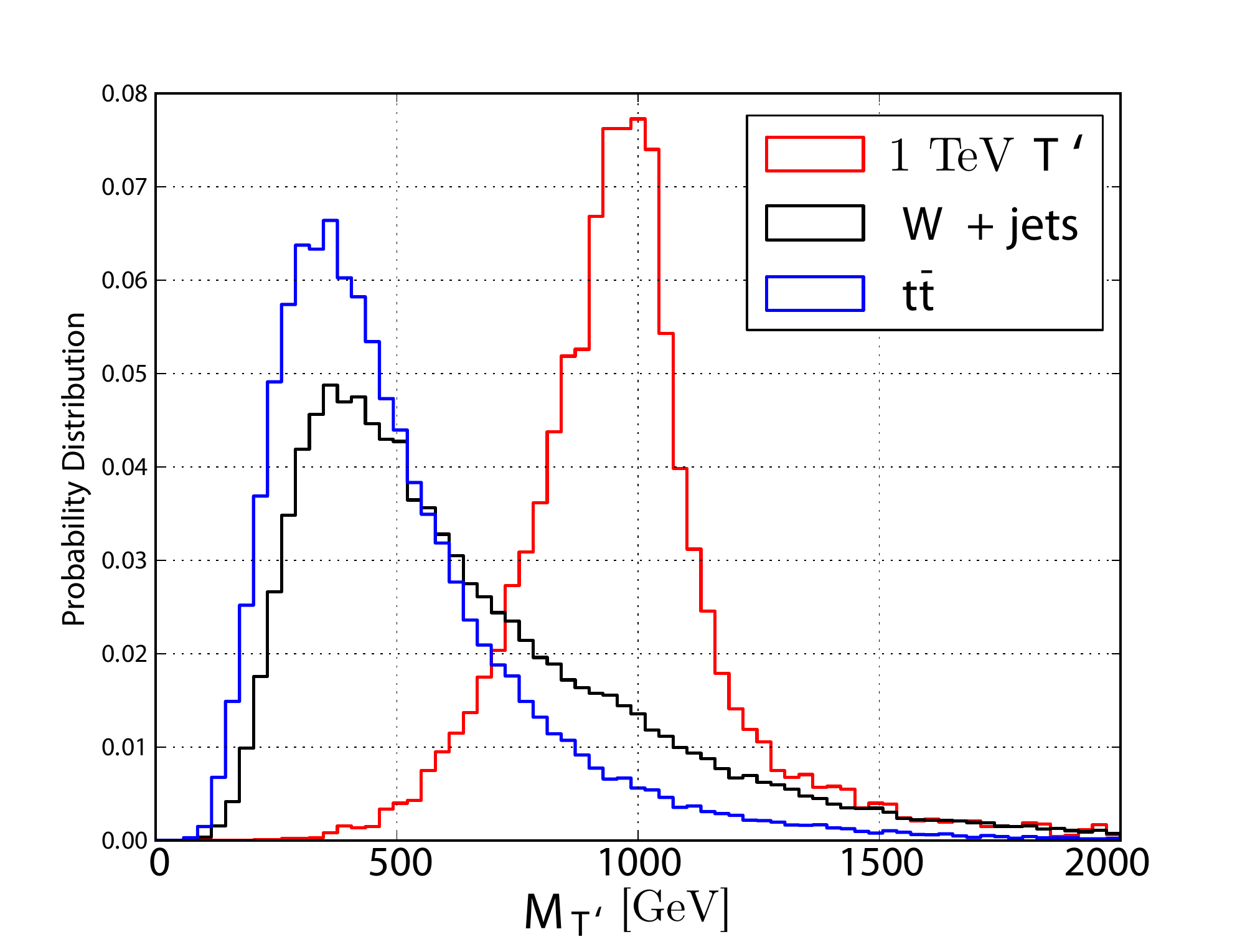}
\includegraphics[scale=0.35]{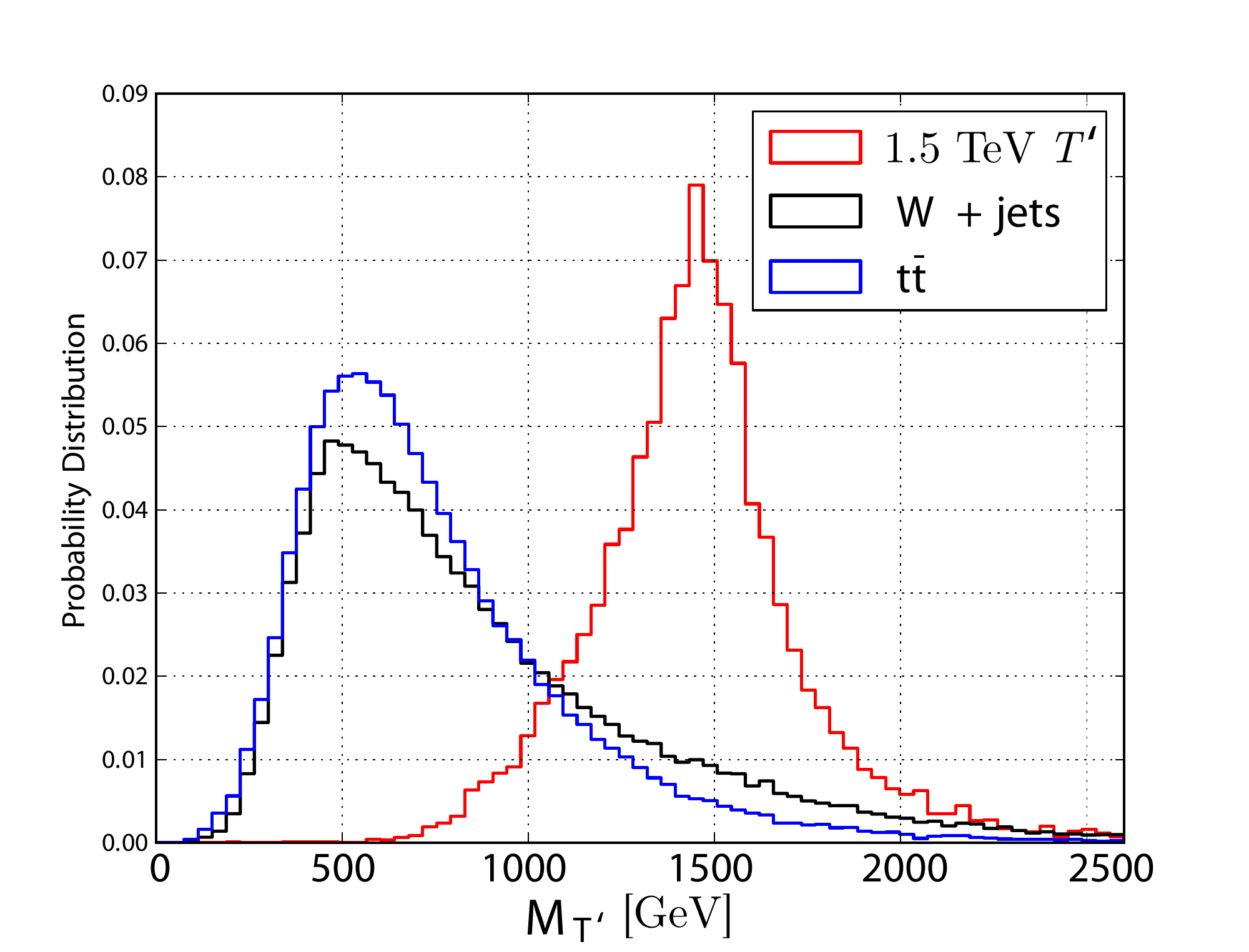}
\caption{Reconstructed mass of $T'$ in signal and background events in the case of a 1 TeV $T'$ (left) and 1.5 TeV $T'$ (right). The mass reconstruction assumes a collinear approximation of $\eta_{\MET} = \eta_l$, due to the expected boosted $W$ boson in the signal events. All plotted events assume Basic Cuts.}
\label{fig:WbMT}
\end{figure}

Table \ref{tab:cutflowwb} shows an example cutflow for a signal benchmark point (simulated for $c^{T'bW}_L = 0.3$ with $M_{T'}=1.0$ TeV and $M_{T'}=1.5$ TeV) and relevant backgrounds. We find that requiring a hard lepton and large missing energy improves $S/B$ by a factor of $\sim 5$ in the case of a 1 TeV top partner and $\sim 8$ for the 1.5 TeV partner event selection, at a $\approx 65 -70\%$ signal efficiency relative to the Basic Cuts. The selections on the large-radius fat jet ($p_T^{\rm fj} > 350 \; (500)$, $m_{\rm fj} < 130$ and presence of a $b$-tagged $r=0.4$ jet inside the fat jet) are particularly effective in reducing $t\bar{t}$ backgrounds in the 1.5 TeV top partner search, where we find an improvement  by a factor of $\sim 25 - 30$ in $S/B$  at an efficiency of an additional $\sim 45-50\%$. 

\begin{widetext}
\begin{center}
\begin{table*}[h]
\begin{tabular}{|c||ccc|cc||ccc|cc|}
\hline
\multirow{2}{*}{$T' \rightarrow W_{\rm lep} b$}&\multicolumn{5}{c||}{$M_{T'}=1.0$ TeV search}&\multicolumn{5}{c|}{$M_{T'}=1.5$ TeV search}\\ \cline{2-11}
                                                & signal       & $W+{\rm jets}$   & \hspace{5pt} $t\bar{t} \hspace{5pt}$  & $S/B$     & $S/\sqrt{B} $& signal   & $W+{\rm jets}$    & \hspace{5pt} $t\bar{t} \hspace{5pt}$  & $S/B$      & $S/\sqrt{B} $\\\hline
preselection                             &31             &$4.1\times 10^4  $       &$2.1\times 10^4 $         &$5.0\times 10^{-4} $  &1.2                                                    &5.8         &$1.0\times 10^4    $       &4200                                                    &$4.1\times 10^{-4}  $  &0.48   \\
Basic Cuts                                 &29            & $3.0\times 10^4   $      &$1.5\times 10^4  $       &$6.6\times 10^{-4}$  &1.4                                                     &5.4         &6500             &2400                                                   &$6.1\times 10^{-4}$    &0.57  \\
$p_T^{l}, \MET >100 \; (150)$ GeV  &19             &3900           &1600                                              &0.0035     &2.6                                                                        &3.6        &550               &200                                                     &0.0048       &1.3  \\
  $\Delta R_{{\rm fj},\, b} < 1.0$, $p_T^b > 50 \GeV$                                   & 13             &88            &400                                                 &0.026         & 5.8                                                                & 2.5        &12                &52                                                     &0.039        & 3.1   \\
$p_T^{\rm fj} > 350 \; (500)$ GeV , $m_{\rm fj} < 130$ GeV&9.3  &60     &48                                    &0.086     &9.0                                                                        & 1.6        &7.4                &2.3                                                     &0.16       &5.0   \\
$M_{T'} > 750 \; (1000)$  GeV              &9.1               &51        &24                                             &0.12     & 10                                                                       & 1.5        &6.9               &1.7                                                      & 0.18      & 5.3          \\
$N_{\mathrm{fwd}}\geq 1$        & 5.9             &9.6           &5.8                                             &0.38        & 15                                                              & 1.0         &1.2                 &0.32                                                     & 0.68      & 8.3       \\

\hline
\end{tabular}
\caption{Example-cutflow for signal events (simulated for $c^{T'bW}_L = 0.3$ with $M_{T'}=1.0$ TeV and $M_{T'}=1.5$ TeV) and background events in the $T \rightarrow W_{\rm lep} b$ channel for $\sqrt{s} = 14 \, \mbox{TeV}$. Cross sections after the respective cuts for signal and backgrounds are given in fb. The $S/\sqrt{B}$ values are given for a luminosity of $100\, \mathrm{fb}^{-1}$ for both the $M_{T'}=1.0$ TeV and $M_{T'}=1.5$ TeV searches. The $\Delta R_{{\rm fj}, \, b}$ includes the $b$-tagging efficiencies, as well as requirement that the $b$-jet is isolated from the lepton.}
\label{tab:cutflowwb}
\end{table*}
\end{center}
\end{widetext}

Finally, the requirement on high reconstructed $M_{T'}$ modestly improves $S/B$ while the forward jet tag results in a factor of 3-4 improvement in $S/B$ at an additional $\approx 65\%$ signal efficiency.

As another potentially interesting candidate for $T'$ detection in the $Wb$ decay channel we studied $T'\rightarrow W_{\rm had} b$ (\cf Appendix~\ref{app:Wb} for details). However, we find it to yield approximately five times lower significances  than the $T'\rightarrow W_{\rm had} b$ channel presented here.

\begin{figure}[htb]
\includegraphics[scale=0.35]{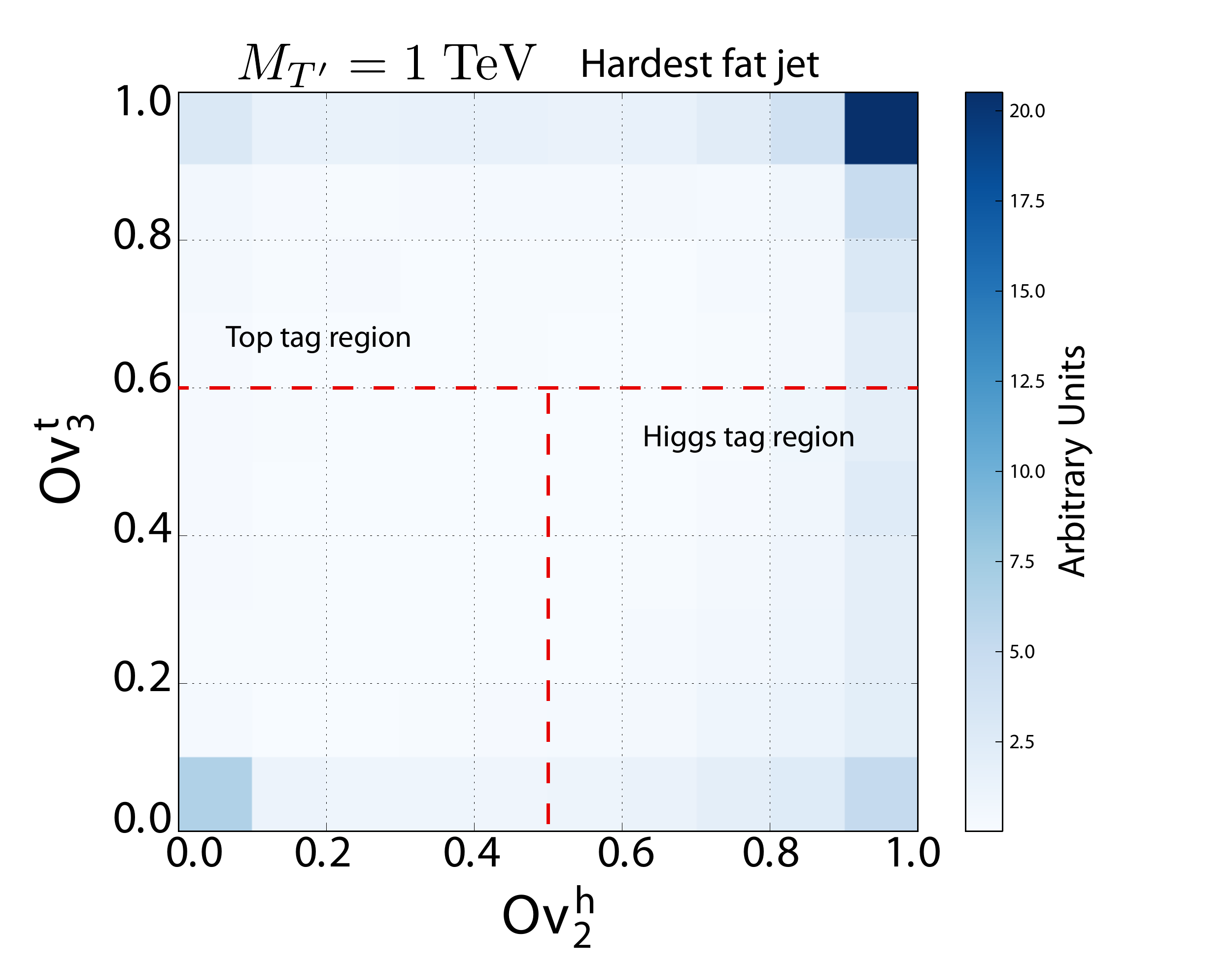}
\includegraphics[scale=0.35]{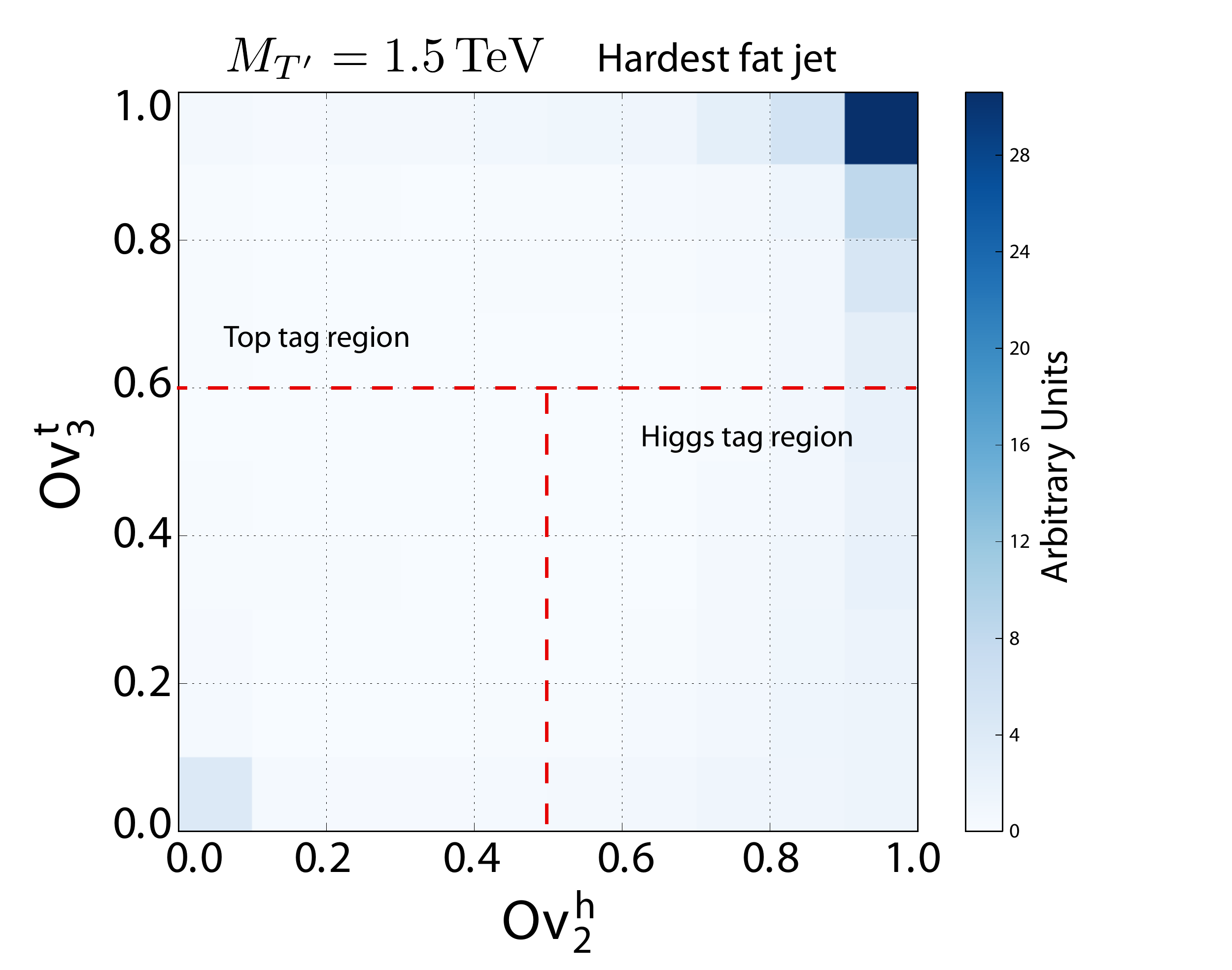}
\includegraphics[scale=0.35]{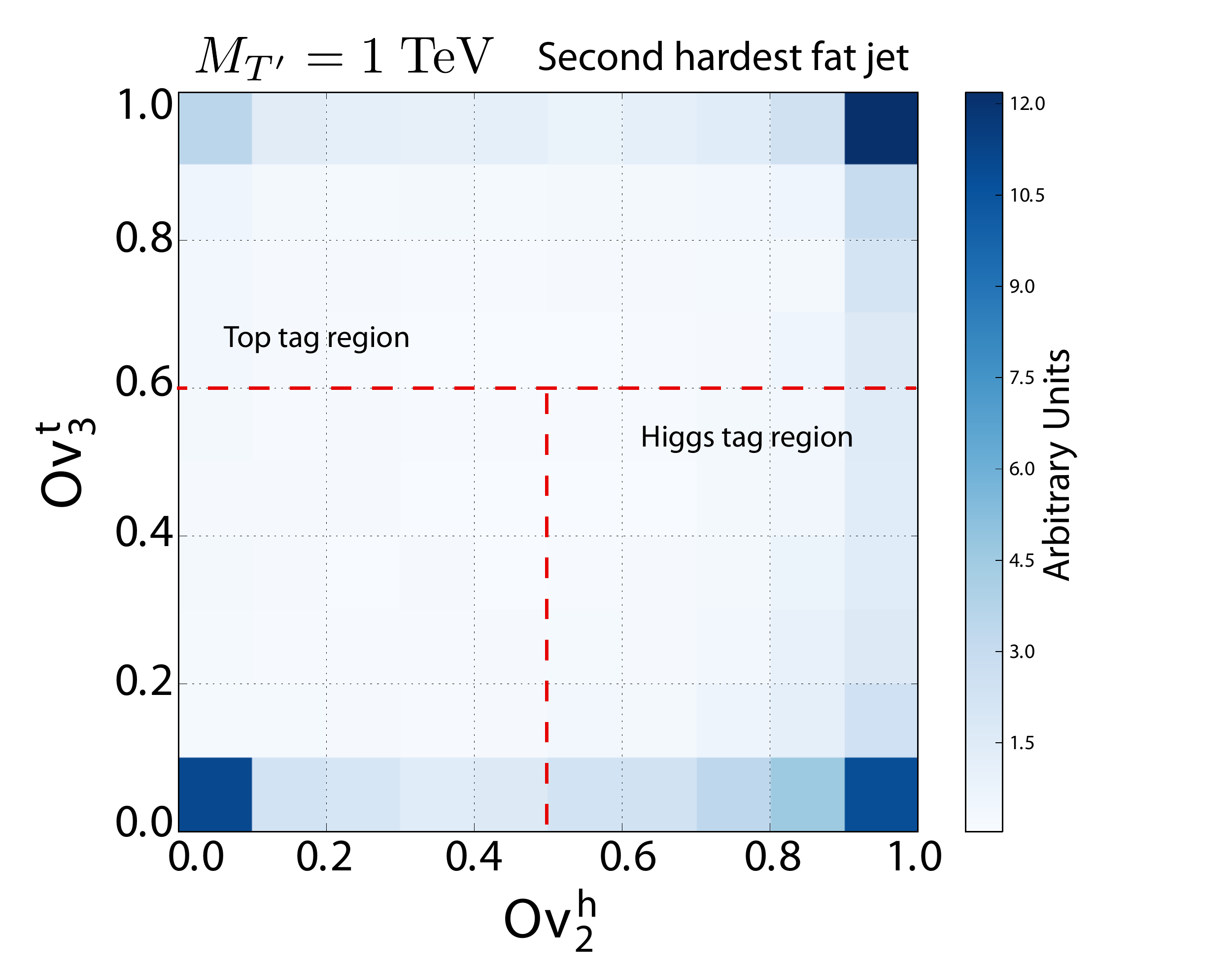}
\includegraphics[scale=0.35]{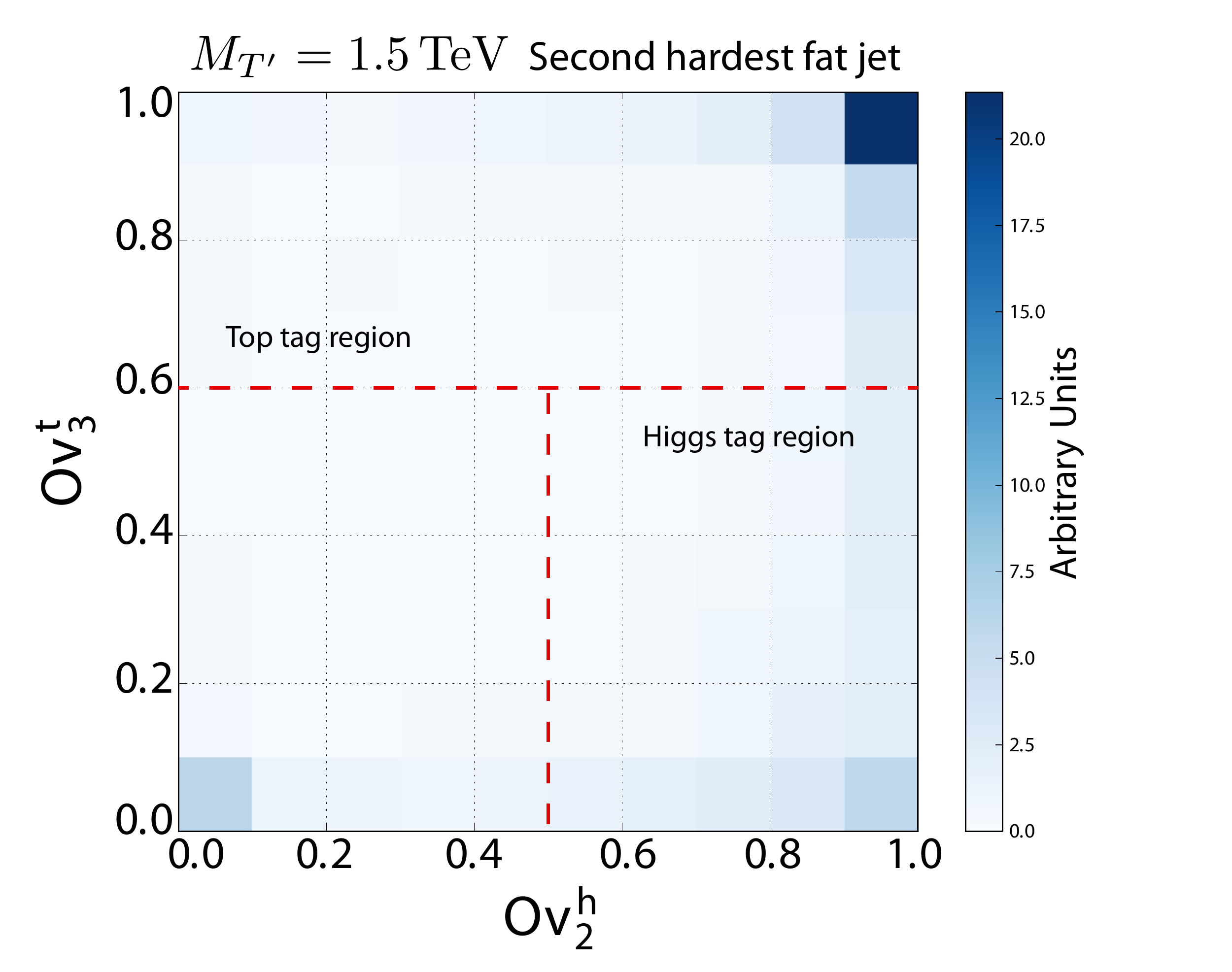}
\caption{Multi-dimensional $Ov$ distributions of (left column) 1TeV and (right column) 1.5 TeV data in the $T' \rightarrow t_{\rm had} h_{bb}$ channels after Basic Cuts. Panels in the first and second rows represent $Ov$ distributions of the first and second hardest fat jets in the signal events, respectively.}
\label{fig:HtMultiOvSignal}
\end{figure}
\begin{figure}[htb]
\includegraphics[scale=0.35]{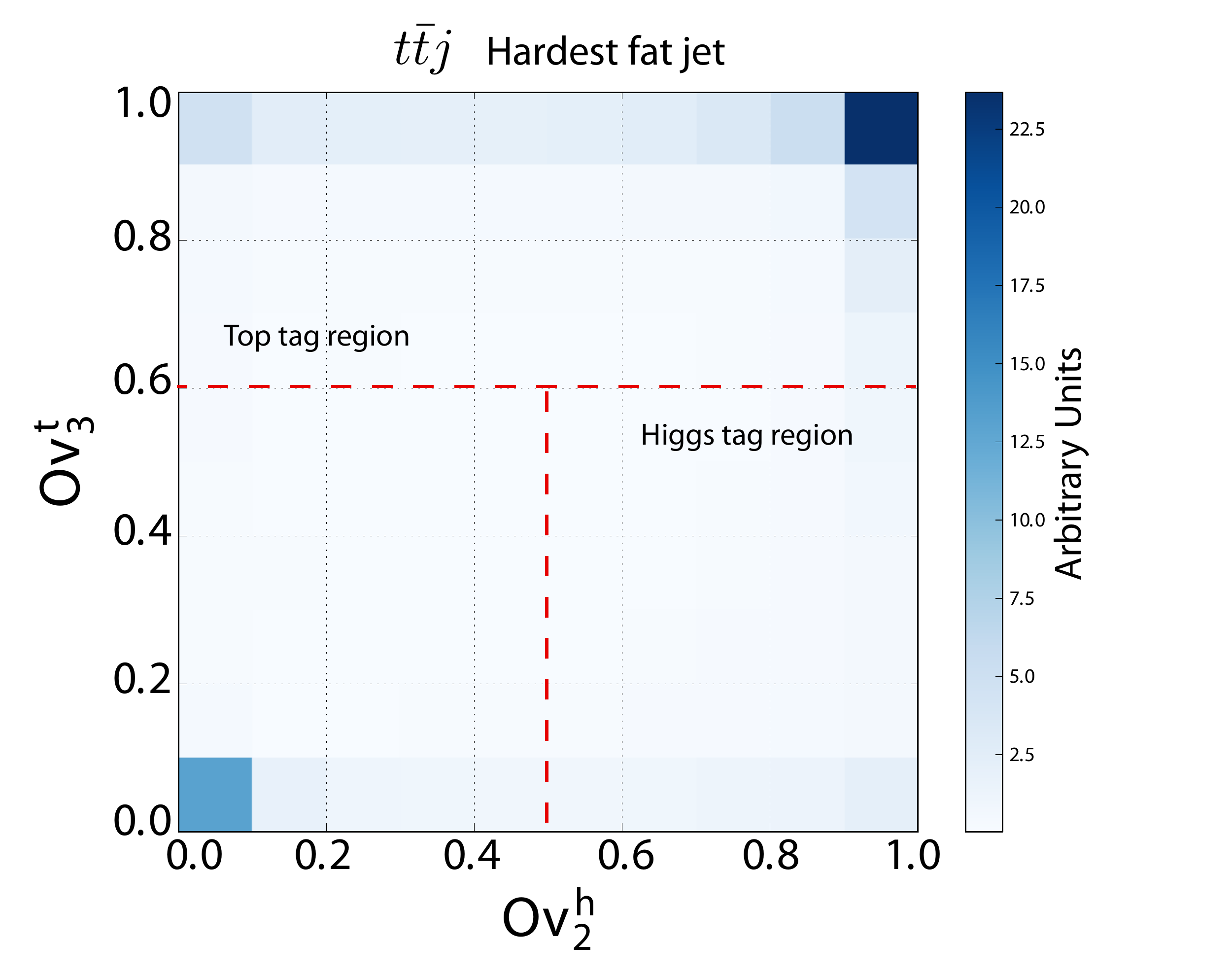}
\includegraphics[scale=0.35]{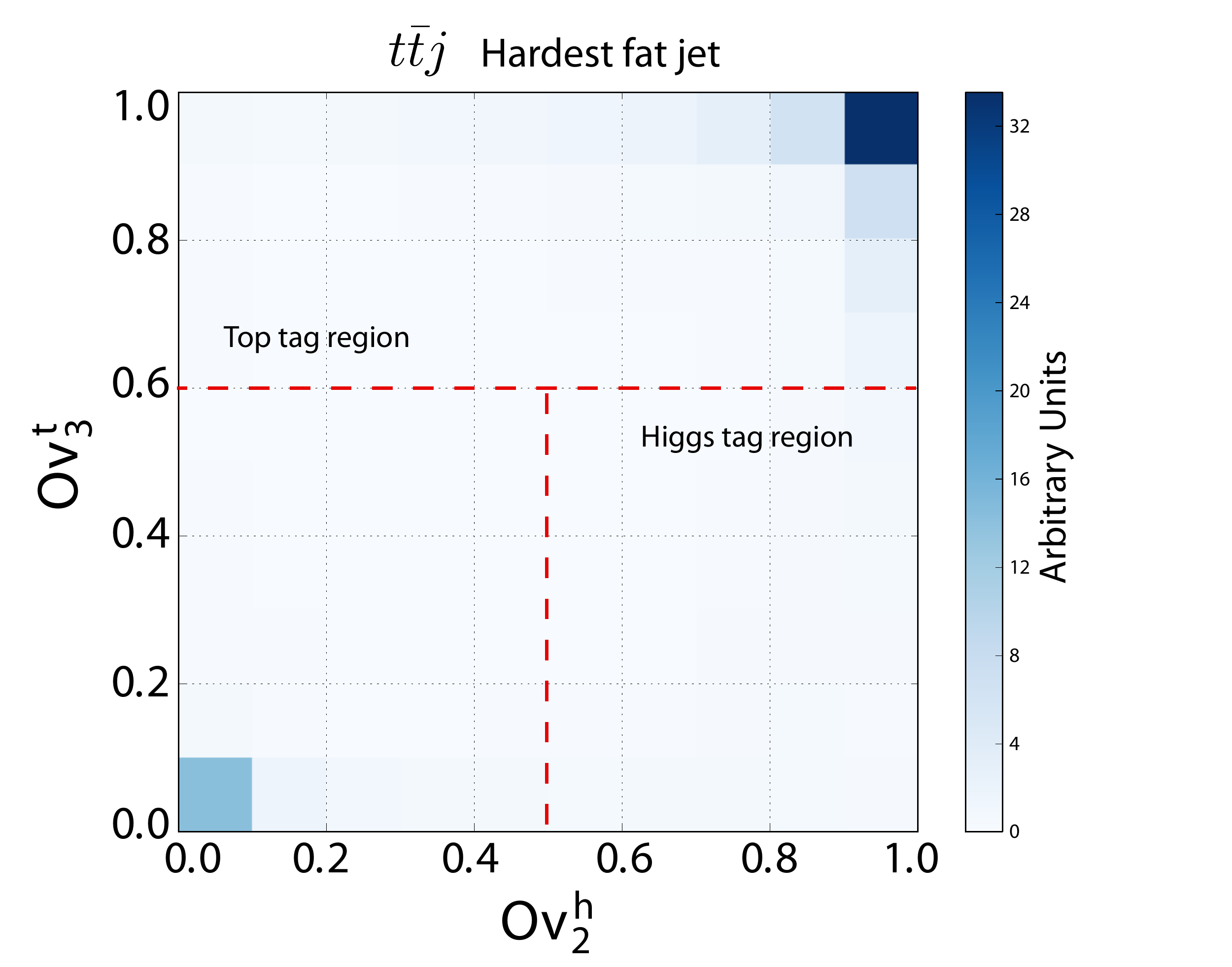}
\includegraphics[scale=0.35]{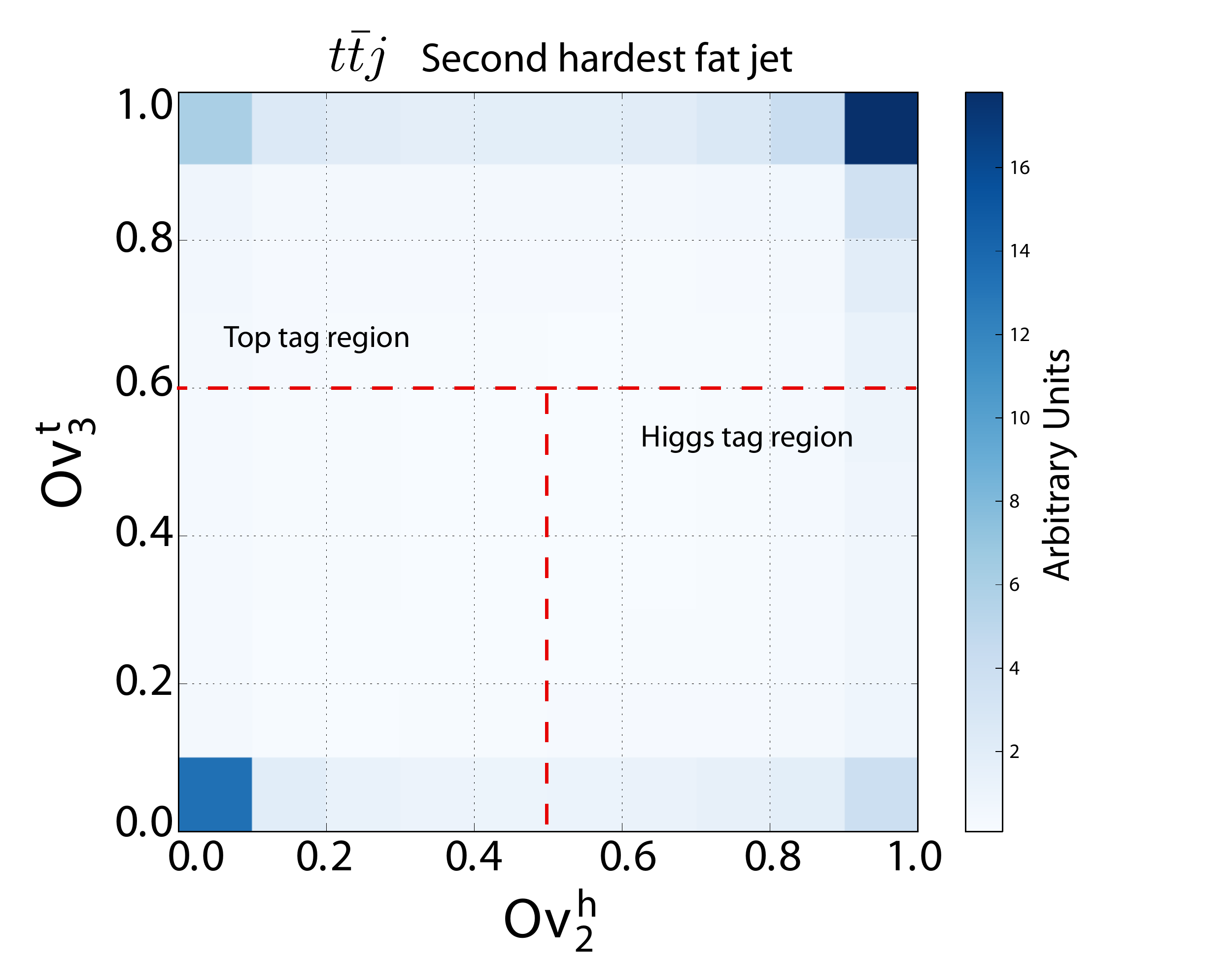}
\includegraphics[scale=0.35]{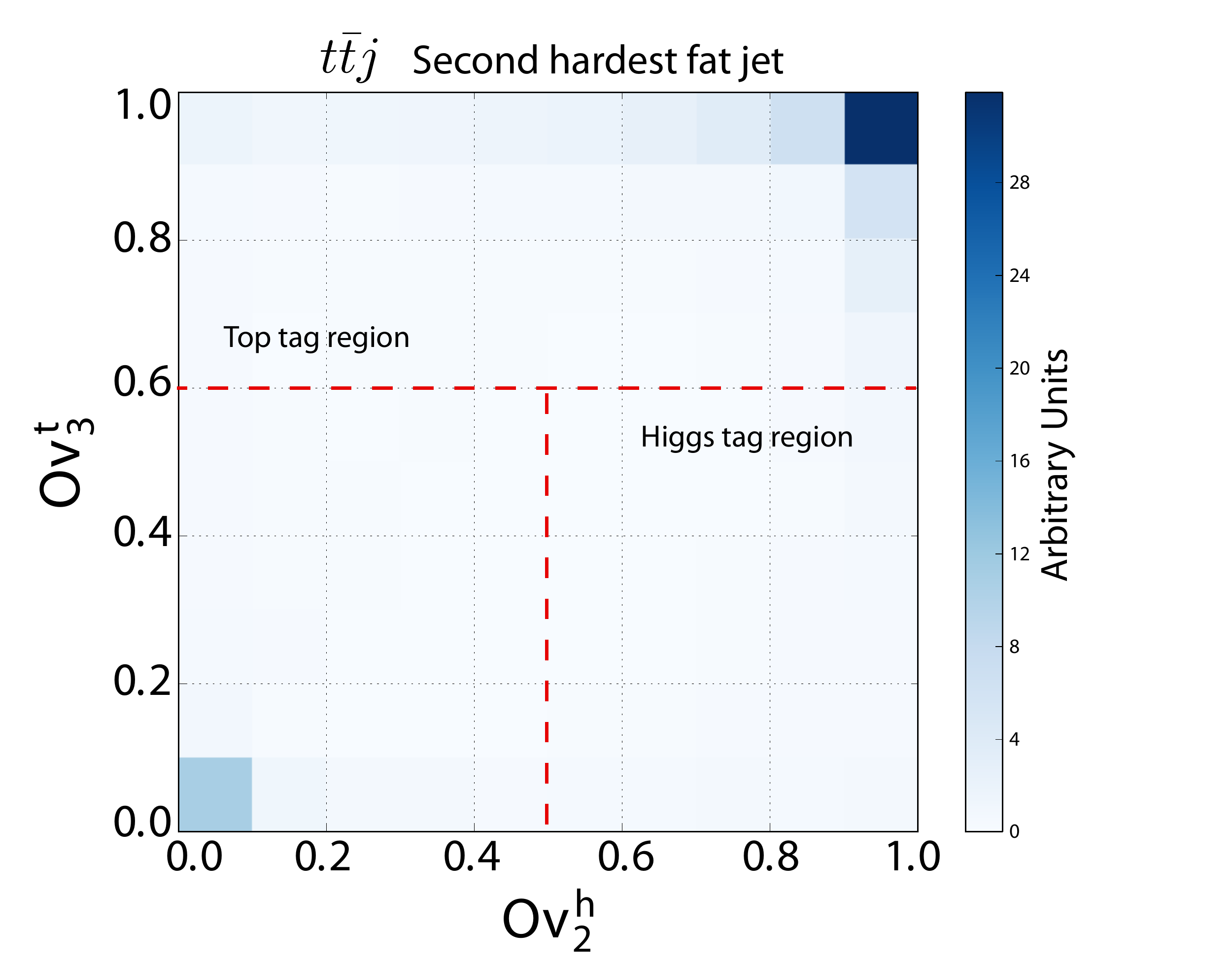}
\caption{Multi-dimensional $Ov$ distributions of (left column) 1TeV and (right column) 1.5 TeV data on $T' \rightarrow t_{\rm had} h_{bb}$ channels after Basic Cuts. Panels in the first and second rows represent $Ov$ distributions of the first and second hardest fat jets in the $t\bar{t}$ SM background events respectively.}
\label{fig:HtMultiOvTop}
\end{figure}


\subsection{$T' \rightarrow t_{\rm had} h_{bb}$ Channel }
\label{sec:th}

Concerning the $T'$ decay into $th$ we focus on search channels in which the Higgs decays into $b\bar{b}$ which yield by far the largest signal cross sections. We investigate both the $T' \rightarrow t_{\rm had} h_{bb}$ (presented here) and the $T' \rightarrow t_{\rm lep} h_{bb}$ (presented in Appendix~\ref{sec:thlep}). 

Searches for new particles in the fully hadronic channels are  challenging as the large QCD backgrounds are orders of magnitude larger than the signal and often difficult to suppress. Yet, we find that a combination of boosted jet techniques and a multi-$b$-tagging strategy is able to reduce the background channels in the $T'\rightarrow t_{\rm had} h_{bb}$ to a manageable level, while maintaining sufficient signal efficiency.

The dominant SM backgrounds for the $t_{\rm had} h_{bb}$ channel are QCD multi-jets, $b\bar{b}$ + jets \footnote{Here we include $b\bar{b}$ + jets originating from a pure QCD process as well as $Z_{b b}$ + jets.} and $t\bar{t}$(\rm had) + jets.  We generate all backgrounds with the pre-selection cuts described in Sec. \ref{sec:PreCuts} where we demand $H_T > 850~ (1350)$~GeV for a hypothetical top partner mass of $1\,(1.5)$ TeV. Table~\ref{tab:TotalBackGroundsht} summarizes the background cross sections including a conservative NLO K-factor of 2.

\begin{table}[h]
\begin{center}
\begin{tabular}{|c|c|c|c|}
\hline
		Channels									&	Backgrounds                     		 				& $\sigma (H_T > 850\GeV) [{\rm fb}]$  & $\sigma (H_T > 1350\GeV) [{\rm fb}]$ \\ \hline												
\multirow{3}{*}{$T' \rightarrow t_{\rm had} h_{bb}$}                    & multi-jet                                                           &     $ 4.2   \times 10^6 $	                        &  $ 3.8     \times 10^5 $	                             \\
											&$b\bar{b}$ + jets                                               &    $ 4.8     \times 10^4 $	                        &  $  5.4    \times 10^3 $		                      \\
											&$t\bar{t}$(had) + jets                                         &	$  8.4    \times 10^3 $	                 &  $  8.9    \times 10^2 $		                         \\ 
\hline		
\end{tabular}
\end{center}
\caption{The simulated cross sections of SM backgrounds (including a conservative estimate of NLO K-factor of 2 after preselection cuts described in Sec.~\ref{sec:PreCuts}.}
\label{tab:TotalBackGroundsht}
\end{table}

Since the final state of interest contains a boosted top and a boosted Higgs, our Basic Cuts consist of requiring at least two fat jets ($R=1.0$) with $p_T^{\rm fj} > 300 \;(400)\GeV$ for 1 (1.5) TeV $T'$ searches respectively and $|\eta_{{\rm fj}} |< 2.5$ (see Table~\ref{tab:BasicCutsth} for summary), as well as requiring no isolated hard leptons in the event. 

Next, the Complex Cuts, summarized in Table~\ref{tab:ComplexCutsth}, begin with the jet substructure selection on candidate events, where the overlap analysis is applied to all fat jets with $|\eta_{\rm fj}| < 2.5$ and $p_T > 300\,(400) \GeV$, for 1 (1.5) TeV top partners. We demand exactly one Higgs (the hardest jet satisfying $Ov$-selection criterion: $Ov_{2}^h > 0.5$ and $Ov_{3}^t < 0.6$), and exactly one top (the hardest fat jet satisfying $Ov$-selection criterion, $Ov_{3}^t > 0.6$).

\begin{table}[h]
\setlength{\tabcolsep}{2em}
{\renewcommand{\arraystretch}{1.8}
\begin{tabular}{c|c}
							& $T' \rightarrow t_{\rm had} h_{bb}$      	    	    	    	    	   \\ \cline{1-2}
\multirow{3}{*}{ \textbf{Basic Cuts} }	& $N_{\rm fj}  \geq 2$ ($R=1.0$) ,          \\ 
		 					& $p_T^{\rm fj} > 300 \;(400) \GeV$, $|\eta_{\rm fj} |< 2.5$ , 	    	  	     \\
						        &$N_{\rm lepton}^{\rm iso} = 0$  .    
\end{tabular}}\par
\caption{Summary of Basic Cuts for $T' \rightarrow t_{\rm had} h_{bb}$ channel. ``fj" stands for the fat jet and $N^{\rm iso}_{\rm lepton}$ represents the number of isolated leptons with mini-ISO $> 0.7$, $p_T^{l} > 25\GeV$ and $|\eta_{l} |< 2.5$. The values outside (inside) the parenthesis refer to 1 (1.5) TeV $T'$ searches respectively.} \label{tab:BasicCutsth} 
\end{table}

\begin{table}[htb]
\setlength{\tabcolsep}{2em}
{\renewcommand{\arraystretch}{1.8}
\begin{tabular}{c|c}

							    & $T' \rightarrow t_{\rm had} h_{bb}$      	    	     \\ \cline{1-2}
\multirow{4}{*}{ \textbf{Complex Cuts} } &$N_{h,t} = 1$, (for $h$: $Ov_3^t < 0.6$ \& $Ov_2^h > 0.5$, for $t$ $Ov_3^t > 0.6$) ,       \\ 
		 					  & $M_{T'} > 750 \; (1000) \GeV$ , 	    	    	   	    \\ 
		 					  & $N_{\mathrm{fwd}}\geq 1$ ,			            \\ 
							  & at least 1 $b$-tag on $t$ and exactly 2b-tags on $h$ .	  	\\ 
\end{tabular}}\par
\caption{The summary of Complex Cuts for $T' \rightarrow t_{\rm had} h_{bb}$ channel. Here $b$-tag refers to the simplified $b$-tagging procedure of Section~\ref{sec:b-tagging}, whle ``$Ov$'' selection applies to the two highest $p_T$ fat jets ($R=1.0$) in the event, and $N_{t,h}$ is the number of tagged top and Higgs fat jets respectively. } \label{tab:ComplexCutsth} 
\end{table}

Simultaneously tagging fat jets with both a Higgs and a top tagger is particularly useful in reducing the rate of boosted tops being mis-tagged as boosted Higgs jets. Ref. \cite{Backovic:2014ega} already utilized such a strategy in a proposal for light quark composite partner searches at LHC14. Figs. \ref{fig:HtMultiOvSignal} and \ref{fig:HtMultiOvTop} illustrate the results of the boosted object tagging procedure. Fig. \ref{fig:HtMultiOvSignal} shows the two-dimensional distribution of Higgs overlap ($x$-axis) and top overlap ($y$-axis) scores, where the top panels refer to the hardest fat jet and the bottom panels refer to the second hardest jet, while left (right) panels refer to 1 (1.5) TeV $T'$ partner. The top panels of Fig. \ref{fig:HtMultiOvSignal} show that the hardest fat jet is typically a good boosted top candidate, both in the case of 1 TeV and 1.5 TeV partner searches. The second hardest jet (lower panels) on the other hand has a roughly same probability of being a boosted top and a boosted Higgs. The fraction of events which are mis-tagged as light jets ($Ov^{t,h} \sim 0$) is relatively small, except in the case of a second hardest jet in a search for 1 TeV $T'$. This is likely due to a fact that at lower boost, there is a greater chance that the decay products of the top or the Higgs are not properly clustered into a single fat jet. 

 Conversely, Fig. \ref{fig:HtMultiOvTop} shows the same distributions for the SM $t\bar{t}$ background, where we find that in most cases, either the hardest or the second hardest jet will pass the top tagging criteria, but only a small fraction of jets will pass our Higgs tagging criteria. Notice that if our tagging strategy involved only $Ov_2^h$, and not the "anti $Ov_3^t$'' requirement,  we would end up with a significantly higher $t\bar{t}$ background.

\begin{figure}[htb]
\includegraphics[scale=0.35]{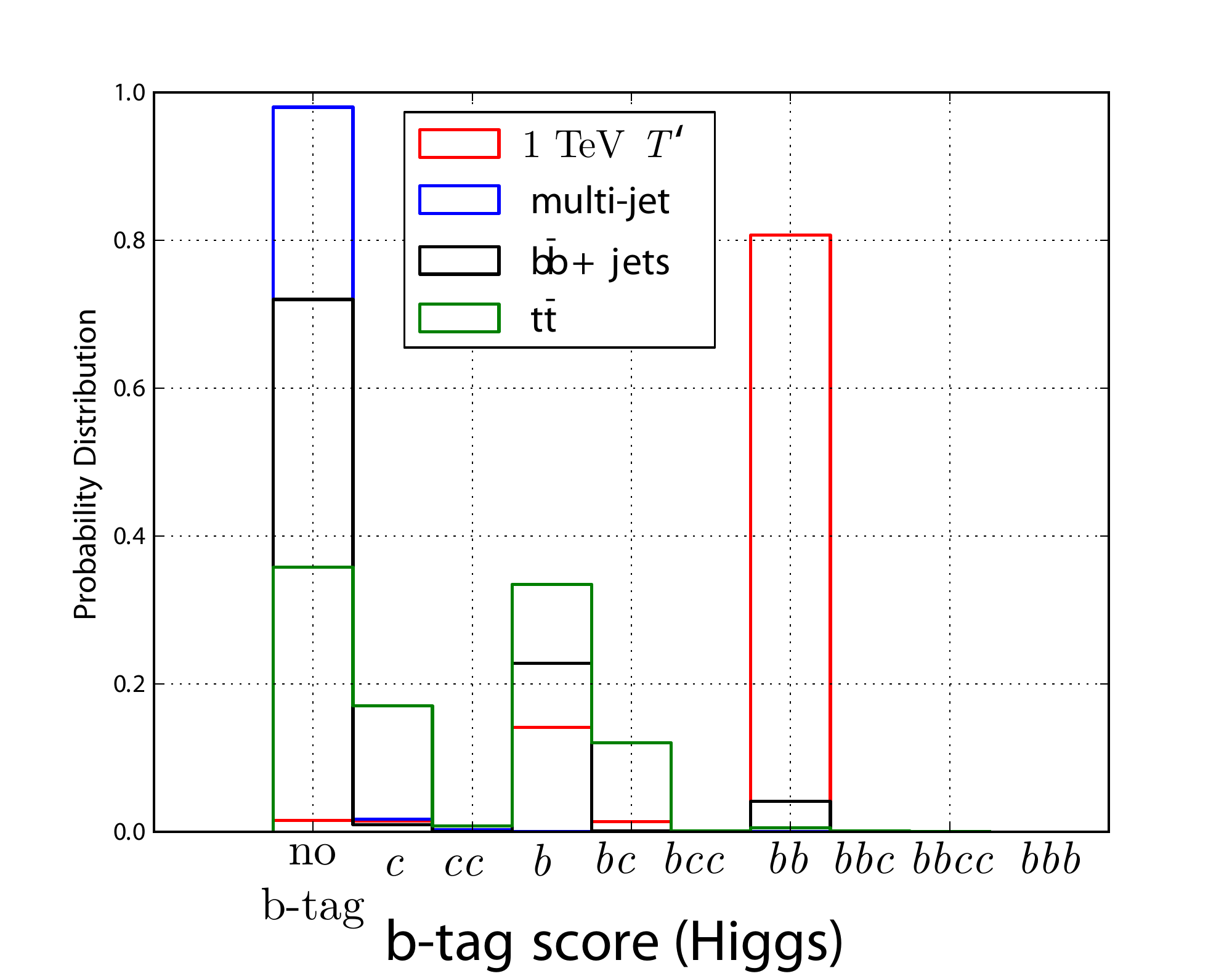}
\includegraphics[scale=0.35]{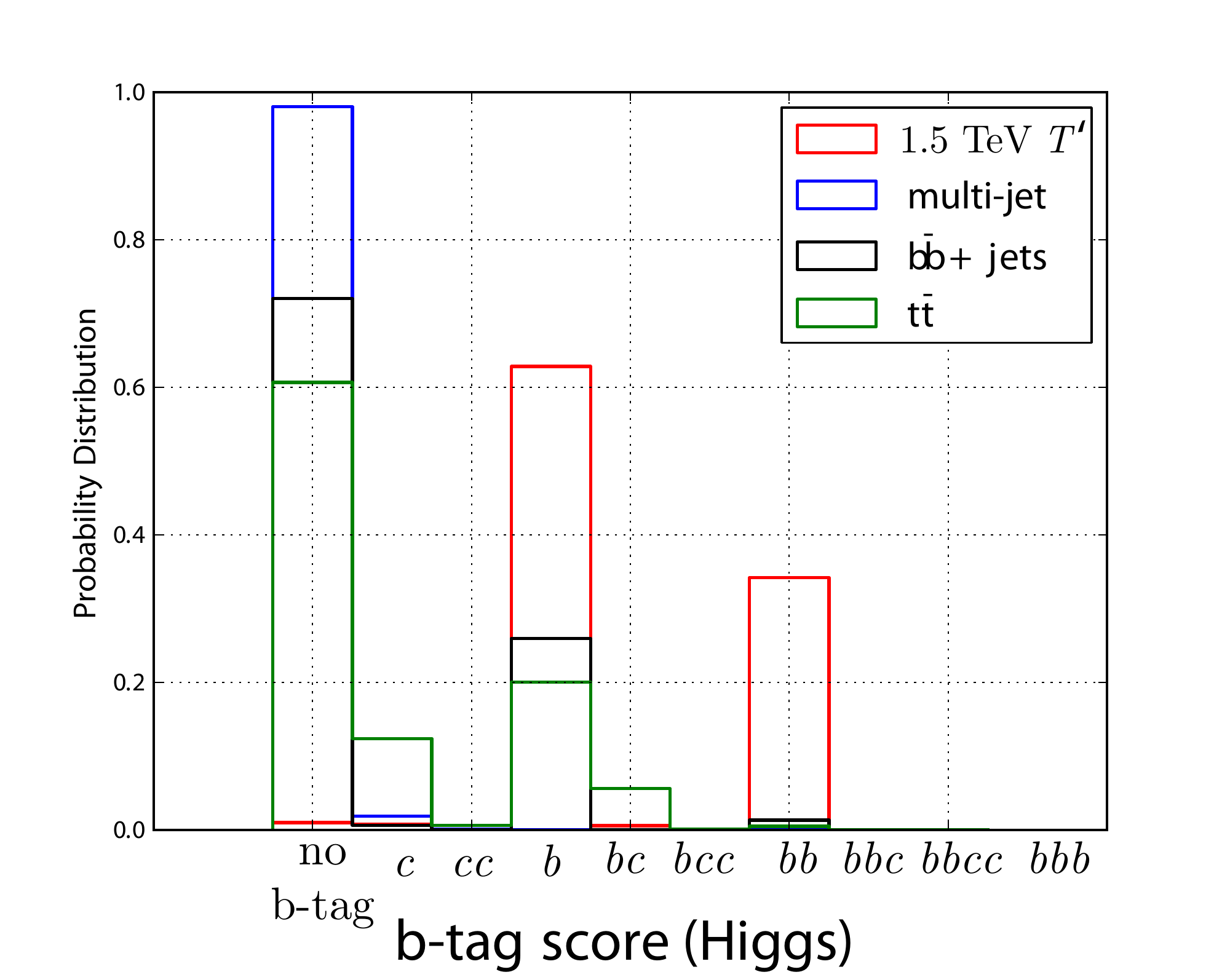}
\includegraphics[scale=0.35]{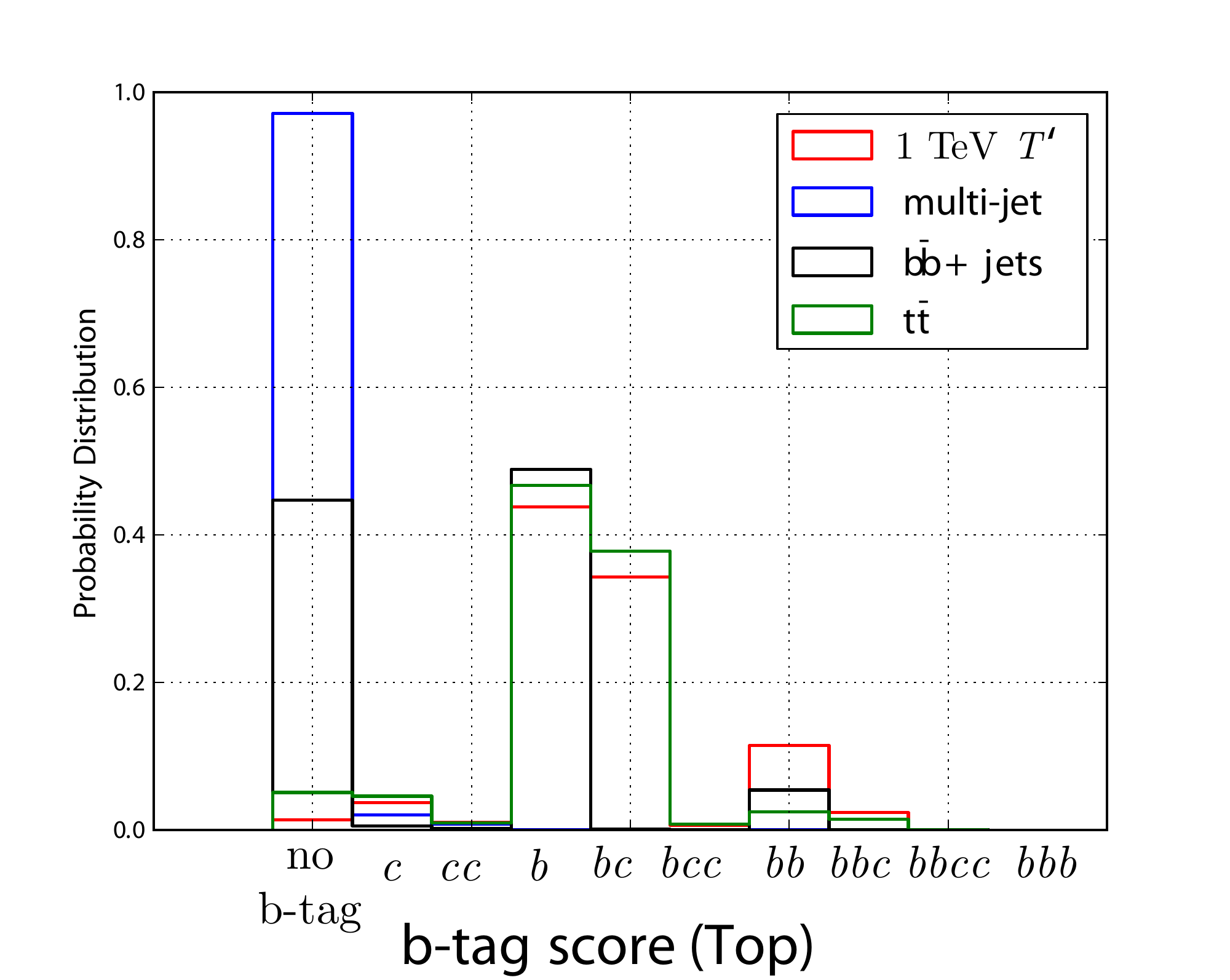}
\includegraphics[scale=0.35]{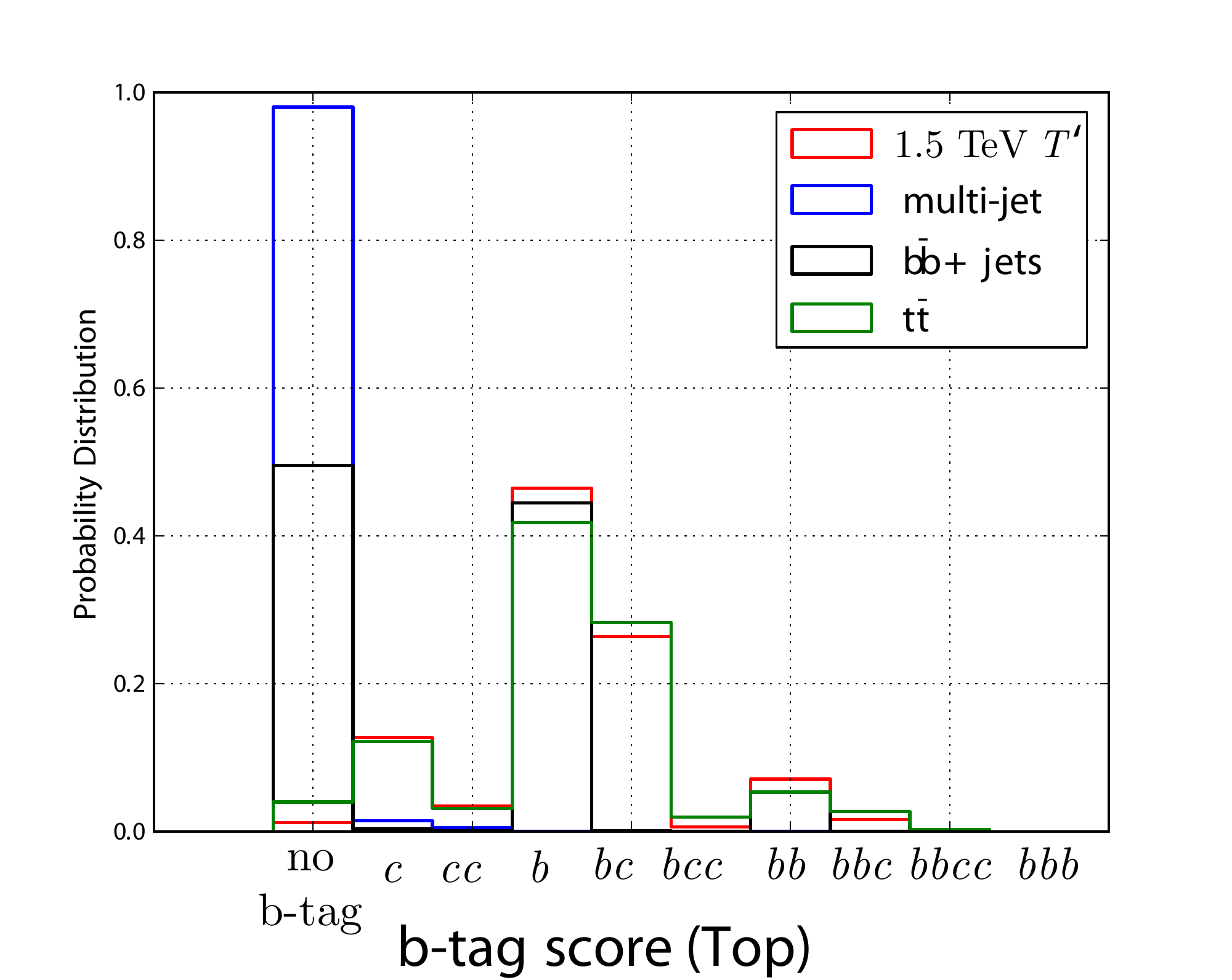}
\caption{$b$-tag scores for 1TeV (left) and 1.5 TeV  (right column)  data for $T' \rightarrow t_{\rm had} h_{bb}$ channel searches after Basic Cuts and boosted object selection. Panels in the first and second rows represent $b$-tag scores of a Higgs tagged  and top tagged fat jet candidates respectively. No $b$-tagging efficiencies have been applied to the data in the plots.}
\label{fig:HtBtag}
\end{figure}

While the boosted jet selection in the $t_{\rm had} h_{bb}$ provides significant background rejection power, we find that the $b$-tagging strategy plays the central role in the prospects for detecting and measuring the $T'$ partner in this channel. The $b$-jet content we expect in signal events is complex enough that it requires special attention. Here, we present only a summary of the results on the optimal $b$-tagging strategy in the $th$ final state, while we present a detailed discussion and comparison of different $b$-tagging options in Appendix~\ref{app:ht}. We find that requiring at least 1 $b$-tag on the  top tagged fat jet and exactly 2 $b$-tags on the Higgs tagged jet results in the highest signal sensitivity. Fig. \ref{fig:HtBtag} shows the  $b$-tag content of the Higgs and top tagged fat jets. For the 1 TeV $T'$ search (left panels), we find that 80\% of Higgs tagged jets contain 2 $b$-tags, while over 90\% of background events contain less than two $b$-tags inside a jet which passes the Higgs tagging selection. In addition, we find that over $90\%$ of signal top tagged jets are properly $b$-tagged ($e.g.$ at least one $b$-tag), and while the $t\bar{t}$ displays similar properties, the multi-jet and $b\bar{b}$ background contain a proper $b$-tag in the top tagged fat jet about $\sim 1\%$ and $\sim 50\%$ of times respectively. The $b$-tag properties of signal and background events do not significantly change for the top tagged fat jet in the 1.5 TeV $T'$ search, while we find that the percentage of signal events which contain two proper $b$-tags in the Higgs tagged fat jet is significantly reduced. The reduction in double $b$-tagging efficiency at higher $T'$ masses is likely due to the fact that a 1.5 TeV $T'$ decays into a Higgs boson with a characteristic $p_T\sim 700 \GeV. $ The decay products of a highly boosted Higgs will be collimated into a cone of roughly  $\Delta R_{bb} \sim 2m_h / p_T \sim 0.3$, implying that the showers of the two $b$-quarks will have a significant overlap (assuming the event is clustered with $r=0.4$ which we use for $b$-tagging), hence degrading the $b$-tagging efficiency.

Finally, as in other channels, we also require at least one $r=0.2$ forward jet ($p_T^{\rm fwd} > 25 \GeV$ and $2.5 < \eta^{\rm fwd} < 4.5$).

\begin{widetext}
\begin{center}
\begin{table*}[h]
\begin{tabular}{|c||cccc|cc||cccc|cc|}
\hline
\multirow{2}{*}{$T' \rightarrow t_{\rm had} h_{bb}$}&\multicolumn{6}{c||}{$M_{T'}=1.0$ TeV search}&\multicolumn{6}{c|}{$M_{T'}=1.5$ TeV search}\\ \cline{2-13}
                                                & signal       & $4\; {\rm jets}$  & $b\bar{b}+{\rm jets}$ & \hspace{5pt} $t\bar{t} \hspace{5pt}$  & $S/B$         & $S/\sqrt{B}$ & signal   &$4\; {\rm jets}$ & $b\bar{b}+{\rm jets}$   & \hspace{5pt} $t\bar{t} \hspace{5pt}$  & $S/B$       & $S/\sqrt{B} $\\\hline
preselection                             &27            &$4.2\times 10^6$    &  $5.1\times 10^4$              &8400                                                    &$6.5\times 10^{-6}$    &0.13                                                   &4.5        &$3.8\times 10^5$     & 5800                  &900                                                     &$1.2\times 10^{-5}$    &0.072   \\
Basic Cuts                                 &21           & $2.6\times 10^6$   &   $3.2\times 10^4$             &6400                                                    &$7.8\times 10^{-6}$       &0.13                                                   &4.1        &$3.0\times 10^5$    & 4700                  &850                                                      &$1.4\times 10^{-5}$    &0.074  \\
$Ov$ selection                                  &9.1             &$8.7\times 10^4$       &  1300                &1200                                                    &$1.0\times 10^{-4}$      &0.30                                                        &1.9        &$2.1\times 10^4$      & 340                   &110                                                       &$8.7\times 10^{-5}$    &0.13  \\
$N_{\mathrm{fwd}}\geq 1$       & 5.5           &$1.4\times 10^4$      &    270               &280                                                       & $3.7\times 10^{-4}$      & 0.45                                                     & 1.2       &3800         &  77                     &27                                                        & $3.2\times 10^{-4}$    & 0.20      \\
$b$-tag                                     & 1.6          & 0.12           &     0.15            & 4.1                                                     &0.37             & 7.7                                                 & 0.15      &0.029            & 0.018                &0.18                                                    &0.66        & 3.2   \\
\hline
\end{tabular}
\caption{Example-cutflow for signal events (simulated for $c^{T'bW}_L = 0.3$ with $M_{T'}=1.0$ TeV and $M_{T'}=1.5$ TeV) and background events in the $T' \rightarrow t_{\rm had} h_{bb}$ channel for $\sqrt{s} = 14 \, \mbox{TeV}$. Cross sections after the respective cuts for signal and backgrounds are given in fb. The $S/\sqrt{B}$ values are given for a luminosity of $100\, \mathrm{fb}^{-1}$ for the $M_{T'}=1.0$ TeV and $M_{T'}=1.5$ TeV searches.}
\label{tab:cutflowTH}
\end{table*}
\end{center}
\end{widetext}

Table \ref{tab:cutflowTH} shows an example cutflow for two benchmark model points (as in the previous sections we use $c^{T'bW}_L = 0.3$ with $M_{T'}=1.0$ TeV and $M_{T'}=1.5$ TeV). The two dimensional $Ov$ selection, including both the Higgs and the top tagging, delivers a factor of 12 (6) improvement in $S/B$ for a 1 (1.5) TeV $T'$ partner, at 40 (45)\% signal efficiency. The forward jet tagging requirement improves $S/B$ by an additional factor of $\sim 4$ at an additional 60\% signal efficiency. The largest improvement in $S/B$ comes from our $b$-tagging strategy, where we find that requiring exactly 2 $b$-tags on the Higgs tagged jet in addition to at least one $b$-tag on the top tagged jet delivers an improvement of 3 orders of magnitude. The dramatic enhancement in $S/B$ is solely due to the fact that properly $b$-tagging the fat jets nearly eliminates the QCD background while also substantially reducing the $b\bar{b}$ and $t\bar{t}$ backgrounds which contain $b$ jets but rarely two strongly collimated $b$'s which fall into one fat jet. 

Apart from the $T' \rightarrow t_{\rm had} h_{bb}$ channel, we also performed an analysis for the  $T' \rightarrow t_{\rm lep} h_{bb}$ for which we find similar but slightly weaker significances. The details can be found in Appendix~\ref{sec:thlep}.

\section{Combined Results} \label{sec:combres}
Our results of previous sections have been obtained from simulations of one specific $T'$ model implementation which in particular fixes the branching fractions between the different $T'$ decay channels (which in this model are $\sim 2~:~1~:~1$ for decays into $Wb$, $Zt$ and $ht$ up to subleading corrections). In more general $T'$ models, the branching ratios can be altered. In this section, we relax the assumption on the branching ratio and present comprehensive predictions for the reach of LHC Run II searches for $T'$ as a function of $Br(T'\rightarrow ht), Br(T'\rightarrow Wb)$ and $Br(T'\rightarrow tZ)$. In the following, we find that presenting results in terms of the signal cross section necessary for discovery is particularly useful, as the information can be used to determine whether a particular $T'$ model is discoverable at LHC Run II or not. 

We calculate the signal cross section necessary for discovering $T'$ partners using the likelihood ratio \cite{Cowan:2010js}:
\beq
	\mathcal{LR_{\rm dis}} \equiv \sqrt{-2\, {\rm ln}\left( \frac{L(B | \mu S + B)}{L(\mu S + B| \mu S + B)} \right) }\,,
\eeq
where $S$ and $B$ are the expected number of signal and background events respectively and
\beq
	L(x |n) = \prod_{i=1}^N \frac{x_j^{n_j}}{n_j\!} e^{-x_j} \nn.
\eeq
Here $i$ runs over all the $T'$ decay modes. For simplicity, we consider a signal modifier parameter $\mu = 1$. In order to claim discovery we demand
\beq
	\mathcal{LR_{\rm dis}} \geq 5.
\eeq

Fig. \ref{fig:TriangleDiscovery} shows  the single production cross section of a $T'$ or $\bar{T}'$ ($\sigma_{T'}+\sigma_{\bar{T}'}$)  required to obtain $\mathcal{LR}_{\rm dis} \geq 5$ for $100 \fb^{-1}$ of data with a fixed $M_{T'} = 1$ TeV (left) and $M_{T'} = 1.5$ TeV (right). In this combined likelihood fit we included the three searches performed previously in this Section. The $W_{\rm lep} b$ channel presents the best chance for a discovery of a 1~TeV $T'$, where we find that a  $\sigma_{T'} \sim 70$ fb is necessary to discover a $T'$ which decays exclusively into $W_{\rm lep} b$ final state. The next promising channel is $t_{\rm had} h_{bb}$ where a cross section $\sigma_{T'} \sim 80$ fb is needed for a discovery assuming that $T'$ exclusively decays into $th$, while $Z_{\rm inv} t_{\rm had}$ channel is the least sensitive with $\sigma_{T'}\sim 100 $ fb needed for discovery with $100 \fb^{-1}$ of data. Our results show the worst sensitivity to model parameter regions which give $Br(T' \rightarrow Zt) \sim 0.5$, where up to $\sigma_T' \sim 120 - 140 \fb$ is needed to claim discovery of a $T'$ of  mass 1 TeV with $100 \fb^{-1}$ of data.

\begin{figure}[h]
\includegraphics[scale=0.70]{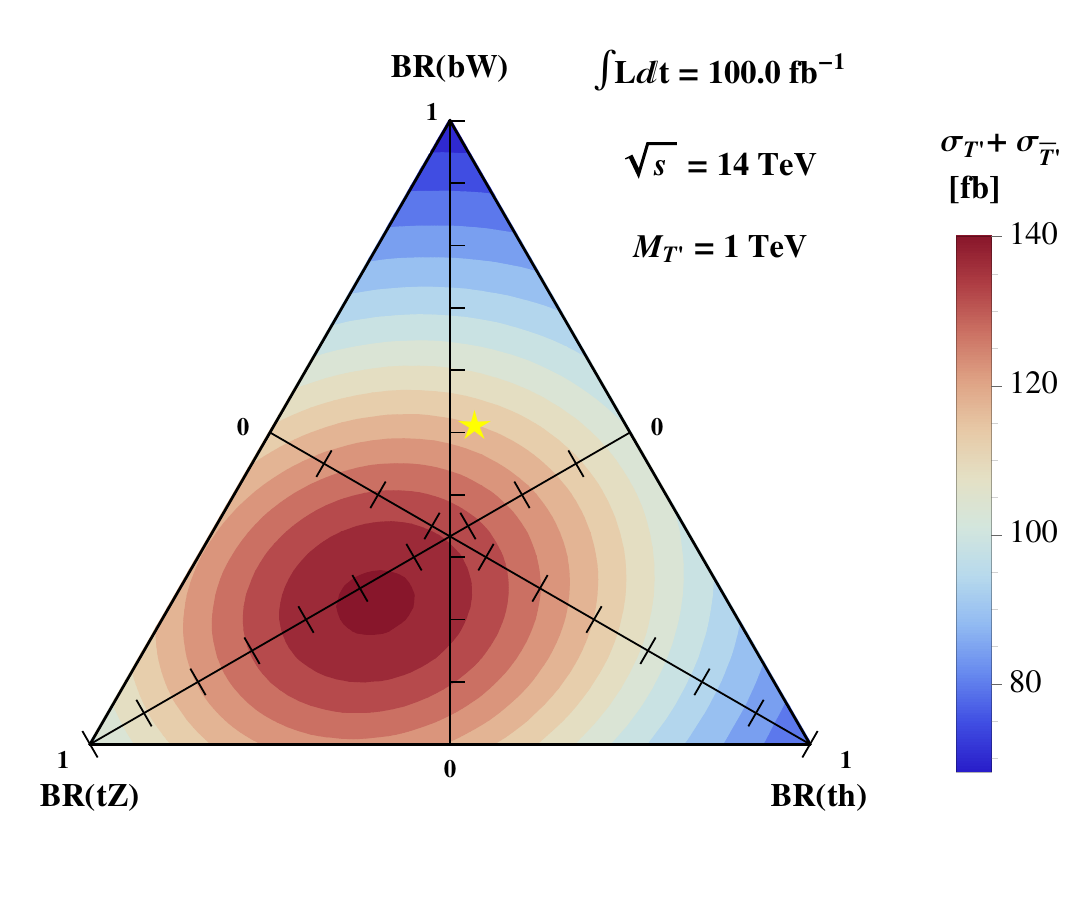}
\includegraphics[scale=0.70]{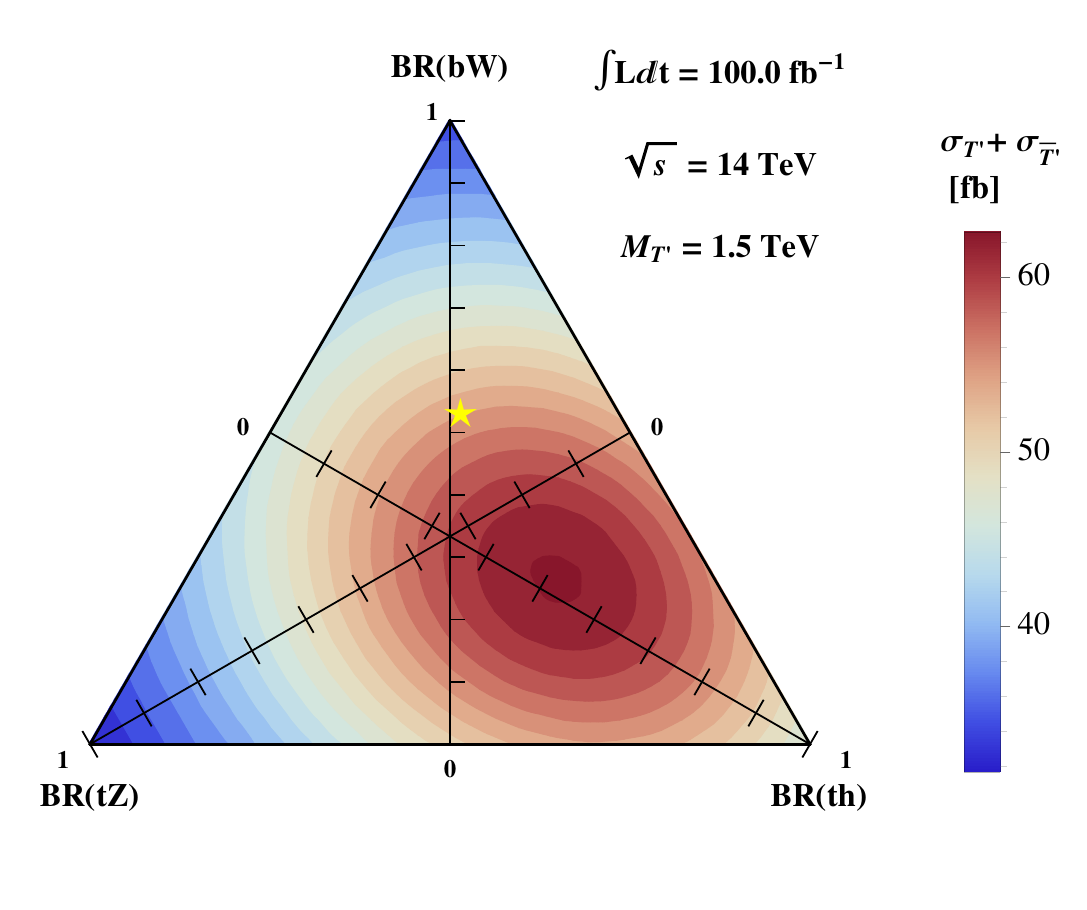}
\caption{  Signal cross section $\sigma_{T'}+\sigma_{\bar{T}'}$ required to obtain 5$\sigma$ using $100 \fb^{-1}$ of data with a fixed $M_{T'} = 1$ TeV (left) and $100 \fb^{-1}$ of data with a fixed $M_{T'} = 1.5$ TeV (right). The yellow stars mark the branching fractions  at the sample points used in our simulations ($c^{T'bW}_L = 0.3$ with $M_{T'}$ = 1.0 TeV and $M_{T'}$ = 1.5 TeV).}
\label{fig:TriangleDiscovery}
\end{figure}

Probing higher $T'$ masses results in a significantly different situation. For $M_{T'} = 1.5 \TeV$ we find that  $Z_{\rm inv} t_{\rm had}$ channel outperforms the competing channels, with a cross section of $\sigma_{T'}=31$ fb needed to detect $T'$ assuming $Br(T' \rightarrow Zt) \sim 1$. The $W_{\rm lep} b$ channel still remains important indicating that lepton signatures as well as $\MET$  remain key probes of $T'$ models at masses higher than 1.5 TeV. The least sensitive channel turns our to be $t_{\rm had} h_{bb}$ , where the loss of sensitivity can be attributed to a significantly lower $b$-tag efficiency on the boosted Higgs (see Figure \ref{fig:HtBtag}), as well as lack of optimization of the TOM procedure for higher masses. The sensitivity of $t_{\rm had} h_{bb}$ channel can potentially be improved by demanding a smaller cone size of a fat jet and optimizing the jet substructure observables for ultra-high $p_T$ range, as well as improving the  $b$-tagging scheme.  The worst sensitivity is to models which predict $Br(T'\rightarrow th) \sim 0.5$ where we find that $\sigma_{T'} \sim 50-60$ fb is needed to claim a discovery.

\begin{figure}[h]
\includegraphics[scale=0.70]{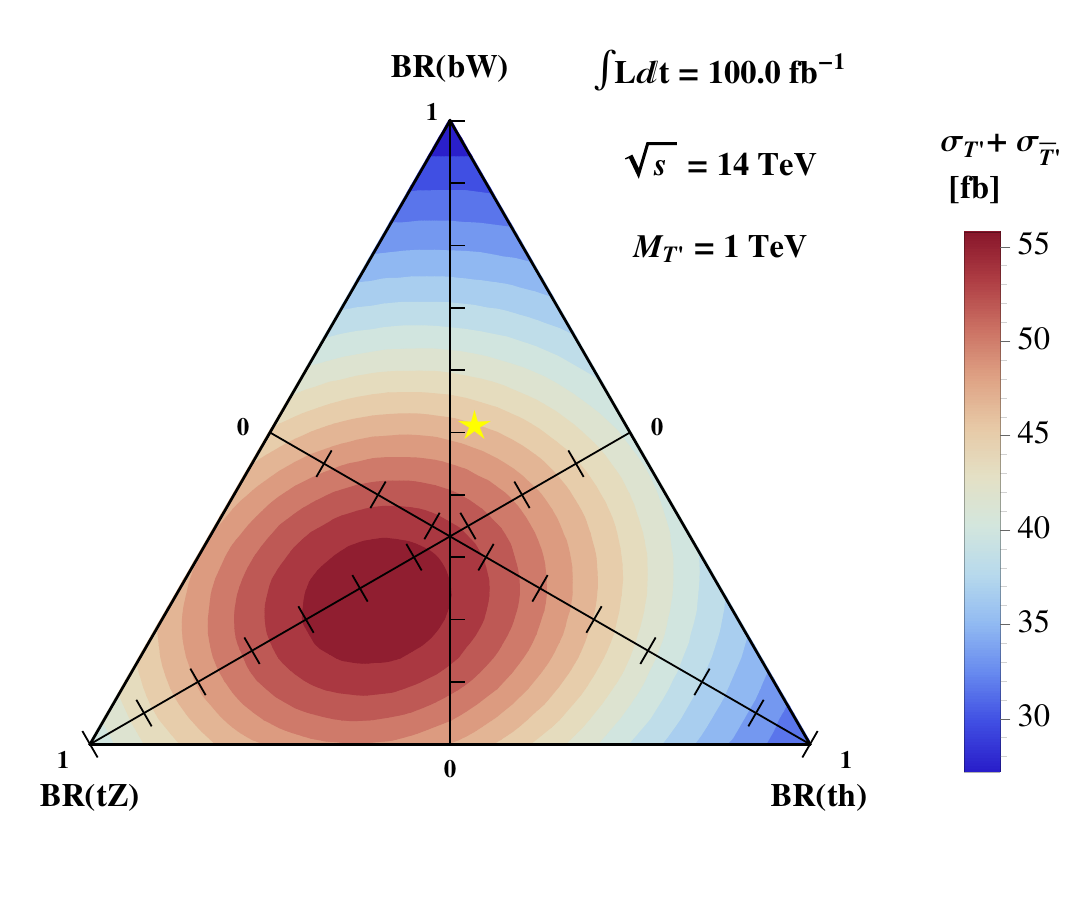}
\includegraphics[scale=0.70]{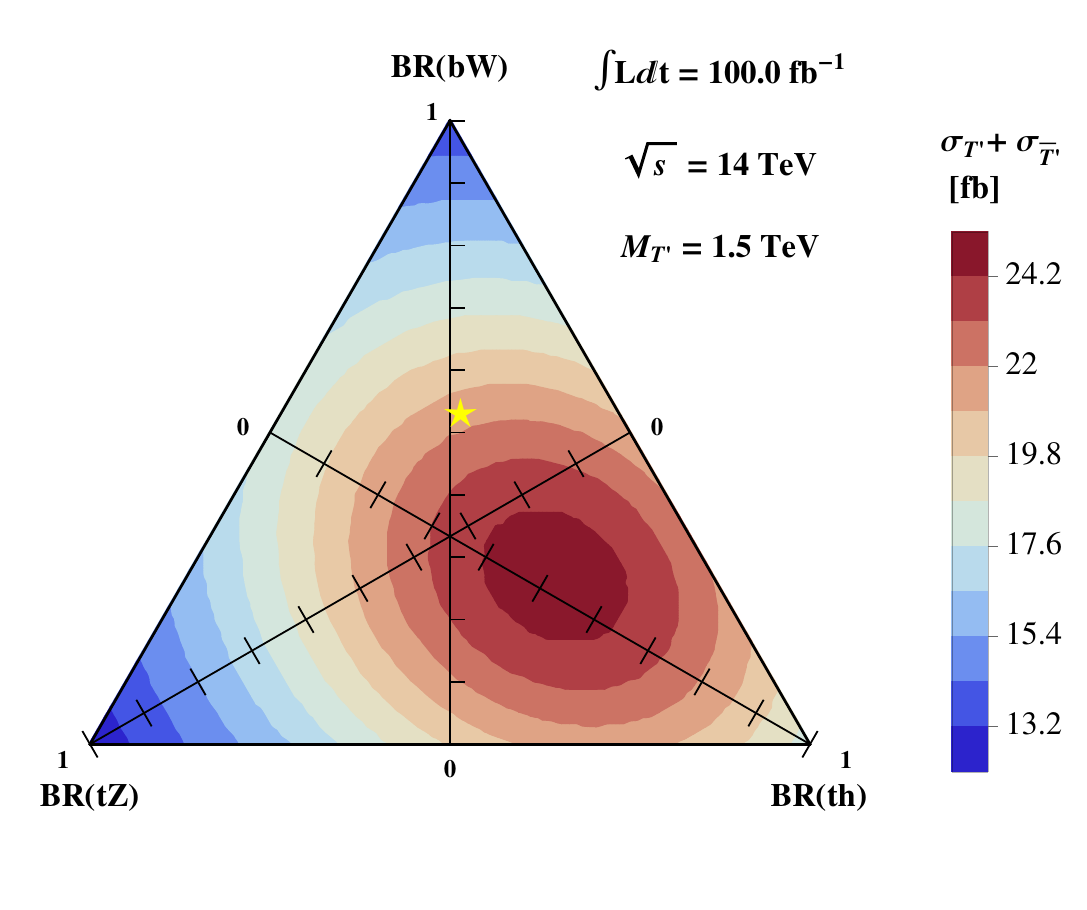}
\caption{  Signal cross section $\sigma_{T'}+\sigma_{\bar{T}'}$ which can be excluded with $100 \fb^{-1}$ of data with a fixed $M_{T'} = 1$ TeV (left) and $100 \fb^{-1}$ of data with a fixed $M_{T'} = 1.5$ TeV (right). The yellow stars mark the branching fractions at the sample points used in our simulations ($c^{T'bW}_L = 0.3$ with $M_{T'}$ = 1.0 TeV and $M_{T'}$ = 1.5 TeV).}
\label{fig:TriangleExclusion}
\end{figure}

In addition to cross sections necessary for discovery, we also calculate the signal cross section which can be excluded with $100 \fb^{-1}$ of LHC Run II data. Similar to the discovery reach analysis, we use the likelihood ratio
\beq
	\mathcal{LR_{\rm exc}} \equiv \sqrt{-2\, {\rm ln}\left( \frac{L(\mu S + B|B)}{L(B|B)} \right) }\,,
\eeq
with a signal strength parameter $\mu = 1$ in order to determine the cross section value which can be excluded. In order to claim 2$\sigma$ exclusion we demand
\beq
	\mathcal{LR_{\rm exc}} \geq 2.
\eeq

Figure \ref{fig:TriangleExclusion} shows single $T'$ production cross section $\sigma_{T'}+\sigma_{\bar{T}'}$ which can be excluded with $100$ fb$^{-1}$ of data with a fixed $M_{T'} = 1$ TeV (left)  and $M_{T'} = 1.5$ TeV  (right). Again, this result is based on the three searches performed previously in this Section. The resulting exclusion bounds  amount to $\sigma_{T'}=$ 27 fb, 30 fb and 38 fb when $T'$ decays exclusively into $Wb$, $th$ and $Zt$ respectively for a 1 TeV search. Similarly, exclusion bounds from 1.5 TeV search are $\sigma_{T'}=$ 13 fb, 17 fb and 12 fb respectively. In a more general sense, we find that LHC Run II will be able to exclude cross sections between 27 fb and 60 fb (depending on the branching ratios) in case of a 1 TeV $T'$, while the exclusion range for the cross sections will range between 13 fb and 24 fb in case of a 1.5 TeV $T'$.

We find that the general trends and features of different $T'$ decay channel sensitivities are similar to the discovery reach analysis ($e.g.$ most sensitive channel etc.).
\section{Conclusions}\label{sec:conclusions}
Searches for fermionic top partner are essential probes of Naturalness and present an important aspect of the BSM experimental program at the LHC. 
In this paper we proposed strategies to search for TeV scale 2/3 charged verctor-like top partners ($T'$), and analyzed the prospects of LHC Run II to discover or rule out $T'$ models. Our analysis spans over all possible decays of the $T'$ ($i.e.$ $th$, $bW$, $tZ$), whereby we discussed in detail  optimal search strategies and LHC Run II sensitivities in each separate channel. 

As the mass limits are pushed to the TeV scale by the null result from LHC Run I, the single production of $T'$ will become one of the primary channels in searches for fermionic top partner at Run II of the LHC. The unique event topology of singly produced $T'$ offers a number of useful handles on SM backgrounds, including boosted heavy jets, large missing energy, $b$-jets and forward jets.  In order to tag heavy boosted SM states in the signal events and reduce the large SM backgrounds, we employed jet substructure techniques based on the Template Overlap Method, including a multi-dimensional TOM implementation, whereby a jet is characterized by a vector of observables quantifying the likelihood that it is a top, Higgs or a heavy SM vector boson. The use of multi-dimensional TOM had great impact in overall improvement of boosted $h \rightarrow b \bar{b}$ tagging and background suppression. We found that jet substructure delivers an improvement of a factor of $5-10$ in overall $S/B$, depending on the channel and $M_{T'}$, while forward jet tagging delivers an overall improvement of factor of $\sim 3$ in $S/B$, across all channels. The most challenging channel, $t_{\rm had}h_{bb} $ demands a complex  $b$-tagging strategy in order to have much hope of being discovered, where we found that $b$-tagging the top tagged jet, and double $b$-tagging the Higgs tagged jet provides the best overall signal significance at 100 $\fb^{-1}$ of integrated luminosity\footnote{Note that the signal in principle contains a fourth $b$ jet (the spectator third family quark) which in principle can be used to demand another isolated $b$-tag.}.

Even though we analyzed a specific implementation of $T'$ partners within the Composite Higgs scenario with a special case where only signet partner is present at the low scale,
 our results contain a minimal amount of model dependence. We presented a detailed analysis of the signal production cross section necessary to claim a discovery at LHC Run II with $100 \fb^{-1}$ as a function of the three $T'$ branching ratios, as well as the value of the cross section which can be ruled out by LHC Run II with $100 \fb^{-1}$. We found that (depending on the branching ratios) $T'$ models which predict a single production cross section between 70 and 140 fb for $M_{T'} = 1 \TeV$ could be discovered at the LHC with $100^{-1}$, while we find that models which predict production cross sections $\sim 30 - 65$ fb for $M_{T'} = 1.5$ TeV could be discovered with the same amount of data. The exclusion limits follow a similar pattern, where we find that LHC Run II will be able to exclude $T'$ models with cross sections $\sim 27 - 60$ fb ($\sim 13 - 24$ fb) for $M_{T'} = 1 (1.5) \TeV$ with $100 \fb^{-1}$ of data.

Both in discovery reach and exclusion, we found that the LHC Run II will be least sensitive to $T'$ models which predict $Br(T' \rightarrow Zt) \sim 0.5$ for $M_{T'} = 1 \TeV$ and $Br(T' \rightarrow th) \sim 0.5$ for $M_{T'} = 1.5 \TeV$. The sensitivity in different parts of the parameter space changes with the increase in $T'$ mass because of the change in kinematics of signal events with the increase in $T'$ mass and the consequent change in selection efficiencies. 

We presented a comprehensive discussion of potential final states of singly produced $T'$ and then focused on the final states deemed most sensitive for discovery of TeV scale $T'$ partners at LHC Run II, while we provide discussions of several sub-leading channels in the Appendix. Note that in some cases, such as the search for 1 TeV $T'$ in the $t_{\rm had} Z$, the analysis would clearly benefit from a combination of the di-lepton and MET channels, as they individually display similar sensitivity. 

Finally we emphasize that the searches for channels including a final state Higgs decaying to $b\bar{b}$ require efficient $b$-tagging of non-isolated $b$-jets. In our analysis we used a simplistic estimate of $b$-tagging efficiencies which is aimed to mimic the recently studied non-isolated $b$-tagging efficiencies of Ref.~\cite{ATLAS:2014} while for future experimental studies of these channels a more detailed investigation and  treatment of $b$-tagging in boosted objects with jet-substructure is required.

Future experimental studies would benefit of inclusion of pileup and detector effects into the analysis. Note however that Ref. \cite{Backovic:2013bga} has shown that performance of TOM for boosted jet tagging is weakly sensitive to pileup up to $\sim 70$ interactions per bunch crossing, while the forward jet tagging algorithm we employ is robust against pileup up to at least $\sim 50$ interactions per bunch crossing \cite{Backovic:2014uma}. Hence, it is likely that no aggressive pileup subtraction/correction technique will be necessary in our proposed search strategies, even at the high instantaneous luminosity expected at the advanced stages of LHC Run II.

\bigskip
\emph{Acknowledgements:}  We would like to thank Patrizia Azzi, Phillip Baringer, Kyongchul Kong, Heiko Lacker, Doug McKay, Devdatta Majumder, John Ralson, Alexander Schmidt, and Graham Wilson for useful discussions and suggestions during the course of this project. We are very grateful to Olivier Mattelaer for help and comments regarding Monte Carlo simulations. The authors would like to thank the Weizmann theory group and the organisers of the Naturalness 2014 workshop for the hospitality during the initial stages of this project. 
This work was supported by the National Research Foundation of Korea (NRF) grant funded by the Korea government (MEST) (No. 2012R1A2A2A01045722), the International Research \& Development Program of the National Research Foundation of Korea (NRF) funded by the Ministry of Science, ICT \& Future Planning (Grant number: 2015K1A3A1A21000234), and by the Basic Science Research Program through the National Research Foundation of Korea (NRF) funded by the ministry of Education, Science and Technology (No. 2013R1A1A1062597). This work is also supported by HTCaaS group of KISTI (Korea Institute of Science and Technology Information). SL and TF are also supported by Korea-ERC researcher visiting program through the National Research Foundation of Korea(NRF) (No. 2014K2a7B044399 and No. 2014K2a7A1044408). JHK is supported by the IBS Center for Theoretical Physics of the Universe. MB is in part supported by the Belgian Federal Science Policy Office through the Inter- university Attraction Pole P7/37.

\bigskip

\appendix

\begin{center}
\textbf{APPENDIX}
\end{center}



In the main text of this article we considered only the final states which showed to provide the highest sensitivity sensitive. As discussed in Sec.~\ref{sec:Production} several other final states could a priory provide similar performance. Our results for these channels are summarized in the  following Appendices.

\section{$T' \rightarrow Z_{l l} t_{\rm had}$ Channel }\label{app:Zt}

Refs. \cite{Backovic:2015lfa, Reuter:2014iya} studied the $Z_{ll} t_{\rm had}$ channel in much detail and showed that it is a promising channel for the  discovery of $T'$ partners of mass $\lesssim 1 \TeV$. Here we will give an overview of main features of the signal topologies and different ways one can suppress the large SM backgrounds. 

The main backgrounds for the $Z_{l l} t_{\rm had}$ channel are processes containing a $Z$ boson in the final state ($Z$, $Z b \bar{b}$, $Z + Z/W$, where we consider all of them with up to two additional jets), and $Z$ boson production associated with top quarks ($Z t$ and $Z t \bar{t}$, also with up to two additional jets). We categorize the background channels as $Z+X$ and $Z+t$ respectively, according to the rule introduced in Section \ref{sec:Zt}, with the exception of the semi-leptonic $t \bar{t}$, which we can be efficiently vetoed by demanding two isolated leptons which reconstruct a $Z$ boson. We simulate all  backgrounds with the preselection cuts described in Section \ref{sec:PreCuts} where we demand $H_T > 500~ (750)\GeV$ at event generation level for a hypothetical mass of the top partner $1~(1.5)$ TeV. Table \ref{tab:TotalBackGroundsZll} summarizes the background cross sections including a conservative K-factor of 2.

\begin{table}[h]
\begin{center}
\begin{tabular}{|c|c|c|c|}
\hline
		Signal Channel							&	Backgrounds 				                                & $\sigma (H_T > 500\GeV) [{\rm fb}]$ & $\sigma (H_T > 750\GeV) [{\rm fb}]$ \\ \hline
\multirow{6}{*}{$T' \rightarrow Z_{ll} t_{\rm had}$}	        &$Z_{ll}$ + jets 			                                        &  4570                 			           &	 1180                            			\\
											&$Z_{ll}$ + $b \bar{b}$ + jets                                       &  126 			                            &	 27.8                   				\\
											&$Z_{ll}$ + $t_{\rm had} \bar{t}_{\rm had}$ + jets       & 4.82 					           &	1.36						\\
											& $Z_{ll}$ + $t_{\rm had}$/$\bar{t}_{\rm had}$ + jets  & 7.83  					           &	2.12						\\
											&$Z_{ll}$ + $Z_{\rm had}$ + jets                                 & 22.5                             			   &	 7.0 						\\
											&$Z_{ll}$ + $W_{\rm had}$ + jets                                & 113                               			   &	 38.4      				\\ 
\hline		
\end{tabular}
\end{center}
\caption{The simulated cross sections of SM backgrounds (including a conservative K-factor estimate of 2 after preselection cuts described in Sec.~\ref{sec:PreCuts}.}
\label{tab:TotalBackGroundsZll}
\end{table}

Our Basic Cuts, summarized in Table \ref{tab:BasicCutstZll}, are based on a search strategy proposed in Ref. \cite{Reuter:2014iya}. We require at least two isolated leptons with $p_T^{l} > 25 \GeV$, and we require that a four vector constructed from the two leptons ($i.e.$ $p_{l_1}+ p_{l_2}$)  gives $p_T(l_1 + l_2) > 225 \GeV$ and $|\eta (l_1 +l_2) |< 2.3$. In addition, we require at least one fat jet ($R=1.0$) with $p_T^{\rm fj} > 400 \;(600)\GeV$ and $|\eta_{{\rm fj}} |< 2.5$, for $M_{T'} = 1 (1.5) \TeV$. 

\begin{table}[h!]
\setlength{\tabcolsep}{2em}
{\renewcommand{\arraystretch}{1.8}
\begin{tabular}{c|c}

							&$T' \rightarrow Z_{ll} t_{\rm had}$                                                      \\ \cline{1-2}
\multirow{3}{*}{ \textbf{Basic Cuts} }	&$N_{\rm fj} \geq 1$ ($R=1.0$), $N^{\rm iso}_{\rm lepton} \geq 2$ ,      \\ 
		 					&$p_T^{\rm fj} > 400 \;(600) \GeV$, $|\eta_{\rm fj} |< 2.5$ ,  		  \\ 
							&$p_T^{l_1 + l_2} > 225 \GeV$,~$|\eta_{l_1 + l_2}|< 2.3$ .	 \\ 
\end{tabular}} \par
\caption{Summary of Basic Cuts for $T' \rightarrow Z_{ll} t_{\rm had}$ channel . ``fj" stands for the fat jet with $|\eta_{\rm fj}| < 2.5$ and $p^{\rm fj}_T > 400\, (600) \GeV$ for $M_{T'} = 1 (1.5) \TeV$. $N^{\rm iso}_{\rm lepton}$ represents the number of isolated leptons with mini-ISO $> 0.7$, $p_T^{l} > 25\GeV$ and $|\eta_{l} |< 2.5$. ``$l_{1, 2}$" stands for two hardest, isolated leptons.} \label{tab:BasicCutstZll} 
\end{table}

As a part of the Complex Cuts (see Table \ref{tab:ComplexCutsZll} for a summary), we demand the hardest fat jet to pass $Ov$ selection of $Ov_{3}^t > 0.6$, to tag the top quark. Since two hardest leptons from the boosted $Z$ boson are collimated, we put  a tight cut on their angular separation of $\Delta R_{l l} < 1.0$, and impose a strict condition on the reconstructed mass $m_{l l}$ of the di-lepton pair to fall into the window of a true $Z$ boson mass, $|m_{ll} - m_Z| < 10\GeV$. Finally, we require at least one $r=0.2$ forward jets ($p_T^{\rm fwd} > 25 \GeV$ and $2.5 < \eta^{\rm fwd} < 4.5$), as well as at least 1 $b$-tag on the fat jet under the simplified $b$-tagging rules described in the section \ref{sec:b-tagging}.

\begin{table}[h]
\setlength{\tabcolsep}{2em}
{\renewcommand{\arraystretch}{1.8}
\begin{tabular}{c|c}
							   &$T' \rightarrow Z_{ll} t_{\rm had}$                   \\ \cline{1-2}
\multirow{5}{*}{ \textbf{Complex Cuts} }&$Ov_3^t > 0.6$ ,                   \\ 
		 					  &$|m_{l l} - m_z| < 10\GeV$ , 	                                        \\ 
		 					  &$\Delta R_{l l} < 1.0$ ,			   \\ 
							  &$N^{\rm fwd} \ge 1$ ,                           	    	\\ 
							  &fat jet $b$-tag.	             \\
\end{tabular}}\par
\caption{Summary of Complex Cuts for $T' \rightarrow Z_{ll} t_{\rm had}$ channel. $Ov_3^t$ refers to the top tagging score with Template Overlap Method, $m_{l l}$ is a reconstructed mass out of two hardest leptons ($\Delta R_{l l}$ is an angular distance between them), $N^{\rm fwd}$ is the multiplicity of forward jets ($p_T^{\rm fwd} > 25 \GeV$ and $2.5 < \eta^{\rm fwd} < 4.5$), and $b$-tag refers to presence of at least one $b$-tagged $r=0.4$ jet inside the fat jet which is tagged as a top.} \label{tab:ComplexCutsZll} 
\end{table}

Table \ref{tab:cutflowZll} shows an example cutflow for a signal benchmark point and relevant backgrounds (we use the same parameter point as in the previous sections). For both 1 TeV and 1.5 TeV top partner searches, jet substructure method combined with fat jet $b$-tagging plays a dominant role improving $S/B$, where we see a  factor of $\sim 10$ improvement at a $ 50\%$  signal efficiency. Moreover, employing forward jet tagging gives an additional improvement in $S/B$ by a factor of $\sim 3-4$. 
Notice that the signal sensitivity  achieved in the $Z_{ll} t_{\rm had}$ channel in searches for a 1 TeV partner is comparable to the sensitivity we obtain in the $Z_{\rm inv} t_{\rm had}$ channel (see Table \ref{tab:cutflowZt}), the performance of di-lepton channels in searches for a 1.5 TeV partner is clearly inferior compared to the invisible $Z$ channel. The decrease in sensitivity is primarily due to the fact that $Br(Z\rightarrow l^+ l^-)$ is roughly three times smaller than $Br(Z\rightarrow \nu \bar{\nu}), $ which severely limits the observable cross section in the di-lepton channel and hence the sensitivity at fixed luminosity. Nonetheless, the $Z_{ll} t_{\rm had}$ channel has a strong virtue of offering one of the cleanest ways to reconstruct the top  partner mass and will hence always remain important in searches for $T'$ partners. In addition, the combination of the di-lepton and missing energy channels in searches for $T' \rightarrow Zt$ has good prospects of enhancing the signal sensitivity in the 1 - 1.5 TeV mass range. 

\begin{widetext}
\begin{center}
\begin{table*}[h]
\begin{tabular}{|c||ccc|cc||ccc|cc|}
\hline
\multirow{2}{*}{$T' \rightarrow Z_{ll} t_{\rm had}$}&\multicolumn{5}{c||}{$M_{T'}=1.0\TeV$ search}&\multicolumn{5}{c|}{$M_{T'}=1.5\TeV$ search}\\\cline{2-11}
&signal&$Z+X$&$Z+t$&$S/B$&$S/\sqrt{B}$& signal &$Z+X$&$Z+t$&$S/B$&$S/\sqrt{B}$\\\hline
preselection            	         	&1.6             &4800   &13         &$3.3 \times 10^{-4}$       &0.23             &0.42           &1300        &3.5               &$3.3 \times 10^{-4}$     &0.12     \\
Basic Cuts                			&1.1             &750       &1.3      &0.0014        &0.39          &0.30          &170              &0.36             &0.0018      &0.23     \\
$Ov^t_{3} > 0.6$        		&0.71             &71        &0.61    &0.010        &0.85             &0.24         &19                &0.14               &0.012    &0.54     \\
$b$-tag                    			&0.49           &2.6       &0.40     &0.16         &2.8                &0.14         &0.64             &0.082            &0.19       &1.7   \\
$\Delta R_{ll}<1.0$	                 &0.49           &2.6       &0.39      &0.16       &2.8               &0.14         &0.64              &0.081          &0.20     &1.7      \\
$|m_{ll}-m_Z| < 10$ GeV		&0.44     &2.4       &0.35      &0.16 &2.7     &0.13  &0.58               &0.074           &0.19      &1.6\\

$N_\mathrm{fwd}\geq 1$      	&0.28     &0.38     &0.10      &0.58   &4.0     &0.084   &0.098            &0.018           &0.72      &2.5\\\hline
\end{tabular}

\caption{Example-cutflow for signal and background events in the $T' \rightarrow Z_{ll} t_{\rm had}$ channel for $\sqrt{s} = 14 \, \mbox{TeV}$. Cross sections after the respective cuts for signal and backgrounds are given in fb. The $S/\sqrt{B}$ values are given for a luminosity of $100\, \mathrm{fb}^{-1}$ for both the $M_{T'}=1.0\TeV$ and $M_{T'}=1.5\TeV$ searches.}\label{tab:cutflowZll}
\end{table*}
\end{center}
\end{widetext}

\section{$T' \rightarrow W_{\rm had} b$ Channel }\label{app:Wb}

The $W_{\rm had} b$ channel is perhaps the most challenging final state in $T'$ searches to observe. The fully hadronic final state suffers from enormous SM backgrounds, with only the single $b$-tag offering prospects for a significant improvement in $S/B$. The main backgrounds for the $W_{\rm had} b$ channel consist of QCD multi-jet, $b\bar{b}$ + jets \footnote{Here we include $b\bar{b}$ + jets originating from a pure QCD process as well as $Z_{b b}$ + jets.} and $t_{\rm had}\bar{t}_{\rm had}$ + jets. We simulate all the backgrounds with the preselection cuts described in Section \ref{sec:PreCuts} where we demand $H_T > 850~ (1350)\GeV$ at event generation level for a hypothetical mass of the top partner $1~(1.5)\TeV$. Table \ref{tab:TotalBackGroundsWhad} summarizes the background cross sections including a conservative K-factor of 2.

\begin{table}[h]
\begin{center}
\begin{tabular}{|c|c|c|c|}
\hline
		Channels								&	Backgrounds 						& $\sigma (H_T > 850\GeV) [{\rm fb}]$& $\sigma (H_T > 1350\GeV) [{\rm fb}]$ \\ \hline
\multirow{3}{*}{ $T' \rightarrow W_{\rm had} b$}                	& 4 jets                                                               &     $4.2 \times 10^{6}$                         &   $3.8 \times 10^{5}$	                   \\
											&$b\bar{b}$ + jets                                               &    $5.1 \times 10^{4}$                        &  5800		                                   \\
											&$t_{\rm had}\bar{t}_{\rm had}$ + jets                &	8400	                                              &  900          		                         \\ 
\hline		
\end{tabular}
\end{center}
\caption{The simulated cross sections of SM backgrounds (including a conservative  NLO K-factor of 2) after preselection cuts described in Section~\ref{sec:PreCuts}.}
\label{tab:TotalBackGroundsWhad}
\end{table}

In this particular search, we consider a fat jet cone size of $R=0.7$ so as to improve the rejection of QCD backgrounds. As a part of the Basic Cuts we demand at least one fat jet ($R=0.7$) with $p_T^{\rm fj} > 400 \;(600)\GeV$ and $|\eta_{{\rm fj}} |< 2.5$  (see Table \ref{tab:BasicCutsWhad} for summary).

\begin{table}[h]
\setlength{\tabcolsep}{2em}
{\renewcommand{\arraystretch}{1.8}
\begin{tabular}{c|c}

							 &  $T' \rightarrow W_{\rm had} b$	   			                                     \\ \cline{1-2}
\multirow{2}{*}{ \textbf{Basic Cuts} }	 &  $N_{\rm fj}  \geq 1$ ($R=0.7$)  ,     $N^{iso}_{lepton} = 0 $  ,       \\ 
		 					 &  $p_T^{\rm fj} > 400 \;(600) \GeV$, $|\eta_{\rm fj} |< 2.5$.       \\ 
\end{tabular}}\par
\caption{Summary of Basic Cuts for $T' \rightarrow W_{\rm had} b$ channel. ``fj" stands for the fat jet with $|\eta_{\rm fj}| < 2.5$ and $p_T > 200 \,(400) \GeV$ for $M_{T'} = 1 (1.5) \TeV$ and $N^{\rm iso}_{\rm lepton}$ represents the number of isolated leptons with mini-ISO $> 0.7$, $p_T^{l} > 25\GeV$ and $|\eta_{l} |< 2.5$. } \label{tab:BasicCutsWhad} 
\end{table}

Our event selection continues with the Complex Cuts (see Table \ref{tab:ComplexCutsWhad}), first by demanding exactly one $W$ fat jet (the hardest jet satisfying $Ov$-selection criterion: $Ov_{2}^W > 0.5$ and $Ov_{3}^t < 0.6$), and then by requiring the hardest light jet ($r=0.4$) that is isolated from the $W$ fat jet (by $\Delta R > r + R $) to pass a cut of $p_T^{j} > 150\;(300) \GeV$ and $| \eta^{j} |< 2.5$ for $M_{T'} = 1 (1.5) \TeV$ and to be  $b$-tagged under the criteria described in the Section \ref{sec:b-tagging}. The top partner mass can be reconstructed by combining the $W$ fat jet and the hardest $r=0.4$ $b$ jet (isolated from the $W$ fat jet), where we impose a lower mass bound of $M_{T'} > 600\; (900) \GeV$. Finally, we require at least one $r=0.2$ forward jet ($p_T^{\rm fwd} > 25 \GeV$ and $2.5 < \eta^{\rm fwd} < 4.5$).

\begin{table}[h]
\setlength{\tabcolsep}{2em}
{\renewcommand{\arraystretch}{1.8}
\begin{tabular}{c|c}

							    &  $T' \rightarrow W_{\rm had} b$   			  \\ \cline{1-2}
\multirow{5}{*}{ \textbf{Complex Cuts} }&  $N_{W} = 1$ ($Ov_3^t < 0.6$ \& $Ov_2^W > 0.5$) ,                                \\ 
		 					   & $p_T^{j} > 150 \; (300) \GeV$, $| \eta^{j} |< 2.5$ ,        \\ 
		 					  &   $M_{T'} > 600 \; (900) \GeV$ ,	                		\\ 
							  &   $N^{\rm fwd} \ge 1$,        			\\
							  &   $b$-tag.                                           \\
\end{tabular}} \par
\caption{The summary of Complex Cuts for $T' \rightarrow W_{\rm had} b$ channels. ``$Ov$'' selection applies to the highest $p_T$ fat jet ($R=0.7$) in the event, and $N_{W}$ is the number of tagged $W$ fat jet. The label ``$j$'' refers to the hardest $r=0.4$ jet isolated from the $W$ fat jet, $N^{\rm fwd}$ is the multiplicity of forward jets ($p_T^{\rm fwd} > 25 \GeV$ and $2.5 < \eta^{\rm fwd} < 4.5$), and $b$-tag refers to at least 1 $b$-tag on the hardest $r=0.4$ jet. The values outside (inside) parenthesis show the choice of cuts for $1\; (1.5) \TeV$ $T'$ searches. } \label{tab:ComplexCutsWhad} 
\end{table}

Table \ref{tab:cutflowWhad} shows an example cutflow for a signal benchmark point and relevant backgrounds (we use the same parameter point as in the previous sections). In a nutshell, it is hard to avoid a severe contamination from QCD multi-jet backgrounds, as the  $W$ tagger without an aid of $b$-tagging shows weak improvement in $S/B$. Note that the relatively weak performance of $W$ taggers (compared to for example boosted top taggers) is not an artifact of Template Overlap, but of boosted $W$ tagging in general. Requirements on the reconstructed $M_{T'}$ and $p^j_T$ of the isolated $r=0.4$ jet do not provide much rejection power as requiring a highly boosted $W$ and a high $p_T$ $b$-jet already selects background events in the high $Wb$ mass regime. $b$-tagging provides the greatest improvement in $S/B$, while the forward jet-tagging gives an improvement of $\sim 3$ in $S/B$ as in all other channels. The present analysis indicates that probing the $W_{\rm had} b$ channel will likely be very challenging until late stages of the LHC Run II. 

\begin{widetext}
\begin{center}
\begin{table*}[!]
\begin{tabular}{|c||cccc|cc||cccc|cc|}
\hline
\multirow{2}{*}{$T' \rightarrow W_{\rm had} b$}&\multicolumn{6}{c||}{$M_{T'}=1.0\TeV$ search}&\multicolumn{6}{c|}{$M_{T'}=1.5\TeV$ search}\\ \cline{2-13}
                                                & signal       & $4\; jets$                    & $b\bar{b}+jets$ & \hspace{5pt} $t\bar{t} \hspace{5pt}$  & $S/B$         & $S/\sqrt{B}$ & signal   &$4\; jets$ & $b\bar{b}+jets$   & \hspace{5pt} $t\bar{t} \hspace{5pt}$  & $S/B$       & $S/\sqrt{B}$\\\hline
preselection                             &96            &$4.2 \times 10^{6}$     &$5.1 \times 10^{4}$              &8400                                                    &$2.3 \times 10^{-5}$   &0.47                                                     &18          &$3.8 \times 10^{5}$     & 5800                  &900                                                     &$4.6 \times 10^{-5}$    &0.29   \\
Basic Cuts                                 &83           &$2.6 \times 10^{6}$     &$3.4 \times 10^{4}$              &4800                                                    &$3.1 \times 10^{-5}$  &0.51                                                      &16         &$2.7 \times 10^{5}$     & 4200                 &750                                                      &$5.9 \times 10^{-5}$    &0.31  \\
$Ov$ selection                         &55             &$1.2 \times 10^{6}$     &$1.4 \times 10^{4}$             &2600                                                    &$4.6 \times 10^{-5}$     &0.50                                                     &10        &$1.4 \times 10^{5}$    & 2000                   &340                                                     &$7.3 \times 10^{-5}$    &0.27  \\
$p_T^{j} > 150  \; (300)\GeV$  &54            &$1.2 \times 10^{6}$     &$1.3 \times 10^{4}$            &2500                                                    &$4.6 \times 10^{-5}$    &0.50                                                     &9.8       &$1.3 \times 10^{5}$      & 1900                  &300                                                      &$7.1 \times 10^{-5}$   &0.26   \\
$M_{T'} > 600 \; (900)\GeV$       & 54           &$1.2 \times 10^{6}$     &$1.3 \times 10^{4}$           &2300                                                   &$4.6 \times 10^{-5}$   &0.50                                                     & 9.7       &$1.3 \times 10^{5}$       &  1900                  &300                                                      &$7.1 \times 10^{-5}$  & 0.26       \\
$N_{\mathrm{fwd}}\geq 1$      & 33         &$1.9 \times 10^{5}$       &    3000                             &570                                                     & $1.7 \times 10^{-4}$      & 0.74                                                    & 6.2       &$2.2 \times 10^{4}$        &  410                   &70                                                     & $2.7 \times 10^{-4}$    & 0.41      \\
$b$-tag                                    & 20          & 4000                            &     790                              &152                                                     &0.0042                         & 2.9                                                & 3.7       &460                                  & 100                    &22                                                       &0.0064                        & 1.5   \\
\hline
\end{tabular}
\caption{Example cutflow for signal and background events in the $T' \rightarrow W_{\rm had} b$ search for $\sqrt{s} = 14 \, \mbox{TeV}$. Cross sections after the respective cuts for signal and backgrounds are given in fb. The $S/\sqrt{B}$ values are given for a luminosity of $100\, \mathrm{fb}^{-1}$ for both the $M_{T'}=1.0\TeV$ and $M_{T'}=1.5\TeV$ searches.}\label{tab:cutflowWhad}
\end{table*}
\end{center}
\end{widetext}


\section{$T' \rightarrow t_{\rm lep} h_{bb}$ Channel }
\label{sec:thlep}

In searches for TeV scale BSM physics, decays of the Higgs boson into states other than $b\bar{b}$ are unlikely to yield significant signal sensitivity, as the branching ratio to other states is so small that an enormous amount of integrated luminosity is likely to be required in order to see a sufficient number of signal events, even in the case of the very clean di-gamma signature of the Higgs decay. Hence, the only $T' \rightarrow th$ final state other than $h_{\rm had} h_{bb}$ that has a potential of giving significant signatures of $T'$ decays at the LHC Run II is the $t_{\rm lep} h_{bb}$.

The only dominant SM background for the $t_{\rm lep} h_{bb}$ channel is SM semi-leptonic $t \bar{t}$ + jets. We neglect all other background channels, as we have checked that in the signal region they are effectively vetoed by our event selection.

\begin{table}[h]
\setlength{\tabcolsep}{0.5em}
{\renewcommand{\arraystretch}{1.7}
\begin{tabular}{|c|c|c|c|c|}
\hline
		Channels								&	Backgrounds 						& $\sigma (H_T > 500\GeV) [{\rm fb}]$& $\sigma (H_T > 750\GeV) [{\rm fb}]$	\\ \hline
$T' \rightarrow t_{\rm lep} h_{bb}$                   		&$t\bar{t}$(semi-leptonic) + jets                                 &	$2.1 \times 10^{4}$	                     &   4200			   \\ 
\hline		
\end{tabular}}\par
\caption{The simulated cross sections of SM backgrounds (including a conservative  NLO K-factor of 2) after preselection cuts described in Section~\ref{sec:PreCuts}.}
\label{tab:TotalBackGrounds}
\end{table}

The Basic Cuts in the $t_{\rm lep } h_{b b}$ channel consist of requiring exactly one isolated lepton with $p_T^{l} > 25 \GeV$, as well as demanding at least one fat jet ($R=1.0$) with $p_T^{\rm fj} > 400 \;(600)\GeV$ and $|\eta_{{\rm fj}} |< 2.5$ (see Table \ref{tab:BasicCutsthlep} for summary).

\begin{table}[h]
\setlength{\tabcolsep}{2em}
{\renewcommand{\arraystretch}{1.8}
\begin{tabular}{c|c}
							& $T' \rightarrow t_{\rm lep} h_{bb}$      	    	    	    	    	   \\ \cline{1-2}
\multirow{3}{*}{ \textbf{Basic Cuts} }	& $N_{\rm fj}  \geq 1$ ($R=1.0$), $N^{\rm iso}_{\rm lepton} = 1$ ,          \\ 
		 					& $p_T^{\rm fj} > 400 \;(600) \GeV$, $|\eta_{\rm fj} |< 2.5$.	  \\ 
\end{tabular}}\par
\caption{Summary of Basic Cuts for $T' \rightarrow t_{\rm lep} h_{bb}$ channel. ``fj" stands for the fat jet and $N^{\rm iso}_{\rm lepton}$ represents the number of isolated leptons with mini-ISO $> 0.7$, $p_T^{l} > 25\GeV$ and $|\eta_{l} |< 2.5$. The values outside (inside) the parenthesis refer to $1\; (1.5) \TeV$ $T'$ searches respectively.} \label{tab:BasicCutsthlep} 
\end{table}

As a part of the Complex Cuts, we demand exactly one Higgs (the hardest jet satisfying $Ov$-selection criterion: $Ov_{2}^h > 0.5$ and $Ov_{3}^t < 0.6$). As the main background channel contains significant missing energy and a hard lepton, we impose only a minimal cut on the isolated lepton and $\MET$ of  $p_T^{l} > 25, \MET > 20$. Analogous to Section \ref{sec:Wb}, we can reconstruct the mass of $T'$ by using the Higgs tagged fat jet, the isolated lepton, $\MET$ and a hardest $r=0.4$ jet with $p_T > 50 \GeV$ (while simultaneously demanding that the jet be located within $\Delta R_{j,\,l} < 1.0$ from the lepton) in the collinear approximation of $\eta_\nu = \eta_l$. We impose a lower mass bound by $M_{T'} > 750\; (1000) \GeV$. 

As already noted in Section \ref{sec:th}, the $b$-tagging strategy takes on the greatest part in obtaining a better signal sensitivity in the $T'\rightarrow th$ channel. This still holds true for the $t_{\rm lep} h_{bb}$ because in spite of not being plagued by QCD background (which is eliminated by requiring a final state lepton) the $b$-tagging still is important for the reduction of the $t\bar{t}$ background. For a leptonic top to be $b$-tagged, we require that $b$-tagged $r=0.4$ jets land in $\Delta R_{j,\,l} < 1.0$ from the lepton axis (in the similar manner described in Section \ref{sec:b-tagging}). We present three ways of $b$-tagging: at least 1 $b$-tag on the top and at least 1 $b$-tag on the Higgs (Case 1) ,  at least 1 $b$-tag on the top and exactly 2 $b$-tags on the Higgs (Case 2) and  exactly 2 $b$-tags on the Higgs (Case 3).

As in other channels, we also require at least one $r=0.2$ forward jets ($p_T^{\rm fwd} > 25 \GeV$ and $2.5 < \eta^{\rm fwd} < 4.5$) (see Table \ref{tab:ComplexCutsthlep} for summary).

\begin{table}[htb]
\setlength{\tabcolsep}{2em}
{\renewcommand{\arraystretch}{1.8}
\begin{tabular}{c|c}

							    & $T' \rightarrow t_{\rm lep} h_{bb}$      	    	     \\ \cline{1-2}
\multirow{4}{*}{ \textbf{Complex Cuts} } & $N_{h} = 1$ ($Ov_3^t < 0.6$ \& $Ov_2^h > 0.5$) ,        \\ 
		 					  & $p_T^{l} > 25, \MET > 20$ ,	    	    	   	    \\ 
		 					  & $M_{T'} > 750 \; (1000)\GeV$ ,			            \\ 
							  & $N_{\mathrm{fwd}}\geq 1$ ,		  	\\ 
							  & $b$-tag (Case 1,2,3).                                              \\
\end{tabular}}\par
\caption{The summary of Complex Cuts for $T' \rightarrow t_{\rm lep} h_{bb}$ channel. ``$Ov$'' selection applies to the highest $p_T$ fat jet ($R=1.0$) in the event, and $N_{h}$ is the number of tagged Higgs fat jet. $b$-tag refers to (Case 1) at least 1 $b$-tag on the top and at least 1 $b$-tag on the Higgs, (Case 2) at least 1 $b$-tag on the top and exactly 2 $b$-tags on the Higgs and (Case 3) exactly 2 $b$-tags on the Higgs.} \label{tab:ComplexCutsthlep} 
\end{table}

\begin{widetext}
\begin{center}
\begin{table*}[h]
\begin{tabular}{|c||cc|cc||cc|cc|}
\hline
\multirow{2}{*}{$T' \rightarrow t_{\rm lep} h_{bb}$}&\multicolumn{4}{c||}{$M_{T'}=1.0$ TeV search}&\multicolumn{4}{c|}{$M_{T'}=1.5$ TeV search}\\ \cline{2-9}
                                                & signal       & \hspace{5pt} $t\bar{t} \hspace{5pt}$  & $S/B$     & $S/\sqrt{B}$ & signal     & \hspace{5pt} $t\bar{t} \hspace{5pt}$  & $S/B$      & $S/\sqrt{B}$\\\hline
preselection                             &13            &$2.1 \times 10^{4}$                            &$6.3 \times 10^{-4}$  &0.92                                                    &2.7           &4200                                                       &$6.6 \times 10^{-4}$    &0.43   \\
Basic Cuts                                 &5.9            &3700                                                  &0.0016                     &0.97                                                     &1.1            &650                                                     &0.0017       &0.44  \\
$Ov$ cut                                   &3.2            &520                                                   &0.0061                       &1.4                                                     &0.52            &76                                                     &0.0068       &0.59  \\
$p_T^{l} > 25\GeV, \; \MET > 20\GeV$       &3.0            &490                                                    &0.0061                      &1.4                                                      &0.51            &73                                                      &0.0069       &0.59  \\
$M_{T'} > 750 \; (1000)\GeV$               & 2.1          &290                                                     &0.0072                   &1.2                                                      & 0.45           &50                                                     & 0.0089       & 0.63       \\
$N_{\mathrm{fwd}}\geq 1$       & 1.3          &60                                                       & 0.021                      & 1.6                                                      & 0.29          &11                                                        & 0.026        & 0.86      \\\hline
\bf{Case 1}                               &                 &                                                            &                              &                                                           &                  &                                                             &                  &        \\\hline
at least 1 $b$-tag on  $t$ and at least 1 $b$-tag on $h$ & 0.78    &     12                                 & 0.068                  &2.3                                         &  0.15    &1.2                                                        & 0.13         & 1.4   \\\hline
\bf{Case 2}                               &                &                                                             &                              &                                                          &                  &                                                             &                 &        \\\hline
at least 1 $b$-tag on $t$ and exactly 2 $b$-tags on $h$ & 0.39  &     0.74                                 & 0.52                 &4.5                                               & 0.037  &0.036                                                   & 1.0            & 1.9   \\\hline
\bf{Case 3}                               &                &                                                              &                            &                                                          &                   &                                                             &                 &        \\\hline
exactly 2 $b$-tags on $h$                     & 0.56  &     1.1                                                 & 0.53                    &5.5                                              & 0.053   &0.057                                                    & 0.93         & 2.2   \\

\hline
\end{tabular}
\caption{Example-cutflow for signal and background events in the $T' \rightarrow t_{\rm lep} h_{bb}$ for $\sqrt{s} = 14 \, \mbox{TeV}$. Cross sections after the respective cuts for signal and backgrounds are given in fb. The $S/\sqrt{B}$ values are given for a luminosity of $100\, \mathrm{fb}^{-1}$ for the $M_{T'}=1.0$ TeV and $M_{T'}=1.5$ TeV searches.}\label{tab:cutflowthlep}

\end{table*}
\end{center}
\end{widetext}

Table \ref{tab:cutflowthlep} shows an example cutflow for a signal benchmark point and relevant backgrounds (we use the same parameter point as in the previous sections). We find that $Ov$ selection cut gives a factor of $\sim 4$ improvement in $S/B$ at the cost of $\sim 50\%$ signal efficiency, while forward jet tagging improves $S/B$ by an additional factor of 3. The largest improvement in $S/B$ comes from the $b$-tagging strategy, where we find that the Case 3 strategy (exactly 2 $b$-tags on the Higgs) provides the greatest signal significance at $100 \fb^{-1}$. This is in contrast with our analysis of $t_{\rm had} h_{bb}$ where we found that 2 $b$-tags on the Higgs and 1 $b$-tag on the top was the optimal strategy. The other cases which involve a 1 $b$-tag on the top yield lower improvements on overall signal significance as the the $t \bar{t}$ background also contains the same top. Although the $t_{\rm had} h_{bb}$ channel out-performs the $t_{\rm lep} h_{bb}$ channel, we find that a signal significance of $5\sigma$, sufficient for discovery, is still possible in the case of $\sim 1$~TeV top partner. It hence might be important to include the $t_{\rm lep} h_{bb}$ channel into the analysis, as it could significantly improve the overall significance of the possible $T'$ signal.


\section{Comparison of different b-tagging strategies in the $t_{\rm had} h_{bb}$ Channel}\label{app:ht}

When considering search strategies, it is important to keep in mind that the signal significance at fixed luminosity in searches for TeV scale new physics with femto-barn cross sections is ultimately limited by the tiny magnitude of the signal cross sections. Hence, ``overcutting'' the signal, even though it might suppress the backgrounds, could ultimately lead to diminishing the signal to the point where not a sufficient number of signal events could be observed. As a single proper $b$-tag in our proposal comes with a $70\%$ signal efficiency, it is not obvious which $b$-tagging strategy should perform the best in searches for $T' \rightarrow t_{\rm had} h_{bb}$. 

Continuing the $b$-tagging discussion in Section \ref{sec:th}, here we compare performance of different $b$-tagging strategies defined as: at least 1 $b$-tag on the top and at least 1 $b$-tag on the Higgs  (Case 1), at least 1 $b$-tag on the top and exactly 2 $b$-tags on the Higgs (Case 2)  and exactly 2 $b$-tags on the Higgs  (Case 3).  For completeness, we also provide the information on performance of other handles on SM backgrounds, such as jet substructure and forward jet tagging, a detailed discussion of which can be found in Section \ref{sec:th}. Our results are shown in Table~\ref{tab:cutflowbb}. 

In case of $t_{\rm had} h_{bb}$, the most aggressive $b$-tagging strategy (Case 2)  gives the best signal significance, both in the case of 1 and 1.5 TeV. Note however, that if we instead considered $t_{\rm lep} h_{bb}$, which is characterized by a significantly smaller signal cross section, the highest significance comes from the Case 3 $b$-tagging scheme, as requiring 3 $b$-tags in this case would ``overcut'' the cross signal cross section.

Even though we have not explicitly checked the performance of the $b$-tagging schemes for $T'$ masses of $\gtrsim 1.5 \TeV$, the pattern of differences between the $t_{\rm had} h_{bb}$ and the $t_{\rm lep} h_{bb}$ channels suggests that it is likely that the aggressive $b$-tagging strategy of Case 3 will not be optimal for higher masses. 

\begin{widetext}
\begin{center}
\begin{table*}[h]
\begin{tabular}{|c||cccc|cc|}
\hline
\multirow{2}{*}{$T' \rightarrow t_{\rm had} h_{bb}$}&\multicolumn{6}{c|}{$M_{T'}=1.0$ TeV search} \\ \cline{2-7}
                                                & signal       & $4\; jets$   & $b\bar{b}+jets$ & \hspace{5pt} $t\bar{t} \hspace{5pt}$  & $S/B$         & $S/\sqrt{B}$ \\\hline
preselection                             &27            &$4.2 \times 10^{6}$      &$5.1 \times 10^{4}$    &8400                                                    &$6.5 \times 10^{-6}$    &0.13                                                     \\
Basic Cuts                                 &21           &$2.6 \times 10^{6}$      &$3.2 \times 10^{4}$     &6400                                                    &$7.8 \times 10^{-6}$       &0.13                                                   \\
$Ov$ cut                                  &9.1             &$8.7 \times 10^{4}$     &  1300                         &1200                                                    &$1.0 \times 10^{-4}$      &0.30                                                      \\
$M_{T'} > 750 \; (1000)\GeV$   & 9.0           &$8.7 \times 10^{4}$     &   1300                         &1200                                                     &$1.0 \times 10^{-4}$      &0.30                                                   \\
$N_{\mathrm{fwd}}\geq 1$       & 5.5           &$1.4 \times 10^{4}$     &    270                             &280                                                       &$3.7 \times 10^{-4}$     & 0.45                                                    \\\hline
\bf{Case 1}                               &                 &                   &                          &                                                            &                   &                                                              \\\hline
at least 1 $b$-tag t and at least 1 $b$-tag h & 3.4          & 7.8             &     3.0                & 67                                         &0.043           & 3.8                                                    \\\hline
\bf{Case 2}                               &                 &                   &                          &                                                            &                   &                                                              \\\hline
at least 1 $b$-tag t and exactly 2 $b$-tags h & 1.6          & 0.12              &     0.15            & 4.1                                        &0.37             & 7.7                                                    \\\hline
\bf{Case 3}                               &                 &                   &                          &                                                            &                   &                                                              \\\hline
exactly 2 $b$-tags h                      & 2.3          & 4.0             &     5.5               & 6.4                                                     &0.14              & 5.7                                             \\
\hline
\end{tabular}

\vspace{10pt}
\begin{tabular}{|c||cccc|cc|}
\hline
\multirow{2}{*}{$T' \rightarrow t_{\rm had} h_{bb}$}&\multicolumn{6}{c|}{$M_{T'}=1.5$ TeV search}\\ \cline{2-7}
                                               & signal   &$4\; jets$         & $b\bar{b}+jets$   & \hspace{5pt} $t\bar{t} \hspace{5pt}$  & $S/B$       & $S/\sqrt{B}$\\\hline
preselection                            &4.5        &$3.8 \times 10^{5}$           & 5800                  &900                                                     &$1.2 \times 10^{-5}$    &0.072   \\
Basic Cuts                               &4.1        &$3.0 \times 10^{5}$           & 4700                  &850                                                      &$1.4 \times 10^{-5}$   &0.074  \\
$Ov$ cut                                 &1.9        &$2.1 \times 10^{4}$          & 340                   &110                                                       &$8.7 \times 10^{-5}$  &0.13  \\
$M_{T'} > 750 \; (1000)\GeV$         & 1.9       &$2.1 \times 10^{4}$      &  340                  &110                                                      &$8.9 \times 10^{-5}$  & 0.13       \\
$N_{\mathrm{fwd}}\geq 1$      & 1.2       &3800                &  77                     &27                                                        &$3.2 \times 10^{-4}$    & 0.20      \\\hline
\bf{Case 1}                               &                 &                   &                          &                                                            &                   &                                                              \\\hline
at least 1 $b$-tag t and at least 1 $b$-tag h & 0.62       &1.8       & 0.71          &3.5                                                     &0.10             & 2.5                                                    \\\hline
\bf{Case 2}                               &                 &                   &                          &                                                            &                   &                                                              \\\hline
at least 1 $b$-tag t and exactly 2 $b$-tags h & 0.15       &0.029           & 0.018     &0.18                                                   &0.66              & 3.2                                                    \\\hline
\bf{Case 3}                               &                 &                   &                          &                                                            &                   &                                                              \\\hline
exactly 2 $b$-tags t                     & 0.24      &1.1              & 0.58                     &0.35                                                      &0.12            & 1.7                                                   \\\hline

\end{tabular}
\caption{Example-cutflow for signal and background events in the $T' \rightarrow t_{\rm had} h_{bb}$ channel for $\sqrt{s} = 14 \, \mbox{TeV}$. Cross sections after the respective cuts for signal and backgrounds are given in fb. The $S/\sqrt{B}$ values are given for a luminosity of $100\, \mathrm{fb}^{-1}$ for the $M_{T'}=1.0$ TeV and $M_{T'}=1.5$ TeV searches.}
\label{tab:cutflowbb}
\end{table*}
\end{center}
\end{widetext}

\newpage
\bibliography{draft_paper}
\end{document}